%% file: phd_thesis.tex
\documentclass[ce,plainnat,dissertation]{puthesis}

\usepackage{amsmath}
\usepackage{amssymb}
\usepackage{multicol}
\usepackage{subfigure}

\title{Markov Switching Models:\\
    an Application to Roadway Safety}

\author{Nataliya V. Malyshkina}{Malyshkina, Nataliya V.}

\pudegree{Doctor of Philosophy}{Ph.D.}{December}{2008}

\majorprofs{Fred L. Mannering and Andrew P. Tarko}

\campus{West Lafayette}

\def  \betabf  {{\mbox{\boldmath$\beta$}}}
\def  \gammabf  {{\mbox{\boldmath$\gamma$}}}
\def  \gammabfscript  {{\mbox{\scriptsize\boldmath$\gamma$}}}
\def  \thetabf  {{\mbox{\boldmath$\theta$}}}

\begin{document}

%BEGIN OUR FORMAT
\makeatletter
\def\@@makechapterhead#1{\protect\uppercase{CHAPTER}~\thechapter. #1}
\makeatother
%END OUR FORMAT

\volume

% Front matter (dedication, etc.).
\include{ch_front}

% Include chapters
\include{ch_intro}

\include{ch_literature}

\include{ch_model_specif}

\include{ch_model_estim}
\include{ch_MCMC_simul}

\include{ch_results_freq}

\include{ch_results_sev}

\include{ch_conclusions}
\include{ch_bibliography}

\include{ch_appendix}

\include{ch_vita}

\end{document}

%% file: ch_front.tex
%
% This is the front page text and the abstract
%

\begin{dedication}
To my husband Leonid and my parents Nadezhda and Vladimir
\end{dedication}

\begin{acknowledgments}
First of all, I would like to thank my advisor, Professor Fred Mannering. Without his interest, encouragement and financial assistance none of this research would be possible. He supported me during all my three and a half years at Purdue. He also gave me a lot of freedom in research. I feel very lucky to be his student.

I would like to thank Professor Andrew Tarko, my co-advisor, for his very helpful comments and encouragement. I am especially grateful to him for the accident frequency data that he provided for the study reported in this dissertation.

In addition, I would like to thank Jose Thomaz for preparing the accident severity data that was used in this study.

I would like to thank Professor Kristofer Jennings and Professor Jon Fricker for their helpful comments and for carefully reading this dissertation. I also would like to thank my colleagues and friends for their help and support during my stay on Purdue campus.

Finally, I feel infinite love and gratitude to my wonderful family -- my husband Leonid, my mother Nadezhda and my father Vladimir. I owe everything I have to them and to their love and support.

\end{acknowledgments}

\tableofcontents

\listoftables

\listoffigures

\begin{abbreviations}
AADT& Average Annual Daily Traffic\cr
AIC& Akaike Information Criterion\cr
BIC& Bayesian Information Criterion\cr
BTS& Bureau of Transportation Statistics\cr
i.i.d.& independent and identically distributed\cr
MCMC& Markov Chain Monte Carlo\cr
M-H& Metropolis-Hasting\cr
ML& Multinomial logit\cr
MLE& Maximum Likelihood Estimation\cr
MS& Markov Switching\cr
MSML& Markov Switching Multinomial Logit\cr
MSNB& Markov Switching Negative Binomial\cr
MSP& Markov Switching Poisson\cr
NB& Negative Binomial\cr
PDO& Property Damage Only\cr
ZINB& Zero-inflated Negative Binomial\cr
ZIP& Zero-inflated Poisson\cr
\end{abbreviations}

\begin{abstract}
\vspace{3.5pt}
In this research, two-state Markov switching models are proposed to study accident frequencies and severities. These models assume that there are two unobserved states of roadway safety, and that roadway entities (e.g., roadway segments) can switch between these states over time. The states are distinct, in the sense that in the different states accident frequencies or severities are generated by separate processes (e.g., Poisson, negative binomial, multinomial logit). Bayesian inference methods and Markov Chain Monte Carlo (MCMC) simulations are used for estimation of Markov switching models. To demonstrate the applicability of the approach, we conduct the following three studies.

In the first study, two-state Markov switching count data models are considered as an alternative to zero-inflated models, in order to account for preponderance of zeros typically observed in accident frequency data. In this study, one of the states of roadway safety is a zero-accident state, which is perfectly safe. The other state is an unsafe state, in which accident frequencies can be positive and are generated by a given counting process -- a Poisson or a negative binomial. Two-state Markov switching Poisson model, two-state Markov switching negative binomial model, and standard zero-inflated models are estimated for annual accident frequencies on selected Indiana interstate highway segments over a five-year time period. An important advantage of Markov switching models over zero-inflated models is that the former allow a direct statistical estimation of what states specific roadway segments are in, while the later do not.

In the second study, two-state Markov switching Poisson model and two-state Markov switching negative binomial model are estimated using weekly accident frequencies on selected Indiana interstate highway segments over a five-year time period. In this study, both states of roadway safety are unsafe. Thus, accident frequencies can be positive and are generated by either Poisson or negative binomial processes in both states. It is found that the more frequent state is safer and it is correlated with better weather conditions. The less frequent state is found to be less safe and to be correlated with adverse weather conditions.

In the third study, two-state Markov switching multinomial logit models are estimated for severity outcomes of accidents occurring on Indiana roads over a four-year time period.  It is again found that the more frequent state of roadway safety is correlated with better weather conditions. The less frequent state is found to be correlated with adverse weather conditions.

One of the most important results found in each of the three studies, is that in each case the estimated Markov switching models are strongly favored by accident frequency and severity data and result in a superior statistical fit, as compared to the corresponding standard (single-state) models.

\end{abstract}

%% file: ch_intro.tex
%
%  This is the introduction chapter.
%

\chapter{Introduction}

%\phantom{$\int\limits^|_|$}

This chapter explains the motivation and objectives of the present research, and the organization of this dissertation.

%\addtocounter{section}{+1}
%\begin{center}
%\section*{\normalfont\arabic{chapter}.\arabic{section}~~
%\underline{Motivation and research objectives}}
%\end{center}
%\addcontentsline{toc}{section}
%{\arabic{chapter}.\arabic{section}~~
%Motivation and research objectives}

%\begin{center}
\section{Motivation and research objectives}
%\end{center}
According to Bureau of Transportation Statistics \citep[][]{BTS}, in 2006, $99.55\%$ of all transportation related accidents (including air, railroad, transit, waterborne and pipeline accidents) were motor vehicle accidents on roadways. Motor vehicle accidents result in fatalities, injuries and property damage, and represent high cost not only for involved individuals but also for our society as a whole. In particular, on average, about one-quarter of the costs of crashes is paid directly by the party involved, while the society pays the rest. As an example of the economic burden related to motor vehicle crashes, in the year 2000 the estimated cost of accidents occurred in the United States was 231 billion dollars, which is about 820 dollars per person or 2 percent of the gross domestic product \citep[][]{BTS}. These numbers show that roadway vehicle travel safety has an enormous importance for our society and for the national economy. As a result, extensive research on roadway safety is ongoing, in order to better understand the most important factors that contribute to vehicle accidents.

In general, there are two measures of roadway safety that are commonly considered:
\begin{enumerate}
    \item
    The first measure evaluates accident frequencies on roadway segments. Accident frequency on a roadway segment is obtained by counting the number of accidents occurring on this segment during a specified period of time. Then count data statistical models (e.g. Poisson, negative binomial models and their zero-inflated counterparts) are estimated for accident frequencies on different roadway segments. The explanatory variables used in these models are the roadway segment characteristics (e.g. roadway segment length, curvature, slope, type, pavement quality, etc).
    \item
    The second measure evaluates accident severity outcomes as determined by the injury level sustained by the most severely injured individual (if any) involved into the accident. This evaluation is done by using data on individual accidents and estimating discrete outcome statistical models (e.g. ordered probit and multinomial logit models) for the accident severity outcomes. The explanatory variables used in these models are the individual accident characteristics (e.g. time and location of an accident, weather conditions and roadway characteristics at the accident location, characteristics of the vehicles and drivers involved, etc).
\end{enumerate}

These two measures of roadway safety are complementary. On one hand, an accident frequency study provides a statistical model of the probability of an accident occurring on a roadway segment. On the other hand, an accident severity study provides a statistical model of the conditional probability of a severity outcome of an accident, given the accident occurred. The unconditional probability of the accident severity outcome is the product of its conditional probability and the probability of the accident.

The main objective of this research study is to propose a new statistical approach to modeling accident frequencies and severities, which may provide new guidance to theorists and practitioners in the area of roadway safety. Our approach is based on application of
two-state Markov switching models of accident frequencies and severities. These models assume an existence of two unobserved states of roadway safety. The roadway entities (e.g., roadway segments) are assumed to be able to switch between these states over time, and the switching process is assumed to be Markovian. Accident frequencies and severity outcomes are assumed to be generated by two distinct data-generating processes in the two states. Two-state Markov switching models avoid several drawbacks of the popular conventional models of accident frequencies and severities. We estimate Markov switching models and compare them to the conventional models. We find that the former are strongly favored by accident frequency and severity data and provide a superior statistical fit as compared to the later. Because of the complexity of Markov switching models, this research employs Bayesian inference and Markov Chain Monte Carlo (MCMC) simulations for their statistical estimation.

\section{Organization}
An overview of the previous research on accident frequency and severity is presented in Chapter \ref{LIT}. Chapter~\ref{MODEL_SPECIF} gives specification of the two-state Markov switching and conventional models that are proposed, considered and estimated in this study. Bayesian inference methods are given in Chapter~\ref{MODEL_ESTIM}. Chapter~\ref{MODEL_SIMUL} presents Markov Chain Monte Carlo (MCMC) simulation techniques used for Bayesian inference and model estimation in this study. The model estimation results for accident frequencies are presented in Chapter~\ref{RESULTS_FREQ}. The model estimation results for accident severities are given in Chapter~\ref{RESULTS_SEV}. Finally, we discuss our results and give conclusions in Chapter~\ref{SUMMARY}. Some of the results are given in the Appendix at the end.

%% file: ch_literature.tex
%
% This is the literature review chapter
%

\chapter{Literature review}
\label{LIT}
This chapter includes a brief overview of the previous roadway safety studies of accident frequencies and severities. First, we give an overview of accident frequency studies and standard statistical models used for accident frequencies. Then we review previous work on severities of accidents. Finally, we discuss studies that consider both accident frequencies and accident severities. The literature review of this chapter does not claim to be full or exhaustive. A more detailed literature review, as well as a comprehensive description of conventional methodologies commonly used in roadway safety studies, can be found in \citet[][]{WKM_03}.

\section{Accident frequency studies}
Considerable research has been conducted on understanding and predicting accident frequencies (the number of accidents occurring on roadway segments over a given time period). Because accident frequencies are non-negative integers, count data models are a reasonable statistical modeling approach. Simple modeling approaches include Poisson models and negative binomial (NB) models. These models assume a single process for accident data generation (a Poisson process or a negative binomial process) and involve a nonlinear regression of the observed accident frequencies on various roadway-segment characteristics (such as roadway geometric and environmental factors). Selected previous research on accident frequencies, conducted by application of count data models, is as follows:
\begin{itemize}
    \item
    \citet[][]{HALW_95}
    used negative binomial models to estimate the effect of cross section roadway design elements (e.g. presence of curb, lane width) and traffic volume on accident frequencies for different types of highways. The authors found that some cross section design elements can influence accident rates (e.g. lane width, interchange presence, speed limit) and that some other do not have any effect on number of accidents (e.g. type of friction course material).
    \item
    \citet[][]{SMB_95} applied a negative binomial model to an accident data collected in Washington State. Roadway geometries of fixed-equal length roadway segments (e.g. horizontal and vertical alignments), weather, and other seasonal effects were analyzed along with overall accident frequencies of specific accident types (e.g., rear-end and same direction accidents). This research concluded that highway segments with challenging geometries as well as areas that frequently experience adverse weather conditions are important determinants of accident frequency.
    \item
    \citet[][]{PM_96} estimated a negative binomial regression of the frequencies of accidents at intersection approaches in Seattle suburban areas. The authors of this paper considered traffic volume, geometric characteristics of intersection approaches (e.g. approach sight-distance, speed limit) and approach signalization characteristics (e.g. eight-phase signal) as the model explanatory variables. Authors found a significant influence of some of these variables on accident frequencies at intersection approaches. In particular, they found that high left-turn and opposite traffic volumes considerably increase numbers of accidents at intersection approaches.
    \item
    \citet[][]{ML_03}, based on accident data collected in Toronto, examined generally accepted statistical models (Poisson and NB) applied to accident frequencies intersections. By using the empirical Bayes method, mathematical properties and performance of different popular model functional forms were considered. The authors questioned invariability of the dispersion parameter, given the complexity of the traffic interaction in an intersection area. In addition, the full Bayes statistical approach was used for model specification and estimation.
    \item
    \citet{PL_08} recently considered finite mixture Poisson and negative binomial models of accident frequencies, in order to account for heterogenous populations of accident data. Accident data heterogeneity can result from data generation by distinct (Poisson or NB) processes that operate in different unobserved states of roadway safety. \citet{PL_08} suggested a two-component finite mixture negative binomial model as the best model to account for accident data heterogeneity in their data sample.
    \item
    Recently, \citet{AM_08} applied random parameters count models to the analysis of accident frequencies. The authors found these models to be beneficial for accident frequency prediction. Random parameter models can potentially define unique parameters for each roadway segment, but these models still assume a single state for each segment. This single-state assumption would also be true for count models with random effects \citep[see][]{SAMM_98}.
    \item
    \citet{A_08} were the first to use tobit regression models for prediction of accident rates (accident rates are number of accidents happened per unit roadway segment length and per unit averaged annual daily traffic volume). They considered five-year accident data and found that international roughness index (of the pavement), pavement rutting, the pavement's condition rating, median types and width, shoulder widths, number of ramps and bridges, horizontal and vertical curves, rumble strips, annual average daily travel and the percent of combination truck in the traffic stream have a significant impact on accident rates.
\end{itemize}

Because a preponderance of zero-accident observations is often observed in empirical data, some researchers have applied zero-inflated Poisson (ZIP) and zero-inflated negative binomial (ZINB) models for predicting accident frequencies. Zero-inflated models assume a two-state process for accident data generation. One state is assumed to be perfectly safe with zero accidents (over the duration of time being considered). The other state is assumed to be unsafe with a possibility of nonzero accident frequencies in which accidents can happen and accident frequencies are generated by some given counting process (Poisson or negative binomial). Below are selected studies that are based on an application of zero-inflated count data models:
\begin{itemize}
    \item
    \citet[][]{M_94} applied Poisson regression, zero-inflated Poisson (ZIP) regression, and NB regression to determine a relationship between geometric design characteristics of roadway segments and the number of truck accidents. Results suggest that under the maximum likelihood estimation (MLE) method, all three models perform similarly in terms of estimated truck-involved accident frequencies across roadway segments. To model the relationship, the author recommended the use of a Poisson regression as an initial model, then the use of a negative binomial model if the accident frequency data is overdispersed, and the use of a zero-inflated Poisson model if the data contains an excess of zero observations.
    \item
    \citet[][]{SMM_97} studied the distinction between safe and unsafe roadway segments by estimating zero-inflated Poisson and zero-inflated negative binomial models for accident frequencies in Washington State. The authors established the underlying principles of zero-inflated models, based on a two-state data-generating process for accident frequencies. The two states are a safe state that corresponds to the zero accident likelihood on a roadway segment, and an unsafe state. The results show that two-state zero-inflated structure models provide a superior statistical fit to accident frequency data as compared to the conventional single-state models (without zero-inflation). Thus, the authors found that zero-inflated models are helpful in revealing and understanding important factors that affect accident frequencies with preponderance of zeros.
    \item
    \citet[][]{LWI_05,LWI_07} addressed the question of choosing the best approach to the modeling of roadway accident data by using count data models (e.g. whether to use standard single-state or zero-inflated models). Authors argued that an application of zero-inflated models to the analysis of accident data with a preponderance of zeros is not a defensible modeling approach. They argued that an excess of zeros can be caused by an inappropriate data collection and by many other factors, instead of due to a two-states process. In addition, they claimed that it is unreasonable to expect some roadway segments to be always perfectly safe and questioned ``safe'' and ``unsafe'' state definitions. The authors also argued that zero-inflated models do not explicitly account for a likely possibility for roadway segments to change in time from one state to another. \citet[][]{LWI_05,LWI_07} concluded that, while an application of zero-inflated models often provides a better statistical fit to an observed accident frequency data, the applicability of these models can be questioned.
\end{itemize}

\section{Accident severity studies}
Research efforts in predicting accident severity, such as property damage, injury and fatality, are clearly very important. In the past there has been a large number of studies that focused on modeling accident severity outcomes. The probabilities of severity outcomes of an accident are conditioned on the occurrence of the accident. Common modeling approaches of accident severity include multinomial logit models, nested logit models, mixed logit models and ordered probit models. All accident severity models involve nonlinear regression of the observed accident severity outcomes on various accident characteristics and related factors (such as roadway and driver characteristics, environmental factors, etc). Some of the past accident severity studies are as follows:
\begin{itemize}
    \item
    \citet[][]{OC_96} explored severity of motor vehicle accidents in Australia by estimating the parameters of ordered multiple choice models: ordered logit and probit models. By studying driver, passengers and vehicle characteristics (e.g. vehicle type, seating position of vehicle occupants, blood alcohol level of a driver), the authors found the effects of these characteristics on the probabilities of different types of severity outcomes. For example, they found that the older the victims are and the higher the vehicle speeds are, the higher the probabilities of serious injuries and deaths are.
    \item
    \citet[][]{SM_96} estimated the likelihoods of motorcycle rider accident severity outcomes. In their research work, a multinomial logit model was applied to a 5-year Washington state data for single-vehicle motorcycle collisions. It was found that a helmeted-riding is an effective means of reducing injury severity in any types of collisions, except in fixed-object collisions. At the same time, alcohol-impaired riding, high age of a motorcycle rider, ejection of a rider, wet pavement, interstate as a roadway type, speeding and rider inattention were found to be the factors that increase roadway motorcycle accident severity.
    \item
    \citet[][]{SMB_96} used a nested logit model for statistical analysis of accident severity outcomes on rural highways in Washington State. They found that environment conditions, highway design, accident type, driver and vehicle characteristics significantly influence accident severity. They found that overturn accidents, rear-end accidents on wet pavement, fixed-object accidents, and failure to use the restraint belt system lead to higher probabilities of injury or/and fatality accident outcomes, while icy pavement and single-vehicle collisions lead to higher probability of property damage only outcomes.
    \item
    \citet[][]{DKC_98} applied an ordered probit model to injury severity outcomes in truck-passenger car rear-end collisions in North Carolina. They found that injury severity is increased by darkness, high speed differentials, high speed limits, wet grades, drunk driving, and being female.
    \item
    \citet[][]{CM_99} focused on the effects of trucks and vehicle occupancies on accident severities. They estimated nested logit models for severity outcomes of truck-involved and non-truck-involved accidents in Washington State and found that accident injury severity is noticeably worsened if the accident has a truck involved, and that the effects of trucks are more significant for multi-occupant vehicles than for single-occupant vehicles.
    \item
    \citet[][]{K_01} estimated ordered probit models for severity outcomes of multi-vehicle rear-end accidents in North Carolina. In particular, the results of his research indicate that in two-vehicle collisions the leading driver is more likely to be severely injured, in three-vehicle collisions the driver in the middle is more likely to be severely injured, and being in a newer vehicle protects the driver in rear-end collisions.
    \item
    \citet[][]{U_01}, \citet[][]{UM_04} focused on male and female differences for accident severity outcomes. They used multinomial logit models and accident data from Washington State. They found significant behavioral and physiological differences between genders, and also found that probability of fatal and disabling injuries is higher for females as compared to males.
    \item
    \citet[][]{KK_02} applied ordered probit models to modeling of driver injury severity outcomes. They used a nationwide accident data sample and found that pickups and sport utility vehicles are less (more) safe than passenger cars in single-vehicle (two-vehicle) collisions.
    \item
    \citet[][]{KPSH_02} focused on the safety of aged drivers in the United States. Nine-year Iowa-statewide accident data was considered and the ordered probit modeling technique was implemented for accident severity modeling. Authors inspected vehicle, roadway, driver, collision, and environmental characteristics as factors that may potentially effect accident severity of aged drivers. The modeling results were consistent with a common sense, for example, an animal-related accident tends to have severe consequences for elderly drivers. Also, it was found that accidents with farm vehicles involved are highly severe for elderly drivers in Iowa.
    \item
    \citet[][]{A_03} used ordered probit models for analysis of driver injury severity outcomes at different road locations (roadway segments, signalized intersections, toll plazas) in Central Florida. He found higher probabilities of severe accident outcomes for older drivers, male drivers, those not wearing seat belt, drivers who speed, those who drove vehicles struck at the driver's side, those who drive in rural areas, and drivers using electronic toll collection device (E-Pass) at toll plazas.
    \item
    \citet[][]{YS_04} applied bivariate ordered probit models to an analysis of driver's and passenger's injury severities in collisions with fixed objects. They considered a 4-year accident data sample from Washington State and found that collisions with leading ends of guardrail and trees tend to cause more severe injuries, while collisions with sign posts, faces of guardrail, concrete barrier or bridge and fences tend to cause less severe injuries. They also found that proper use of vehicle restraint system strongly decreases the probability of severe injuries and fatalities.
    \item
    \citet[][]{KNSM_05} explored the differences of driver injury severities in rural and urban accidents involving large trucks. Using four years of California accident data and multinomial logit model approach, they found considerable differences between rural and urban accident injury severities. In particular, they found that the probability of severe/fatal injury increases by $26\%$ in rural areas and by $700\%$ in urban areas when a tractor-trailer combination is involved, as opposed to a single-unit truck being involved. They also found that in accidents where alcohol or drug use is identified, the probability of severe/fatal injury is increased by $250\%$ and $800\%$ in rural and urban areas respectively.
    \item
    \citet[][]{IM_06} studied driver aging and its effect on male and female single-vehicle accident injuries in Indiana. They employed multinomial logit models and found significant differences between different genders and age groups. Specifically, they found an increase in probabilities of fatality for young and middle-aged male drivers when they have passengers, an increase in probabilities of injury for middle-aged female drivers in vehicles 6 years old or older, and an increase in fatality probabilities for males older than 65 years old.
    \item
    \citet[][]{M_06}, \citet[][]{MM_06} focused on the relationship between speed limits and roadway safety. Their research explored the influence of the posted speed limit on the causation and severity of accidents. Multinomial logit statistical models were estimated for causation and severity outcomes of different types of accidents on different road classes. The results showed that speed limits do not have a statistically significant adverse effect on unsafe-speed-related causation of accidents on all roads. At the same time higher speed limits generally increase the severity of accidents on the majority of roads other than interstate highways (on interstates speed limits were found to have statistically insignificant effect on accident severity).
    \item
    \citet[][]{S_06}, \citet[][]{SM_07} focused on the important topic of motorcycle safety on Indiana roads. They used multinomial and nested logit models and found that poor visibility, unsafe speed, alcohol use, not wearing a helmet, right-angle and head-on collisions, and collisions with fixed objects increase severity of motorcycle-involved accidents.
    \item
    \citet[][]{MSM_08}, by using accident severity data from Washington State, estimated a mixed logit model with random parameters. This approach allows estimated model parameters to vary randomly across roadway segments to account for unobserved effects that can be related to other factors influencing roadway safety. Authors found that, on one hand, some roadway characteristic parameters (e.g. pavement friction, number of horizontal curves) can be taken as fixed. On the other hand, other model parameters, such as weather effects and volume-related model parameters (e.g. truck percentage, average annual snowfall), are random and normally-distributed.
    \item
    \citet[][]{EB_07} modeled a seat belt use endogeneity to accident severity due to unsafe driving habits of drivers not using seat belts. For severity outcomes, the authors considered a system of two mixed probit models with random coefficients estimated jointly for seat belt use dummy and severity outcomes. The probit models included random variables that moderate the influence of the primary explanatory attributes associated with drivers. The estimation results highlight the importance of moderation effects, seat belt use endogeneity and the relation of between failure to use seat belt and unsafe driving habits.
\end{itemize}

\section{Mixed studies}
Several previous research studies considered modeling of both accident frequencies and accident severity outcomes. It is beneficial to look at both frequencies and severities simultaneously because, as mentioned above, an unconditional probability of the accident severity outcome is the product of its conditional probability and the accident probability. Several mixed studies, which consider both accident frequency and severity, are as follows.
\begin{itemize}
    \item
    \citet[][]{CM_01} studied the effect of ice warning signs on ice-accident frequencies and severities in Washington State. They modeled accident frequencies and severities by using zero-inflated negative binomial and logit models respectively. They found that the presence of ice warning signs was not a significant factor in reducing ice-accident frequencies and severities.
    \item
    \citet[][]{LM_02} estimated zero-inflated count-data models and nested logit models for frequencies and severities of run-off-roadway accidents in Washington State. They found that run-off-roadway accident frequencies can be reduced by avoiding cut side slopes, decreasing (increasing) the distance from outside shoulder edge to guardrail (light poles), and decreasing the number of isolated trees along roadway. The results of their research also show that run-off-roadway accident severity is increased by alcohol impaired driving, high speeds, and the presence of a guardrail.
    \item
    \citet[][]{KK_03} studied probabilities of accidents and accident severity outcomes for a given fixed driver exposure (defined as the total miles driven). They used Poisson and ordered probit models, and considered a nationwide accident data sample. After normalization of accident rates by driver exposure, the results of their study indicated that young drivers are far more crash prone than other drivers, and that sport utility vehicles and pickups are more likely to be involved into rollover accidents.
\end{itemize}

%% file: ch_model_specif.tex
%
% This is the model specification chapter
%

\chapter{Model specification}
\label{MODEL_SPECIF}

In this chapter we specify the statistical models that are used and estimated in the present study. First, we consider standard (conventional) models commonly used in accident studies. These are count data models for accident frequencies (Poisson, negative binomial models and their zero-inflated counterparts) and discrete outcome models for accident severity outcomes (multinomial logit models). Then we explain Markov process for the state of roadway safety. Finally, we present two-state Markov switching models for accident frequencies and severities. In each of the two states the data is generated by a standard process (such as a Poisson or a negative binomial in the case of accident frequencies, and a multinomial logit in the case of accident severities). Our presentation of Markov switching models is similar to that of Markov switching autoregressive models in econometrics \citep[][]{MT_94,T_02}.

All statistical models that we consider here, either for accident frequencies or for severity outcomes, are parametric and can be fully specified by a likelihood function $f({\bf Y}|{\bf\Theta},{\cal M})$, which is the conditional probability distribution of the vector of all observations ${\bf Y}$, given the vector of all parameters ${\bf\Theta}$ of model ${\cal M}$. If accident events are assumed to be independent, the likelihood function is
\begin{eqnarray}
f({\bf Y}|{\bf\Theta},{\cal M})
=\prod\limits_{t=1}^T\prod\limits_{n=1}^{N_t} P({Y_{t,n}|{\bf\Theta},{\cal M}}).
\label{EQ_MS_L}
\end{eqnarray}
Here, $Y_{t,n}$ is the $n^{\rm th}$ observation during time period $t$, and  $P({Y_{t,n}|{\bf\Theta},{\cal M}})$ is the probability (likelihood) of $Y_{t,n}$. The vector of observations ${\bf Y}=\{Y_{t,n}\}$ includes all observations $n=1,2,...,N_t$ over all time periods $t=1,2,...,T$. Number $N_t$ is the total number of observations during time period $t$, and $T$ is the total number of time periods. In the case of accident frequencies, observation $Y_{t,n}$ is the number of accidents observed on the $n^{\rm th}$ roadway segment during time period $t$ (note that $N_t$ is the number of roadway segments in this case). In the case of accident severity, observation $Y_{t,n}$ is the observed outcome of the $n^{\rm th}$ accident occurred during time period $t$ (note that $N_t$ is the number of accidents in this case). Vector ${\bf\Theta}$ is the vector of all unknown model parameters to be estimated from accident data ${\bf Y}$. We will specify the parameter vector ${\bf\Theta}$ separately for each statistical model presented below. Finally, model ${\cal M}=\{M,{\bf X}_{t,n}\}$ includes the model's name $M$ (e.g. $M=\mbox{``negative binomial''}$ or $M=\mbox{``multinomial logit''}$) and the vector ${\bf X}_{t,n}$ of all characteristic attributes (i.e. values of all explanatory variables in the model) that are associated with the $n^{\rm th}$ observation during time period $t$.

\section{Standard count data models of accident frequencies}
\label{STANDARD_FREQUENCY}
The most popular count data models used for predicting accident frequencies are Poisson and negative binomial (NB) models \citep[][]{WKM_03}. These models are usually estimated by the maximum likelihood estimation (MLE) method, which is based on the maximization of the model likelihood function $f({\bf Y}|{\bf\Theta},{\cal M})$ over the values of the model estimable parameters ${\bf\Theta}$.

Let the number of accidents observed on the $n^{\rm th}$ roadway segment during time period $t$ be $A_{t,n}$. Thus, our observations are $Y_{t,n}=A_{t,n}$, where $n=1,2,...,N_t$ and $t=1,2,...,T$. Here $N_t$ is the number of roadway segments observed during time period $t$, and $T$ is the total number of time periods. The likelihood function for the Poisson model of accident frequencies is specified by equation~(\ref{EQ_MS_L}) and the following equations \citep[][]{WKM_03}:
\begin{eqnarray}
&& P(Y_{t,n}|{\bf\Theta},{\cal M})=P(A_{t,n}|{\bf\Theta},{\cal M})=
{\cal P}(A_{t,n}|{\betabf}),
\\
&& {\cal P}(A_{t,n}|{\betabf})=\frac{\lambda_{t,n}^{A_{t,n}}}{A_{t,n}!}
\exp(-\lambda_{t,n}),
\label{EQ_MLE_POISSON}
\\
&& \lambda_{t,n}=\exp(\betabf'{\bf X}_{t,n}),
\quad t=1,2,...,T,\quad n=1,2,...,N_t.
\label{EQ_MLE_LAMBDA}
\end{eqnarray}
Here, $\lambda_{t,n}$ is the Poisson accident rate for the $n^{\rm th}$ roadway segment, this rate is equal to the average (mean) accident frequency on this segment over the time period $t$. The variance of a Poisson-distributed accident frequency is the same as its average and is equal to $\lambda_{t,n}$. Parameter vector $\betabf$ consists of unknown model parameters to be estimated. Prime means transpose, so $\betabf'$ is the transpose of $\betabf$. In the Poisson model the vector of all model parameters is ${\bf\Theta}=\betabf$. Vector ${\bf X}_{t,n}$ includes characteristic variables for the $n^{\rm th}$ roadway segment during time period $t$. For example, ${\bf X}_{t,n}$ may include segment length, curve characteristics, grades, and pavement properties. Henceforth, the first component of vector ${\bf X}_{t,n}$ is chosen to be unity, and, therefore, the first component of vector $\betabf$ is the intercept.

The likelihood function for the negative binomial (NB) model of accident frequencies is specified by equation~(\ref{EQ_MS_L}) and the following equations \citep[][]{WKM_03}:
\begin{eqnarray}
&& P(Y_{t,n}|{\bf\Theta},{\cal M})=P(A_{t,n}|{\bf\Theta},{\cal M})=
{\cal NB}(A_{t,n}|{\betabf,\alpha}),
\\
&& {\cal NB}(A_{t,n}|{\betabf,\alpha})=\frac{\Gamma(A_{t,n}+1/\alpha)}
{\Gamma(1/\alpha)A_{t,n}!}
\left(\frac{1}{1+\alpha\lambda_{t,n}}\right)^{1/\alpha}
\left(\frac{\alpha\lambda_{t,n}}{1+\alpha\lambda_{t,n}}\right)^{A_{t,n}},
\label{EQ_MLE_NB}
\\
&& \lambda_{t,n}=\exp(\betabf'{\bf X}_{t,n}),
\quad t=1,2,...,T,\quad n=1,2,...,N_t.
\label{EQ_MLE_LAMBDA_1}
\end{eqnarray}
Here, $\Gamma(\:)$ is the standard gamma function. The over-dispersion parameter $\alpha\ge0$ is an unknown model parameter to be estimated together with vector $\betabf$. Thus, the vector of all estimable parameters is ${\bf\Theta}=[{\betabf',\alpha}]'$. The average accident rate is equal to $\lambda_{t,n}$, which is the same as in the case of the Poisson model. The variance of the accident rate is $\lambda_{t,n}(1+\alpha\lambda_{t,n})$, which is higher than in the case of the Poisson model (if $\alpha>0$). The negative binomial model reduces to the Poisson model in the limit $\alpha\to0$.

In addition to the Poisson and negative binomial models, we also consider the standard zero-inflated Poisson (ZIP) and zero-inflated negative binomial (ZINB) models. These models account for a possibility of existence of two separate data-generating states: a normal count state and a zero-accident state. The normal state is unsafe, and accidents can occur in it. The zero-accident state is perfectly safe with no accidents occurring in it.\footnote{Note that roadway segments are not required to stay in a particular state all the time and can move from normal count state to zero-accident state and vice versa.} Zero-inflated models are usually used when there is a preponderance of zeros in the data. In the case of accident frequency data with many zeros in it, the probability of $A_{t,n}$ accidents occurring on the $n^{\rm th}$ roadway segment at time period $t$ can be well modeled by a ZIP process or, if the data are over-dispersed, by a ZINB process. The likelihood functions of the ZIP and ZINB models are specified by equation~(\ref{EQ_MS_L}) and the following equations \citep[][]{WKM_03}:
\begin{eqnarray}
P(Y_{t,n}|{\bf\Theta},{\cal M}) & = & P({A_{t,n}|{\bf\Theta},{\cal M}})
\nonumber\\
& = & q_{t,n}{\cal I}(A_{t,n})+(1-q_{t,n}){\cal P}(A_{t,n}|\betabf)
\qquad\qquad \mbox{for ZIP},
\label{EQ_MLE_ZIP}
\\
P(Y_{t,n}|{\bf\Theta},{\cal M}) & = & P({A_{t,n}|{\bf\Theta},{\cal M}})
\nonumber\\
& = & q_{t,n}{\cal I}(A_{t,n})+(1-q_{t,n}){\cal NB}(A_{t,n}|{\betabf,\alpha})
\qquad \mbox{for ZINB},
\label{EQ_MLE_ZINB}
\end{eqnarray}
where
\begin{eqnarray}
{\cal I}(A_{t,n}) & = & \left\{
\mbox{$\,1$ if $A_{t,n}=0$, and $\,0$ if $A_{t,n}>0\,$}\right\},
\label{EQ_MS_I}
\\
q_{t,n}&=&\frac{1}{1+e^{-\tau\log\lambda_{t,n}}},
\label{EQ_Q_TAU}
\\
q_{t,n}&=&\frac{1}{1+e^{-\gammabfscript'{\bf X}_{t,n}}}.
\label{EQ_Q_GAMMA}
\end{eqnarray}
Here we use two different specifications for the probability $q_{t,n}$ that the $n^{\rm th}$ roadway segment is in the zero-accident state during time period $t$. Scalar $\lambda_{t,n}$ is the accident rate that is defined by equation~(\ref{EQ_MLE_LAMBDA}). Probability distribution ${\cal I}(A_{t,n})$ is the probability mass function that reflects the fact that accidents never happen in the zero-accident state. The right-hand-side of equation~(\ref{EQ_MLE_ZIP}) is a mixture of the zero-accident distribution ${\cal I}(A_{t,n})$ and the Poisson distribution ${\cal P}(A_{t,n}|{\betabf})$ given by equation~(\ref{EQ_MLE_POISSON}). The right-hand-side of equation~(\ref{EQ_MLE_ZINB}) is a mixture of ${\cal I}(A_{t,n})$ and the negative binomial distribution ${\cal NB}(A_{t,n}|{\betabf,\alpha})$ given by equation~(\ref{EQ_MLE_NB}).  Scalar $\tau$ and vector $\gammabf$ are estimable model parameters. We call ``\mbox{ZIP-$\tau$}'' and ``\mbox{ZINB-$\tau$}'' the models specified by equations~(\ref{EQ_MLE_ZIP})-(\ref{EQ_Q_TAU}). We call ``\mbox{ZIP-$\gamma$}'' and ``\mbox{ZINB-$\gamma$}'' the models specified by equations~(\ref{EQ_MLE_ZIP})-(\ref{EQ_MS_I}) and~(\ref{EQ_Q_GAMMA}).
The vector of all estimable parameters is ${\bf\Theta}=[\betabf',\tau]'$ for the \mbox{ZIP-$\tau$} model, ${\bf\Theta}=[\betabf',\alpha,\tau]'$ for the \mbox{ZINB-$\tau$} model, ${\bf\Theta}=[\betabf',\gammabf']'$ for the \mbox{ZIP-$\gamma$} model, and ${\bf\Theta}=[\betabf',\alpha,\gammabf']'$ for the \mbox{ZINB-$\gamma$} model. It is important to note that $q_{t,n}$ depends on the estimable model parameters and gives the probability of being in the zero-accident state, but $q_{t,n}$ is not an estimable parameter by itself.

\section{Standard multinomial logit model of accident severities}
\label{MLE_ML}
The severity outcome of an accident is determined by the injury level sustained by the most severely injured individual (if any) involved into the accident. Thus, accident severities are a discrete outcome data. Most common statistical model used for predicting severity outcomes are the multinomial logit model and the ordered probit model. However, there are two potential problems with applying ordered probability models to accident severity outcomes \citep[][]{SM_07}. The first problem is due to under-reporting of non-injury accidents because they are less likely to be reported to authorities. This under-reporting can result in biased and inconsistent model coefficient estimates in an ordered probability model. In contrast, the coefficient estimates of an unordered multinomial logit model are consistent except for the intercept terms \citep[][]{WKM_03}. The second problem is related to undesirable restrictions that ordered probability models place on influences of the explanatory variables \citep[][]{WKM_03}. As a result, in this study we consider only multinomial logit models for accident severity.

Let there be $I$ discrete outcomes observed for accident severity (for example, $I=3$ and these outcomes are fatality, injury and property damage only). Also let us introduce accident severity outcome dummies $\delta^{(i)}_{t,n}$ that are equal to unity if the $i^{\rm th}$ severity outcome is observed in the $n^{\rm th}$ accident that occurs during time period $t$, and to zero otherwise. Then, our individual observations are the severity outcome dummies, $Y_{t,n}=\{\delta^{(i)}_{t,n}\}$, where $i=1,2,...,I$. Note that $n=1,2,...,N_t$ and $t=1,2,...,T$, where $N_t$ is the number of accidents observed during time period $t$, and $T$ is the total number of time periods. The vector of all observations ${\bf Y}=\{\delta^{(i)}_{t,n}\}$ includes all outcomes observed in all accidents that occur during all time periods. The likelihood function for the multinomial logit (ML) model of accident severity outcomes is specified by equation~(\ref{EQ_MS_L}) and the following equations \citep[][]{WKM_03}:
\begin{eqnarray}
P({Y_{t,n}|{\bf\Theta},{\cal M}}) & = & \prod\limits_{i=1}^I
\left[P({i|{\bf\Theta},{\cal M}})\right]^{\delta^{(i)}_{t,n}} =
\prod\limits_{i=1}^I
\left[{\cal ML}(i|{\betabf})\right]^{\delta^{(i)}_{t,n}},
\label{EQ_MLE_L_1}
\\
{\cal ML}(i|{\betabf}) & = & \frac{\exp(\betabf_i'{\bf X}_{t,n})}
{\sum_{j=1}^I\exp(\betabf_j'{\bf X}_{t,n})},
\qquad i=1,2,...,I.
\label{EQ_MLE_ML}
\end{eqnarray}
Parameter vectors $\betabf_i$ consist of unknown model parameters to be estimated, and $\betabf=\{\betabf_i\}$, where $i=1,2,...,I$. Vector ${\bf X}_{t,n}$ contains all characteristic variables for the $n^{\rm th}$ accident that occurs during time period $t$. For example, ${\bf X}_{t,n}$ may include weather and environment conditions, vehicle and driver characteristics, roadway and pavement properties. We set the first component of ${\bf X}_{t,n}$ to unity, and, therefore, the first components of vectors $\betabf_i$ ($i=1,2,...,I$) are the intercepts. In addition, without loss of generality, we set all $\beta$-parameters for the last severity outcome to zero, $\betabf_I={\bf 0}$. This can be done without loss of generality because ${\bf X}_{t,n}$ are assumed to be independent of the outcome $i$, and, therefore, the numerator and denominator in equation~(\ref{EQ_MLE_ML}) can be multiplied by the an arbitrary common factor \citep[][]{WKM_03}.

\section{Markov switching process}
\label{Markov}
Let there be $N$ roadway segments (or, more generally, roadway entities or/and geographical areas) that we observe during successive time periods $t=1,2,...,T$.\footnote{\label{Markov_FN_1}In a more general case, we can observe a variable number of roadway segments over successive time periods. Here, for simplicity of the presentation, we do not consider this general case. However, our analysis is straightforward to extend to it.} Markov switching models, which will be introduced below, assume that there is an unobserved (latent) state variable $s_{t,n}$ that determines the state of roadway safety for the $n^{\rm th}$ roadway segment (or roadway entity, or geographical area) during time period $t$. We assume that the state variable $s_{t,n}$ can take on only two values: $s_{t,n}=0$ corresponds to the first state, and $s_{t,n}=1$ corresponds to the second state. The choice of labels ``0'' and ``1'' for the two states is arbitrary and is a matter of convenience. We further assume that, for each roadway segment $n$ the state variable $s_{t,n}$ follows a stationary two-state Markov chain process in time.\footnote{Stationarity of $\{s_{t,n}\}$ is in the statistical sense \citep[][]{B_69}.} The Markov property means that the probability distribution of $s_{t+1,n}$ depends only on the value $s_{t,n}$ at time $t$, but not on the previous history $s_{t-1,n},\,s_{t-2,n},\,...$ \citep[][]{B_69}. The stationary two-state Markov chain process $\{s_{t,n}\}$ can be specified by time-independent transition probabilities as
\begin{eqnarray}
P(s_{t+1,n}=1|s_{t,n}=0)=p^{(n)}_{0\rightarrow1},
\quad P(s_{t+1,n}=0|s_{t,n}=1)=p^{(n)}_{1\rightarrow0},
\label{EQ_MS_P}
\end{eqnarray}
where $n=1,2,...,N$. In this equation, for example, $P(s_{t+1,n}=1|s_{t,n}=0)$ is the conditional probability of $s_{t+1,n}=1$ at time $t+1$, given that $s_{t,n}=0$ at time $t$. Note that $P(s_{t+1,n}=0|s_{t,n}=0)=p^{(n)}_{0\rightarrow0}=1-p^{(n)}_{0\rightarrow1}$ and $P(s_{t+1,n}=1|s_{t,n}=1)=p^{(n)}_{1\rightarrow1}=1-p^{(n)}_{1\rightarrow0}$. Transition probabilities $p^{(n)}_{0\rightarrow1}$ and $p^{(n)}_{1\rightarrow0}$ are unknown parameters to be estimated from accident data ($n=1,2,...,N$). The stationary unconditional probabilities of states $s_{t,n}=0$ and $s_{t,n}=1$ are\footnote{These can be found from the following stationarity conditions: $\quad\bar p^{(n)}_0=[1-p^{(n)}_{0\rightarrow1}]\bar p^{(n)}_0+p^{(n)}_{1\rightarrow0}\bar p^{(n)}_1$,\quad $\bar p^{(n)}_1=p^{(n)}_{0\rightarrow1}\bar p^{(n)}_0+[1-p^{(n)}_{1\rightarrow0}]\bar p^{(n)}_1$\quad and\quad$\bar p^{(n)}_0+\bar p^{(n)}_1=1$\quad \citep[][]{B_69}.}
\begin{eqnarray}
\begin{array}{lcl}
\bar p^{(n)}_0=p^{(n)}_{1\rightarrow0}/(p^{(n)}_{0\rightarrow1}
+p^{(n)}_{1\rightarrow0})\quad\quad\mbox{for state}\quad\quad s_{t,n}=0,
\\
\bar p^{(n)}_1=p^{(n)}_{0\rightarrow1}/(p^{(n)}_{0\rightarrow1}
+p^{(n)}_{1\rightarrow0})\quad\quad\mbox{for state}\quad\quad s_{t,n}=1.
\label{EQ_MS_P_BAR}
\end{array}
\end{eqnarray}

It is noteworthy that the case when (for each roadway segment $n$) the states $s_{t,n}$ are independent and identically distributed in time $t$ is a special case of the Markov chain process. Indeed, this case corresponds to history-independent probabilities of states ``0'' and ``1'', therefore, $p^{(n)}_{0\rightarrow0}\equiv p^{(n)}_{1\rightarrow0}$ and $p^{(n)}_{0\rightarrow1}\equiv p^{(n)}_{1\rightarrow1}$. Thus, we have $p^{(n)}_{0\rightarrow0}=p^{(n)}_{1\rightarrow0}=\bar p^{(n)}_0$ and $p^{(n)}_{0\rightarrow1}=p^{(n)}_{1\rightarrow1}=\bar p^{(n)}_1$, where the last equalities in these two formulas follow from equations~(\ref{EQ_MS_P_BAR}).
\\

\section{Markov switching count data models of annual accident\\
\hspace{0pt} frequencies}
\label{Markov_freq_year}
When considering annual accident frequency data below, we will use and estimate two-state Markov switching Poisson (MSP) and two-state Markov switching negative binomial (MSNB) models that are proposed as follow. Similar to zero-inflated models, these annual-accident-frequency Markov switching models assume that one of the two states of roadway safety is a zero-accident state, in which accidents never happen. The other state is assumed to be an unsafe state with possibly non-zero accidents occurring. MSP and MSNB models respectively assume Poisson and negative binomial (NB) data-generating processes in the unsafe state. Without loss of generality, below we take $s_{t,n}=0$ to be the zero-accident state and $s_{t,n}=1$ to be the unsafe state.

As in the case of the standard count data models of accident frequencies (see Section~\ref{STANDARD_FREQUENCY}), in this section, a single observation is the number of accidents $A_{t,n}$ that occur on the $n^{\rm th}$ roadway segment during time period $t$. There are $T$ time periods, each is equal to a year, and the periods are $t=1,2,...,T$. For simplicity of presentation, we assume that the number of roadway segments is constant over time~\footnote{\label{N_CONSTANT}The analysis is easily extended to the case when we observe a variable number of roadway segments $N_t\ne{\rm const}$ during time periods $t$, see also footnote~\ref{Markov_FN_1} on page~\pageref{Markov_FN_1}. In this case it would be convenient to count all segments as $n=1,2,...,N$ and to count the time periods as $t=T^{(n)}_i,T^{(n)}_i+1,...,T^{(n)}_f$, where the $n^{\rm th}$ segment is assumed to be observed during interval $T^{(n)}_i\le t\le T^{(n)}_f$ of successive time periods.}, $N_t=N={\rm const}$, and, therefore, the segments are $n=1,2,...,N$. The vector of all observations ${\bf Y}=\{Y_{t,n}\}=\{A_{t,n}\}$ includes all accident frequencies $A_{t,n}$ ($t=1,2,...,T$ and $n=1,2,...,N$). For each roadway segment $n$, the state $s_{t,n}$ can change every year. The likelihood functions of the two-state Markov switching Poisson (MSP) and two-state Markov switching negative binomial (MSNB) models of {\it annual} accident frequencies $A_{t,n}$ are specified by equation~(\ref{EQ_MS_L}) with $N_t=N$, and by the following equations:
\begin{eqnarray}
P(Y_{t,n}|{\bf\Theta},{\cal M}) = P({A_{t,n}|{\bf\Theta},{\cal M}}) = \left\{
\begin{array}{ll}
{\cal I}(A_{t,n}) & \quad\mbox{if $s_{t,n}=0$}
\\
{\cal P}(A_{t,n}|{\betabf}) & \quad\mbox{if $s_{t,n}=1$}
\end{array}
\right.\qquad
\label{EQ_MS_P_year}
\end{eqnarray}
for the MSP model of annual accident frequencies, and
\begin{eqnarray}
P(Y_{t,n}|{\bf\Theta},{\cal M}) = P({A_{t,n}|{\bf\Theta},{\cal M}}) = \left\{
\begin{array}{ll}
{\cal I}(A_{t,n}) & \quad\mbox{if $s_{t,n}=0$}
\\
{\cal NB}(A_{t,n}|{\betabf,\alpha}) & \quad\mbox{if $s_{t,n}=1$}
\end{array}
\right.
\label{EQ_MS_NB_year}
\end{eqnarray}
for the MSNB model of annual accident frequencies. Here zero-accident probability distribution ${\cal I}(A_{t,n})$, given by equation~(\ref{EQ_MS_I}), reflects the fact that accidents never happen in the zero-accident state $s_{t,n}=0$. Probability distributions ${\cal P}(A_{t,n}|{\betabf})$ and ${\cal NB}(A_{t,n}|{\betabf,\alpha})$ are the standard Poisson and negative binomial probability mass functions, see equations~(\ref{EQ_MLE_POISSON}) and~(\ref{EQ_MLE_NB}) respectively. Vector $\betabf$ is the vector of estimable model parameters and $\alpha$ is the negative binomial over-dispersion parameter. To ensure that $\alpha$ is non-negative, during model estimation we consider its logarithm instead of it. For each roadway segment $n$ the state variable $s_{t,n}$ follows a stationary two-state Markov chain process as described in Section~\ref{Markov}.

Because the state variables $s_{t,n}$ are unobservable, the vector
of all estimable parameters ${\bf\Theta}$ must include all states ($s_{t,n}$), in addition to all model parameters ($\beta$-s, $\alpha$-s) and all transition probabilities ($p^{(n)}_{0\rightarrow1}$, $p^{(n)}_{1\rightarrow0}$). Thus,
\begin{eqnarray}
{\bf\Theta}=[\betabf',\alpha,p^{(1)}_{0\rightarrow1},...,
p^{(N)}_{0\rightarrow1},p^{(1)}_{1\rightarrow0},...,
p^{(N)}_{1\rightarrow0},{\bf S}']',
\label{EQ_THETA_freq_year}
\end{eqnarray}
where vector ${\bf S}=[(s_{1,1},...,s_{T,1}),...,(s_{1,N},...,s_{T,N})]'$ contains all state values $s_{t,n}$ and has length $T\times N$. Of course, in the case of the MSP model, over-dispersion parameter $\alpha$ does not enter equation~(\ref{EQ_THETA_freq_year}).

Note that, if $p^{(n)}_{0\rightarrow1}<p^{(n)}_{1\rightarrow0}$, then, according to equations~(\ref{EQ_MS_P_BAR}), we have $\bar p^{(n)}_{0}>\bar p^{(n)}_{1}$, and, on average, for the $n^{\rm th}$ roadway segment state $s_{t,n}=0$ occurs more frequently than state $s_{t,n}=1$. On the other hand, if $p^{(n)}_{0\rightarrow1}>p^{(n)}_{1\rightarrow0}$, then state $s_{t,n}=1$ occurs more frequently for the $n^{\rm th}$ segment.

In addition, note that here the choice of a year as the length of the time periods $t=1,2,...,T$ is arbitrary. For example, one can consider quarterly (or other) periods instead.

Finally, it is important to understand that although the MSP and MSNB models given by Equations~(\ref{EQ_MS_P_year}) and~(\ref{EQ_MS_NB_year}) assume state $s_{t,n}=0$ to be perfectly safe and zero-accident, this state can be (and probably should be) viewed as an approximation for nearly safe states, in which accidents rarely occur.~\footnote{Nearly safe states have average accident rates $\lambda_{t,n}\ll 1$ [see Equations~(\ref{EQ_MLE_LAMBDA}) and~(\ref{EQ_MLE_LAMBDA_1})]. In this case, the perfectly safe, zero-accident state, which has $\lambda_{t,n}=0$, serves as a good approximation for these nearly safe states.}

\section{Markov switching count data models of weekly accident\\
\hspace{0pt} frequencies}
\label{Markov_freq_week}
When considering weekly accident frequency data below, we will use and estimate two-state Markov switching Poisson (MSP) and two-state Markov switching negative binomial (MSNB) models that are proposed as follows. In each of the two states ($s_{t,n}=0$ and $s_{t,n}=1$) these weekly-accident-frequency models assume a standard Poisson data-generating process defined by equation~(\ref{EQ_MLE_POISSON}), or a standard negative binomial process defined by equation~(\ref{EQ_MLE_NB}). Thus, both states are assumed to be unsafe for these models. We observe the number of accidents $A_{t,n}$ that occur on the $n^{\rm th}$ roadway segment during time period $t$, which is a week in this case. Let there be $T$ weekly time periods in total. Let us again assume that the number of roadway segments is constant over time, $N_t=N={\rm const}$ (see footnote~\ref{N_CONSTANT} on page~\pageref{N_CONSTANT}). Thus, in equation~(\ref{EQ_MS_L}) the vector of all observations is ${\bf Y}=\{Y_{t,n}\}=\{A_{t,n}\}$, where $t=1,2,...,T$ and $n=1,2,...,N$. In addition, for weekly-accident-frequency Markov switching models, we assume that all roadway segments always have the same state, and, therefore, the state variable $s_{t,n}=s_t$ depends on time period $t$ only. This is because, here, state $s_t$ is intended to capture common unobserved factors influencing roadway safety on all segments. Correspondingly, all roadway segments switch between the states with the same transition probabilities $p^{(n)}_{0\rightarrow1}=p_{0\rightarrow1}$ and $p^{(n)}_{1\rightarrow0}=p_{1\rightarrow0}$.

With this, the likelihood functions for the two-state Markov switching Poisson (MSP) and two-state Markov switching negative binomial (MSNB) models of {\it weekly} accident frequencies $A_{t,n}$ are specified by equation~(\ref{EQ_MS_L}) with $N_t=N$, and by the following equations:
\begin{eqnarray}
&& P(Y_{t,n}|{\bf\Theta},{\cal M}) =
P({A_{t,n}|{\bf\Theta},{\cal M}}) = \left\{
\begin{array}{ll}
{\cal P}(A_{t,n}|{\betabf_{(0)}})\quad\mbox{if $s_t=0$}
\\
{\cal P}(A_{t,n}|{\betabf_{(1)}})\quad\mbox{if $s_t=1$}
\end{array}
\right.\qquad\qquad
\label{EQ_MS_P_week}
\end{eqnarray}
for the MSP model of weekly accident frequencies, and
\begin{eqnarray}
&& P(Y_{t,n}|{\bf\Theta},{\cal M}) =
P({A_{t,n}|{\bf\Theta},{\cal M}}) = \left\{
\begin{array}{ll}
{\cal NB}(A_{t,n}|{\betabf_{(0)},\alpha_{(0)}})\quad\mbox{if $s_t=0$}
\\
{\cal NB}(A_{t,n}|{\betabf_{(1)},\alpha_{(1)}})\quad\mbox{if $s_t=1$}
\end{array}
\right.\quad
\label{EQ_MS_NB_week}
\end{eqnarray}
for the MSNB model of weekly accident frequencies. Here, $t = 1,2,...,T$ and $n=1,2,...,N$. Probability distributions ${\cal P}(\ldots)$ and ${\cal NB}(\ldots)$ are the standard Poisson and negative binomial probability mass functions, see equations~(\ref{EQ_MLE_POISSON}) and (\ref{EQ_MLE_NB}) respectively. Parameter vectors $\betabf_{(0)}$ and $\betabf_{(1)}$, and negative binomial over-dispersion parameters $\alpha_{(0)}\ge0$ and $\alpha_{(1)}\ge0$ are the unknown estimable model parameters in the two states $s_t=0$ and $s_t=1$. To ensure that $\alpha_{(0)}$ and $\alpha_{(1)}$ are non-negative, their logarithms are considered during model estimation. Because, we choose the first component of ${\bf X}_{t,n}$ to be equal to unity, the first components of $\betabf_{(0)}$ and $\betabf_{(1)}$ are the intercepts in the two states. Note that the state variable $s_{t}$ follows a stationary two-state Markov chain process with transition probabilities $p_{0\rightarrow1}$ and $p_{1\rightarrow0}$ as described in Section~\ref{Markov}.

Because the state variables $s_t$ are unobservable, the vector
of all estimable parameters ${\bf\Theta}$ must include all states ($s_t$), in addition to all model parameters ($\beta$-s, $\alpha$-s) and all transition probabilities ($p_{0\rightarrow1}$, $p_{1\rightarrow0}$). Thus,
\begin{eqnarray}
{\bf\Theta}=[\betabf_{(0)}',\alpha_{(0)},\betabf_{(1)}',\alpha_{(1)},
p_{0\rightarrow1},p_{1\rightarrow0},{\bf S}']'.
\label{EQ_THETA_freq_week}
\end{eqnarray}
where vector ${\bf S}=[s_1,...,s_T]'$ has length $T$ and contains all state values. In the case of the MSP model, over-dispersion parameters $\alpha_{(0)}$ and $\alpha_{(1)}$ are absent from equation~(\ref{EQ_THETA_freq_week}).

Without loss of generality, we assume that (on average) state $s_t=0$ occurs more or equally frequently than state $s_t=1$. Therefore, $\bar p_{0}\geq\bar p_{1}$, and from Equations~(\ref{EQ_MS_P_BAR}) we obtain restriction\footnote{Restriction~(\ref{EQ_MS_P_RESTRICT}) is introduced for the purpose of avoiding the problem of switching of state labels, \mbox{$0\leftrightarrow1$}. This problem would otherwise arise because of the symmetry of the likelihood functions given by equations~(\ref{EQ_MS_L}),~(\ref{EQ_MS_P_week}) and~(\ref{EQ_MS_NB_week}) under the label switching.}
\begin{eqnarray}
p_{0\rightarrow1}\leq p_{1\rightarrow0}.
\label{EQ_MS_P_RESTRICT}
\end{eqnarray}
In this case, we can refer to states $s_t=0$ and $s_t=1$ as ``more frequent'' and ``less frequent'' states respectively.

Note that here the choice of a week as the length of the time periods $t=1,2,...,T$ is arbitrary. For example, one can consider daily (or other) periods instead.

\section{Markov switching multinomial logit models of accident \\
\hspace{0pt} severities}
\label{Markov_sever}
When considering accident severity data below, we will use and estimate two-state Markov switching multinomial logit (MSML) model that is proposed as follows. In each of the two states ($0$ and $1$), this model assumes standard multinomial logit (ML) data-generating  process that is defined by equation~(\ref{EQ_MLE_ML}) and described in Section~\ref{MLE_ML}. We observe severity outcome dummies $\delta^{(i)}_{t,n}$ that are equal to unity if the $i^{\rm th}$ severity outcome is observed in the $n^{\rm th}$ accident that occurs during time period $t$, and to zero otherwise. We consider weekly time periods, $t=1,2,...,T$, where $T$ is the total number of periods observed. Then, the vector of all observations ${\bf Y}=\{\delta^{(i)}_{t,n}\}$ includes all outcomes observed in all accidents that occur during all time periods, $i=1,2,...,I$, $n=1,2,...,N_t$ and $t=1,2,...,T$. Here $I$ is the total number of possible severity outcomes, and $N_t$ is the number of accidents observed during weekly time period $t$. For MSML models of accident severities, we again assume that all roadway segments (where accidents happen) always have the same state of roadway safety, and, therefore, the state variable $s_{t,n}=s_t$ depends on time period $t$ only (in this case, state $s_t$ captures common unobserved factors that influence safety on all segments). Correspondingly, all roadway segments switch between the states with the same transition probabilities $p^{(n)}_{0\rightarrow1}=p_{0\rightarrow1}$ and $p^{(n)}_{1\rightarrow0}=p_{1\rightarrow0}$.

The likelihood function for the two-state Markov switching multinomial logit (MSML) model of accident severity outcomes is specified by equation~(\ref{EQ_MS_L}) and the following equations:
\begin{eqnarray}
P(Y_{t,n}|{\bf\Theta},{\cal M}) &=&
\prod\limits_{i=1}^I
\left[P({i|{\bf\Theta},{\cal M}})\right]^{\delta^{(i)}_{t,n}}
\nonumber\\
{} &=& \left\{
\begin{array}{ll}
\prod\limits_{i=1}^I
\left[{\cal ML}(i|{\betabf_{(0)}})\right]^{\delta^{(i)}_{t,n}} &
\quad\mbox{if $s_t=0$}
\\
\prod\limits_{i=1}^I
\left[{\cal ML}(i|{\betabf_{(1)}})\right]^{\delta^{(i)}_{t,n}} &
\quad\mbox{if $s_t=1$}
\end{array}
\right.,
\label{EQ_MS_ML}
\end{eqnarray}
where $n=1,2,...,N_t$ and $t = 1,2,...,T$. Probability distributions ${\cal ML}(i|{\betabf_{(0)}})$ and ${\cal ML}(i|{\betabf_{(1)}})$ are the standard multinomial logit probability mass functions in the two states, see equation~(\ref{EQ_MLE_ML}). Here $\betabf_{(0)}=\{\betabf_{(0),i}\}$ and $\betabf_{(1)}=\{\betabf_{(1),i}\}$, where $i=1,2,...,I$. Parameter vectors $\betabf_{(0),i}$ and $\betabf_{(1),i}$ are unknown estimable model parameters in states $0$ and $1$ respectively. Since we choose the first component of ${\bf X}_{t,n}$ to be equal to unity, the first components of vectors $\betabf_{(0),i}$ and $\betabf_{(1),i}$ are the intercepts. Similar to the case of the standard (single-state) ML model presented in Section~\ref{MLE_ML}, here, we set all $\beta$-parameters for the last severity outcome to zero, $\betabf_{(0),I}=\betabf_{(1),I}={\bf 0}$.

The vector of all estimable parameters ${\bf\Theta}$ includes all states ($s_t$), in addition to all model parameters ($\beta$-s) and all transition probabilities ($p_{0\rightarrow1}$, $p_{1\rightarrow0}$). Thus,
\begin{eqnarray}
{\bf\Theta}=[\betabf_{(0)}',\betabf_{(1)}',p_{0\rightarrow1},
p_{1\rightarrow0},{\bf S}']'.
\label{EQ_THETA_sev_week}
\end{eqnarray}
where vector ${\bf S}=[s_1,...,s_T]'$ has length $T$ and contains all state values.

In analogy with the assumption made in the previous section, here, without loss of generality, we assume that (on average) state $s_t=0$ occurs more or equally frequently than state $s_t=1$. Therefore, $\bar p_{0}\geq\bar p_{1}$, and from equations~(\ref{EQ_MS_P_BAR}) we again obtain restriction
\begin{eqnarray}
p_{0\rightarrow1}\leq p_{1\rightarrow0}.
\label{EQ_MS_P_RESTRICT_1}
\end{eqnarray}
In this case, we can refer to states $s_t=0$ and $s_t=1$ as ``more frequent'' and ``less frequent'' states respectively.

Note that here the choice of a week as the length of the time periods $t=1,2,...,T$ is arbitrary. For example, one can consider daily (or other) periods instead.

%% file: ch_model_estim.tex
%
% This is the model estimation chapter
%

\chapter{Model estimation and comparison}
\label{MODEL_ESTIM}
This chapter presents the basics of Bayesian estimation of standard models and Markov switching models of accident frequencies and severities. We also discuss comparison of different models by using Bayesian approach, and an evaluation of model fit performance.

\section{Bayesian inference and Bayes formula}
Statistical estimation of Markov switching models is complicated by unobservability of the state variables $s_{t,n}$ (or $s_t$).\footnote{\label{FN_S}For example, in the case of Markov switching models of weekly accident frequencies, we will have 260 time periods ($T=260$ weeks of available data). In this case, there are $2^{260}$ possible combinations for value of vector ${\bf S}=[s_1,...,s_T]'$.} As a result, the traditional maximum likelihood estimation (MLE) procedure is of very limited use for Markov switching models. Instead, a Bayesian inference approach is used. Given a model ${\cal M}$ with likelihood function $f({\bf Y}|{\bf\Theta},{\cal M})$, the Bayes formula is
\begin{eqnarray}
f({\bf\Theta}|{\bf Y},{\cal M})=
\frac{f({\bf Y},{\bf\Theta}|{\cal M})}{f({\bf Y}|{\cal M})}=
\frac{f({\bf Y}|{\bf\Theta},{\cal M})\pi({\bf\Theta}|{\cal M})}
{\int f({\bf Y},{\bf\Theta}|{\cal M})\,d{\bf\Theta}}.
\label{EQ_POSTERIOR}
\end{eqnarray}
Here $f({\bf\Theta}|{\bf Y},{\cal M})$ is the posterior probability distribution of model parameters ${\bf\Theta}$ conditional on the observed data ${\bf Y}$ and model ${\cal M}$. Function $f({\bf Y},{\bf\Theta}|{\cal M})$ is the joint probability distribution of ${\bf Y}$ and ${\bf\Theta}$ given model ${\cal M}$. Function $f({\bf Y}|{\cal M})$ is the marginal likelihood function -- the probability distribution of data ${\bf Y}$ given model ${\cal M}$. Function $\pi({\bf\Theta}|{\cal M})$ is the prior probability distribution of parameters that reflects prior knowledge about ${\bf\Theta}$. The intuition behind equation~(\ref{EQ_POSTERIOR}) is straightforward: given model ${\cal M}$, the posterior distribution accounts for both the observations ${\bf Y}$ and our prior knowledge of ${\bf\Theta}$. We use the harmonic mean formula to calculate the marginal likelihood $f({\bf Y}|{\cal M})$ of data ${\bf Y}$ \citep[see][]{KR_95} as,
\begin{eqnarray}
f({\bf Y}|{\cal M})^{-1} & = & f({\bf Y}|{\cal M})^{-1}\int
\pi({\bf\Theta}|{\cal M})\,d{\bf\Theta}=
f({\bf Y}|{\cal M})^{-1}\int\frac{f({\bf\Theta},{\bf Y}|{\cal M})}
{f({\bf Y}|{\bf\Theta},{\cal M})}\,d{\bf\Theta}
\nonumber\\
& = & f({\bf Y}|{\cal M})^{-1}\int\frac{f({\bf\Theta}|{\bf Y},{\cal M})
f({\bf Y}|{\cal M})}{f({\bf Y}|{\bf\Theta},{\cal M})}\,d{\bf\Theta}
\nonumber\\
& = & \int\frac{f({\bf\Theta}|{\bf Y},{\cal M})}{f({\bf Y}|{\bf\Theta},
{\cal M})}\,d{\bf\Theta}=
E\left[\left.f({\bf Y}|{\bf\Theta},{\cal M})^{-1}\right|{\bf Y}\right],
\label{EQ_L_MARGINAL}
\end{eqnarray}
where $E(\ldots|{\bf Y})$ is the posterior expectation (which is calculated by using the posterior distribution).

In our study (and in most practical studies), the direct application of equation~(\ref{EQ_POSTERIOR}) is not feasible because the parameter vector ${\bf\Theta}$ contains too many components, making integration over ${\bf\Theta}$ in equation~(\ref{EQ_POSTERIOR}) extremely difficult (see footnote~\ref{FN_S} on page~\pageref{FN_S}). However, the posterior distribution $f({\bf\Theta}|{\bf Y},{\cal M})$ in equation~(\ref{EQ_POSTERIOR}) is known up to its normalization constant, namely $f({\bf\Theta}|{\bf Y},{\cal M})\propto f({\bf Y},{\bf\Theta}|{\cal M})=f({\bf Y}|{\bf\Theta},{\cal M})\pi({\bf\Theta}|{\cal M})$. As a result, we use Markov Chain Monte Carlo (MCMC) simulations, which provide a convenient and practical computational methodology for sampling from a probability distribution known up to a constant (the posterior distribution in our case). Given a large enough posterior sample of parameter vector ${\bf\Theta}$, any posterior expectation and variance can be found and Bayesian inference can be readily applied. In the next chapter we describe our choice of prior distribution $\pi({\bf\Theta}|{\cal M})$ and the MCMC simulations in detail. The prior distribution is chosen to be wide and essentially noninformative. For the MCMC simulations, we wrote a special numerical code in the MATLAB programming language and tested it (for details see the next chapter).

In the end of this section, let us make a short noteworthy digression. In Bayesian statistics model observations and model parameters are treated on an equal footing. Therefore, for Markov switching models, one can treat the vector of all state values ${\bf S}$ as latent model parameters, or as latent (hidden) observations. We treat ${\bf S}$ as model parameters. As a result, in our approach, the transition probabilities $p^{(n)}_{1->0}$ and $p^{(n)}_{0->1}$ do not enter the likelihood function $f({\bf Y}|{\bf\Theta},{\cal M})$, which is a function of ${\bf S}$ and model coefficients ($\beta$-s, $\alpha$-s, $\gamma$-s, $\tau$) only (refer to the likelihood functions presented in the previous chapter). In this case, the Markov switching property is treated as a prior information, and the prior distribution, given in the next chapter, reflects this property (in other words, we a priori specify that the state variable $s_{t,n}$ follows a Markov process in time). If we treated state values ${\bf S}$ as latent observations, then the vector of all observation would include both {\bf Y} and {\bf S}. In this case, the likelihood function would depend on the transition probabilities and would become $f({\bf Y},{\bf S}|{\bf\Theta}\backslash{\bf S},{\cal M})=f({\bf Y}|{\bf\Theta},{\cal M})f({\bf S}|{\bf\Theta}\backslash{\bf S},{\cal M})$, where ${\bf\Theta}\backslash{\bf S}$ means all components of ${\bf\Theta}$ except ${\bf S}$. In any case, for the purpose of model comparison discussed below, the marginal likelihood should always be defined as $f({\bf Y}|{\cal M})$ [not as $f({\bf Y},{\bf S}|{\cal M})$] because ${\bf Y}$ is the only data that is truly observed.

\section{Comparison of statistical models}
For comparison of different models we use the following Bayesian approach. Let there be two models ${\cal M}_1$ and ${\cal M}_2$ with parameter vectors ${\bf\Theta_1}$ and ${\bf\Theta_2}$ respectively. Assuming that we have equal preferences of these models, their prior probabilities are $\pi({\cal M}_1)=\pi({\cal M}_2)=1/2$. In this case, the ratio of the models' posterior probabilities, $P({\cal M}_1|{\bf Y})$ and $P({\cal M}_2|{\bf Y})$, is equal to the Bayes factor. The later is defined as the ratio of the models' marginal likelihoods \citep[][]{KR_95}. Thus, we have
\begin{eqnarray}
\frac{P({\cal M}_2|{\bf Y})}{P({\cal M}_1|{\bf Y})}=
\frac{f({\cal M}_2,{\bf Y})/f(\bf Y)}
{f({\cal M}_1,{\bf Y})/f(\bf Y)}=
\frac{f({\bf Y}|{\cal M}_2)\pi({\cal M}_2)}
{f({\bf Y}|{\cal M}_1)\pi({\cal M}_1)}=
\frac{f({\bf Y}|{\cal M}_2)}{f({\bf Y}|{\cal M}_1)},
\label{EQ_BAYES_FACTOR}
\end{eqnarray}
where $f({\cal M}_1,{\bf Y})$ and $f({\cal M}_2,{\bf Y})$ are the joint distributions of the models and the data, $f({\bf Y})$ is the unconditional distribution of the data, and the marginal likelihoods $f({\bf Y}|{\cal M}_1)$ and $f({\bf Y}|{\cal M}_2)$ are given by equation~(\ref{EQ_L_MARGINAL}). If the ratio in equation~(\ref{EQ_BAYES_FACTOR}) is larger than one, then model ${\cal M}_2$ is favored, if the ratio is less than one, then model ${\cal M}_1$ is favored. An advantage of the use of Bayes factors is that it has an inherent penalty for including too many parameters in the model and guards against overfitting.\footnote{\label{DIC}There are other frequently used model comparison criteria, for example, the deviance information criterion, ${\rm DIC}=2E[D({\bf\Theta})|{\bf Y}]-D(E[{\bf\Theta}|{\bf Y}])$, where deviance $D({\bf\Theta})\equiv-2\ln[f({\bf Y}|{\bf\Theta},{\cal M})]$ \citep{R_01}. Models with smaller DIC are favored to models with larger DIC. However, DIC is theoretically based on the assumption of asymptotic multivariate normality of the posterior distribution, in which case DIC reduces to AIC \citep{SBCL_02}. As a result, we prefer to rely on a mathematically rigorous and formal Bayes factor approach to model selection, as given by equation~(\ref{EQ_BAYES_FACTOR}).}

\section{Model performance evaluation}
To evaluate the performance of model $\{{\cal M},{\bf\Theta}\}$
in fitting the observed data ${\bf Y}$, we carry out a $\chi^2$
goodness-of-fit test \citep{MS_96,C_98,W_02,PTVF_07}.

In the case of accident frequency models, quantity $\chi^2$ is~\footnote{Note that for a standard Poisson distribution, the variances are equal to the means, $var(Y_{t,n}|{\bf\Theta},{\cal M})=E(Y_{t,n}|{\bf\Theta},{\cal M})$, and equation~(\ref{CHI_SQUARED}) reduces to the Pearson's $\chi^2$.}
\begin{eqnarray}
\chi^2=\sum\limits_{t=1}^T\sum\limits_{n=1}^{N_t}
\frac{[Y_{t,n}-E(Y_{t,n}|{\bf\Theta},{\cal M})]^2}
{var(Y_{t,n}|{\bf\Theta},{\cal M})},
\label{CHI_SQUARED}
\end{eqnarray}
where $E(Y_{t,n}|{\bf\Theta},{\cal M})$ and $var(Y_{t,n}|{\bf\Theta},{\cal M})$ are the expectations and variances of the observations $Y_{t,n}$. In accident frequency studies, the observations are the frequencies, $Y_{t,n}=A_{t,n}$ on roadway segment $n$ during time period $t$. For example, from equations~(\ref{EQ_MLE_NB}),~(\ref{EQ_MLE_LAMBDA_1}) and~(\ref{EQ_MS_NB_week}) for the MSNB model of weekly accident frequencies we find the following formulas for the (unconditional of state) expectations and variances: $E(Y_{t,n}|{\bf\Theta},{\cal M})=
{\bar p}_0\lambda^{(0)}_{t,n}+{\bar p}_1\lambda^{(1)}_{t,n}$ and
$var(Y_{t,n}|{\bf\Theta},{\cal M})=
{\bar p}_0\lambda^{(0)}_{t,n}(1+\alpha_{(0)}\lambda^{(0)}_{t,n})
+{\bar p}_1\lambda^{(1)}_{t,n}(1+\alpha_{(1)}\lambda^{(1)}_{t,n})
+{\bar p}_0{\bar p}_1(\lambda^{(1)}_{t,n}-\lambda^{(0)}_{t,n})^2$,
where $\lambda^{(0)}_{t,n}=\exp(\betabf_{(0)}'{\bf X}_{t,n})$ and
$\lambda^{(1)}_{t,n}=\exp(\betabf_{(1)}'{\bf X}_{t,n})$ are the mean accident rates in the states $s_t=0$ and $s_t=1$ respectively. For the MSNB model of annual accident frequencies one needs to set $\lambda^{(0)}_{t,n}\equiv0$ in these formulas because state $s_t=0$ is the zero-accident state in this case. The appropriate formulas for Poisson models can be obtained by setting the over-dispersion parameters ($\alpha$-s) to zero.

In the limit of asymptotically normal distribution of large accident frequencies, $\chi^2$ has the chi-square distribution with degrees of freedom equal to the number of observations minus the number of
model parameters \citep{W_02}. Because weekly (and even annual) accident frequencies are typically small, in this study, we do not rely on the assumption of their asymptotic normality. Instead, we carry out Monte Carlo simulations to find the distribution of $\chi^2$ \citep{C_98}. This is done by generating a large number of artificial data sets under the hypothesis that the model $\{{\cal M},{\bf\Theta}\}$ is true, computing and recording the $\chi^2$ value for each data set, and then using these values to find the distribution of $\chi^2$. This distribution is then used to find the goodness-of-fit p-value, equal to the probability that $\chi^2$ exceeds the observed value of $\chi^2$ (the later is calculated by using the observed data ${\bf Y}$).\footnote{Note that for this Monte Carlo simulations approach, specification of quantity $\chi^2$ is actually very flexible. For example, one can potentially use $[Y_{t,n}-E(Y_{t,n}|{\bf\Theta},{\cal M})]^4/var(Y_{t,n}|{\bf\Theta},{\cal M})^2$ under the sum in equation~(\ref{CHI_SQUARED}) for the goodness-of-fit test. However, in this case $\chi^2$ would not become chi-square distributed even in the asymptotic limit of large accident frequencies.
}

In the case of accident severity models, we use the Pearson's $\chi^2$, defined as
\begin{eqnarray}
\chi^2=\sum\limits_{t=1}^T\sum\limits_{n=1}^{N_t}
\sum\limits_{i=1}^{I}
\frac{[\delta^{(i)}_{t,n}-P({i|{\bf\Theta},{\cal M}})]^2}
{P({i|{\bf\Theta},{\cal M}})},
\label{CHI_SQUARED_LOGIT}
\end{eqnarray}
where the accident severity outcome dummies $\delta^{(i)}_{t,n}$ are equal to unity if the $i^{\rm th}$ severity outcome is observed in the $n^{\rm th}$ accident that occurs during time period $t$, and to zero otherwise. According to equation~(\ref{EQ_MS_ML}), the theoretical unconditional probability of the $i^{\rm th}$ outcome is $P({i|{\bf\Theta},{\cal M}})={\bar p}_0{\cal ML}(i|{\betabf_{(0)}})+{\bar p}_1{\cal ML}(i|{\betabf_{(1)}})$. 

%% file: ch_MCMC_simul.tex
%
% This is the MCMC simulation methods chapter
%

\chapter{Markov Chain Monte Carlo simulation \\
methods}
\label{MODEL_SIMUL}
We use MCMC simulations for Bayesian inference and model estimation. This chapter presents MCMC simulation methods in detail. First, we describe a hybrid Gibbs sampler and the Metropolis-Hasting algorithm. Next, we explain a general Markov switching model representation that we use for all Markov switching models of accident frequencies and severities. After that we describe our choice of prior probability distribution. Then we give detailed step-by-step algorithm used for our MCMC simulations. Finally, in the end of this chapter, we briefly overview several important computational issues and optimizations that allow us to make Bayesian-MCMC estimation reliable, efficient and numerically accurate. For brevity, in this chapter we omit model specification notation ${\cal M}$ in all equations. For example, in this chapter we write the posterior distribution $f({\bf\Theta}|{\bf Y},{\cal M})$ simply as $f({\bf\Theta}|{\bf Y})$, and etc.

\section{Hybrid Gibbs sampler and Metropolis-Hasting algorithm}
\label{GIBBS}
As we have mentioned in the previous chapter, because the posterior distribution, given by the Bayes formula~(\ref{EQ_POSTERIOR}), is extremely difficult to find exactly, but is relatively easy to find with accuracy up to its normalization constant, we use Markov Chain Monte Carlo (MCMC) simulations. They provide a feasible statistical methodology for sampling from any probability distribution known up to a constant, the posterior distribution in our case.

To obtain draws of the parameters vector ${\bf\Theta}$ from a posterior distribution $f({\bf\Theta}|{\bf Y})$, we use the hybrid Gibbs sampler, which is an MCMC simulation algorithm that involves both Gibbs and Metropolis-Hasting sampling \citep[][]{MT_94,T_02,SAS_06}. Assume that ${\bf\Theta}$ is composed of $K$ components: ${\bf\Theta}=[\thetabf_1',\thetabf_2',...,\thetabf_K']'$ , where $\thetabf_k$ can be scalars or vectors, $k=1,2,...,K$. Then, the hybrid Gibbs sampler works as follows:
\begin{enumerate}
    \item
    Choose an arbitrary initial value of the parameter vector,       ${\bf\Theta}={\bf\Theta}^{(0)}$ , such that $f({\bf\Theta}^{(0)}|{\bf Y})>0\;$ [i.e. $f({\bf\Theta}^{(0)}|{\bf Y})\propto f({\bf Y},{\bf\Theta}^{(0)})=f({\bf Y}|{\bf\Theta}^{(0)})\pi({\bf\Theta}^{(0)})>0$].
    \item
    For each $g=1,2,3,\ldots\;$, parameter vector ${\bf\Theta}^{(g)}$ is generated component-by-component from ${\bf\Theta}^{(g-1)}$ by the following procedure:
    \begin{enumerate}
        \item
        First, draw $\thetabf^{(g)}_1$ from the conditional posterior probability distribution $f(\thetabf^{(g)}_1|{\bf Y},\thetabf^{(g-1)}_2,...,\thetabf^{(g-1)}_K)$. If this distribution is exactly known in a closed analytical form, then we draw $\thetabf^{(g)}_1$ directly from it. This is Gibbs sampling. If the conditional posterior distribution is known up to an unknown normalization constant, then we draw $\thetabf^{(g)}_1$ by using the Metropolis-Hasting ($\mbox{M-H}$) algorithm described below. This is $\mbox{M-H}$ sampling.
        \item
        Second, for all $k=2,3,...,K-1$, draw $\thetabf^{(g)}_k$ from the conditional posterior distribution $f(\thetabf^{(g)}_k|{\bf Y}, \thetabf^{(g)}_1,..., \thetabf^{(g)}_{k-1},\thetabf^{(g-1)}_{k+1},...,\thetabf^{(g-1)}_K)$ by using either Gibbs sampling (if the distribution is known exactly) or $\mbox{M-H}$ sampling (if the distribution is known up to a constant).
        \item
        Finally, draw $\thetabf^{(g)}_K$ from the conditional posterior probability distribution $f(\thetabf^{(g)}_K|{\bf Y},\thetabf^{(g)}_1,...,\thetabf^{(g)}_{K-1})$ by using either Gibbs or $\mbox{M-H}$ sampling.
    \end{enumerate}
    \item
    The resulting Markov chain $\{{\bf\Theta}^{(g)}\}$ converges to the true posterior distribution $f({\bf\Theta}|{\bf Y})$ as $g\rightarrow\infty$.
\end{enumerate}
Note that all conditional posterior distributions are proportional to the joint distribution $f({\bf Y},{\bf\Theta})=f({\bf Y}|{\bf\Theta})\pi({\bf\Theta})$. For example, we have
\begin{eqnarray}
f(\thetabf_k|{\bf Y}, \thetabf_1,..., \thetabf_{k-1},\thetabf_{k+1},...,\thetabf_K)=\frac{f({\bf Y}, \thetabf_1,..., \thetabf_{k-1},\thetabf_k,\thetabf_{k+1},...,\thetabf_K)}{f({\bf Y}, \thetabf_1,..., \thetabf_{k-1},\thetabf_{k+1},...,\thetabf_K)}
\nonumber\\
\propto f({\bf Y}, \thetabf_1,..., \thetabf_{k-1},\thetabf_k,\thetabf_{k+1},...,\thetabf_K)=f({\bf Y},{\bf\Theta}).
\label{EQ_CONDITIONAL_DISTRIBUTION}
\end{eqnarray}

By using the hybrid Gibbs sampler algorithm described above, we obtain a Markov chain $\{{\bf\Theta}^{(g)}\}$, where $g=1,2,\ldots,G_{bi},G_{bi}+1,\ldots,G$. We discard the first $G_{bi}$ ``burn-in'' draws because they can depend on the initial choice ${\bf\Theta}^{(0)}$. Of the remaining $G-G_{bi}$ draws, we typically store every third or every tenth draw in the computer memory. We use these draws for Bayesian inference. We typically choose $G$ ranging from $3\times10^5$ to $3\times10^6$, and  $G_{bi}=G/10$. In our study, a single MCMC simulation run takes from one day to couple weeks on a single computer CPU. We usually use eight different choices of the initial parameter vector ${\bf\Theta}^{(0)}$. Thus, we obtain eight Markov chains of ${\bf\Theta}$, and use them for the Brooks-Gelman-Rubin diagnostic of convergence of our MCMC simulations \citep[][]{BG_98}, for details see Section~\ref{MCMC_COMP_ISSUES} below. We also check convergence by monitoring the likelihood $f({\bf Y}|{\bf\Theta}^{(g)})$ and the joint distribution $f({\bf Y},{\bf\Theta}^{(g)})$.

We use the Metropolis-Hasting ($\mbox{M-H}$) algorithm to sample from conditional posterior distributions known up to their normalization constants.\footnote{In general, the M-H algorithm allows to make draws from any probability distribution known up to a constant. The algorithm converges as the number of draws goes to infinity.} Specifically, our goal here is to draw $\thetabf_k^{(g)}$ from $f(\thetabf_k|{\bf Y},\thetabf^{(g)}_1,...,\thetabf^{(g)}_{k-1},  \thetabf^{(g-1)}_{k+1},...,\thetabf^{(g-1)}_K)$ distribution that is not known exactly, so we cannot use the Gibbs sampling. The $\mbox{M-H}$ algorithm works as follows:
\begin{itemize}
    \item
    Choose a jumping probability distribution $J(\hat\thetabf_k|\thetabf_k)$ of $\hat\thetabf_k$. It must stay the same for all draws $g=G_{bi}+1,...,G$, and we discuss its choice below.
    \item
    Draw a candidate $\hat\thetabf_k$ from $J(\hat\thetabf_k|\thetabf_k^{(g-1)})$.
    \item
    Calculate ratio
    \begin{eqnarray}
    \hat p=\frac{f(\hat\thetabf_k|{\bf Y},\thetabf^{(g)}_1,
    \ldots,\thetabf^{(g)}_{k-1},\thetabf^{(g-1)}_{k+1},
    \ldots,\thetabf^{(g-1)}_K)}{f(\thetabf_k^{(g-1)}|{\bf Y},
    \thetabf^{(g)}_1,...,\thetabf^{(g)}_{k-1},
    \thetabf^{(g-1)}_{k+1},...,\thetabf^{(g-1)}_K)}\times
    \frac{J(\thetabf_k^{(g-1)}|\hat\thetabf_k)}
    {J(\hat\thetabf_k|\thetabf_k^{(g-1)})}.
    \label{EQ_MH_RATIO}
    \end{eqnarray}
    \item
    Set
    \begin{eqnarray}
    \thetabf_k^{(g)}=\left\{
    \begin{array}{ll}
    \hat\thetabf_k & \mbox{with probability $\min(\hat p,1)$},
    \\
    \thetabf_k^{(g-1)} & \mbox{otherwise}.
    \end{array}
    \right.
    \label{EQ_MH_DRAW}
    \end{eqnarray}
\end{itemize}
Note that the unknown normalization constant of $f(\ldots)$ cancels out in equation~(\ref{EQ_MH_RATIO}). Also, if the jumping distribution is symmetric $J(\hat\thetabf_k|\thetabf_k)=J(\thetabf_k|\hat\thetabf_k)$, then the ratio $J(\thetabf_k^{(g-1)}|\hat\thetabf_k)\big/ J(\hat\thetabf_k|\thetabf_k^{(g-1)})$ becomes equal to unity and Metropolis-Hasting algorithm reduces to Metropolis algorithm. The averaged acceptance rate of candidate values in equation~(\ref{EQ_MH_DRAW}) is recommended to range from $15$ to $50\%$. In this study, during the first $G_{bi}$ burn-in draws we make adjustments to the jumping probability distribution $J(\hat\thetabf_k|\thetabf_k)$ in order to achieve a $30\%$ averaged acceptance rate during the Metropolis-Hasting sampling (carried out during the remaining $G-G_{bi}$ draws used for Bayesian inference). The specifics about the choice of the jumping distribution and of its adjustments are given below in Sections~\ref{MCMC_SIMUL_APPL} - \ref{MCMC_COMP_ISSUES}.

\section{A general representation of Markov switching models}
\label{GENERAL_MS_MODEL}
All Markov switching models for accident frequencies and severities, specified in Sections~\ref{Markov_freq_year} - \ref{Markov_sever}, can be represented in a general, unified way. This representation allows us to estimate all models by using the same mathematical notations, computational methods and, most important, the same numerical code. In this section, first, we introduce a convenient general representation of Markov switching models considered in this study. Second, we show how Markov switching models for accident frequencies and severities, specified in Sections~\ref{Markov_freq_year} - \ref{Markov_sever}, are described by using this general representation.

For the general, unified representation of Markov switching between the roadway safety states over time, we would like to make the state variable to be dependent on time only. For this purpose, we introduce an auxiliary time index $\tilde t$, so that the state variable $s_{\tilde t}$ depends only on $\tilde t$. For example, in the case of annual frequencies of accidents occurring on $N$ roadway segments over $T$ annual time periods (this case is considered in Section~\ref{Markov_freq_year}), the auxiliary time is defined as ${\tilde t}\equiv t+(n-1)T$, where the real time is $t=1,2,...,T$ and the roadway segment number is $n=1,2,...,N$. The auxiliary time index runs from one to $N\times T$, that is ${\tilde t}=1,2,...,NT$. As another example, consider the case of weekly accident frequencies observed over $T$ weekly time periods (refer to Section~\ref{Markov_freq_week}). In this case the auxiliary time simply coincides with the real time, ${\tilde t}\equiv t$.

\begin{figure}[t]
\vspace{4.0truecm}
\includegraphics{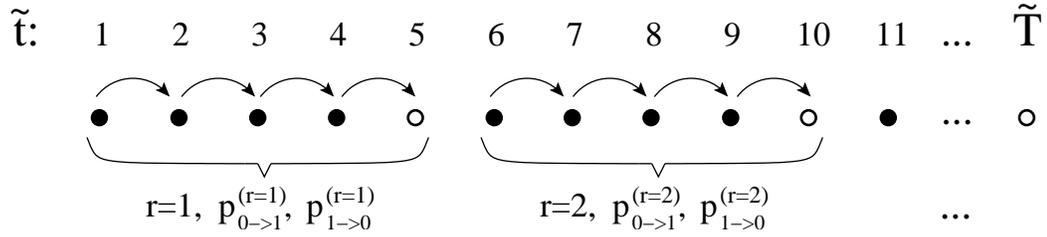}
\caption{Auxiliary time indexing of observations for a general Markov switching process representation.
\label{FIG_MARKOV}
}
\end{figure}

A general scenario of Markov switching between the roadway safety states over auxiliary time $\tilde t$ is schematically demonstrated in Figure~\ref{FIG_MARKOV}. The auxiliary time index runs from one to $\tilde T$, that is ${\tilde t}=1,2,...,{\tilde T}$. During an auxiliary time period $\tilde t$ the system is in state $s_{\tilde t}$ (which can be $0$ or $1$). As the auxiliary time index increases from $\tilde t$ to ${\tilde t}+1$, the state of roadway safety switches from $s_{\tilde t}$ to $s_{{\tilde t}+1}$. We assume that for all $\tilde t\notin{\cal T}_-$ (for all $t$ that do not belong set ${\cal T}_-$) this switching is Markovian, that is the probability distribution of $s_{{\tilde t}+1}$ depends on the value of $s_{\tilde t}$ (see Section~\ref{Markov}). We assume that for those values of $\tilde t$ that belong to the set ${\cal T}_-$, the switching is independent of the previous state, that is for $\tilde t\in{\cal T}_-$ the probability distribution of $s_{{\tilde t}+1}$ is independent of $s_{\tilde t}$ and of the earlier states.\footnote{Independent switching can be view as a special case of Markovian switching, see the discussion that follows equation~(\ref{EQ_MS_P_BAR})} The values ${\tilde t}\in{\cal T}_-$ are shown by white dots in Figure~(\ref{FIG_MARKOV}), the values ${\tilde t}\notin{\cal T}_-$ are shown by black dots, and the Markov switching transitions are shown by concave arrows. In a general case, the transition probabilities for Markov switching $s_{\tilde t}\rightarrow s_{{\tilde t}+1}$, where ${\tilde t}\notin{\cal T}_-$, do not need to be necessarily constant and can depend on the auxiliary time index $\tilde t$. As a result, we assume that there are $R$ auxiliary time intervals ${\cal T}(r)\le\tilde t<{\cal T}(r+1)$, $r=1,2,...,R$, such that the transition probabilities are constant inside each time interval and can differ from one interval to another. Here the set ${\cal T}$ contains, in an increasing order, all left boundaries of the time intervals, the first element of ${\cal T}$ is equal to $1$, and the last element of ${\cal T}$ is equal to $\tilde T+1$. Note that the size of set ${\cal T}$ (i.e. the number of elements in it) is equal to $R+1$. Thus, to repeat, for each value of index $r=1,2,...,R$, the transition probabilities $p^{(r)}_{0\rightarrow1}$ and $p^{(r)}_{1\rightarrow0}$ are constant inside the $r^{\rm th}$ interval ${\cal T}(r)\le\tilde t<{\cal T}(r+1)$. In Figure~(\ref{FIG_MARKOV}) the intervals of constant transition probabilities are shown by curly brackets beneath the dots.

In the real time $t$ all data observations (accident frequencies or severity outcomes) are counted by using the real time index, that is the vector of all observations is ${\bf Y}=\{Y_{t,n}\}$, where $t=1,2,...,T$ and $n=1,2,...,N_t$. When we change to the auxiliary time, all observations are counted by using the auxiliary time index, that is ${\bf Y}=\{Y_{{\tilde t},{\tilde n}}\}$, where ${\tilde t}=1,2,...,{\tilde T}$ and ${\tilde n}=1,2,...,{\tilde N}_{\tilde t}$. Here $N_t$ and ${\tilde N}_{\tilde t}$ are the number of observations during real and auxiliary time periods $t$ and $\tilde t$ respectively. There is always a unique correspondence between the indexing pairs $(t,n)$ and $({\tilde t},{\tilde n})$. Using the auxiliary time indexing, the likelihood function $f({\bf Y}|{\bf\Theta})$, given by equation~(\ref{EQ_MS_L}), becomes
\begin{eqnarray}
f({\bf Y}|{\bf\Theta}) & = & \prod\limits_{\tilde t=1}^{\tilde T}\prod\limits_{\tilde n=1}^{\tilde N_{\tilde t}} P({Y_{\tilde t,{\tilde n}}|{\bf\Theta}})=\prod\limits_{\tilde t=1}^{\tilde T}\prod\limits_{\tilde n=1}^{\tilde N_{\tilde t}}\left\{
\begin{array}{ll}
f({Y_{\tilde t,\tilde n}|{\tilde\betabf_{(0)}}}) & \mbox{if $s_{\tilde t}=0$}
\\
f({Y_{\tilde t,\tilde n}|{\tilde\betabf_{(1)}}}) & \mbox{if $s_{\tilde t}=1$}
\end{array}
\right\}
\nonumber\\
& = & \left[\prod\limits_{\{{\tilde t}:\:s_{\tilde t}=0\}}\prod\limits_{\tilde n=1}^{\tilde N_{\tilde t}}
f(Y_{{\tilde t},{\tilde n}}|{\tilde\betabf}_{(0)})\right]\times
\left[\prod\limits_{\{{\tilde t}:\:s_{\tilde t}=1\}}\prod\limits_{\tilde n=1}^{\tilde N_{\tilde t}}
f(Y_{{\tilde t},{\tilde n}}|{\tilde\betabf}_{(1)})\right]
\label{EQ_MS_L_1}
\end{eqnarray}
where $f(Y_{{\tilde t},{\tilde n}}|{\tilde\betabf}_{(0)})$ and $f(Y_{{\tilde t},{\tilde n}}|{\tilde\betabf}_{(1)})$ are the model likelihoods of single observations $Y_{{\tilde t},{\tilde n}}$ in roadway safety states $s_{\tilde t}=0$ and $s_{\tilde t}=1$ respectively. Set $\{\tilde t\!:s_{\tilde t}=0\}$ is defined as all values of $\tilde t$ such that $1\le{\tilde t}\le{\tilde T}$ and $s_{\tilde t}=0$, and set $\{\tilde t\!:s_{\tilde t}=1\}$ is defined analogously. Vectors $\tilde\betabf_{(0)}$ and $\tilde\betabf_{(1)}$ are the model parameters vectors in the states $0$ and $1$, these vectors are specified by the model type as follows:
\begin{eqnarray}
\tilde\betabf_{(s)} & = & \left\{
\begin{array}{ll}
\betabf_{(s)} & \mbox{for Poisson or multinomial logit},
\\
{}[{\betabf'_{(s)}},\alpha_{(s)}]' & \mbox{for negative binomial},
\\
{}[{\betabf'_{(s)}},\tau_{(s)}]'\:\mbox{or}\:{}[{\betabf'_{(s)}},
\alpha_{(s)},\tau_{(s)}]' & \mbox{for ZIP-$\tau$ or ZINB-$\tau$},
\\
{}[{\betabf'_{(s)}},\gammabf'_{(s)}]'\:
\mbox{or}\:{}[{\betabf'_{(s)}},\alpha_{(s)},\gammabf'_{(s)}]' &
\mbox{for ZIP-$\gamma$ or ZINB-$\gamma$ models},
\end{array}
\right.
\label{EQ_beta_tilde}
\end{eqnarray}
where $s=0,1$ are the state values. Scalar $\tau$ and vector $\gammabf$ are estimable zero-inflated model parameters, and $\alpha$ is the over-dispersion parameter, as defined in Section~\ref{STANDARD_FREQUENCY}.

By defining the auxiliary time $\tilde t$ and sets ${\cal T}_-$ and ${\cal T}$, we specify the general unified representation of the Markov switching models introduced in Chapter~\ref{MODEL_SPECIF}, as follows:
\begin{itemize}
    \item For Markov switching models of annual accident frequencies, introduced in Section~\ref{Markov_freq_year}, we have
        \begin{eqnarray}
        &&{\tilde t}=t+(n-1)T,\qquad{\tilde T}=N\times T,\qquad
        {\tilde n}=1,\qquad{\tilde N_{\tilde t}}=1,
        \label{EQ_general_freq_year_1}\\
        &&{\cal T}_-=\{nT,\;\mbox{ where }n=1,...,N\},
        \label{EQ_general_freq_year_2}\\
        &&{\cal T}=\{1+(r-1)T,\:(1+NT)\},\qquad r=1,...,N,
        \qquad R=N,\qquad
        \label{EQ_general_freq_year_3}\\
        &&n=\lceil{\tilde t}/T\rceil\;\mbox{~and~}\;
        t={\tilde t}-(n-1)T,
        \label{EQ_general_freq_year_4}
        \end{eqnarray}
        where $t=1,2,...,T$ and $n=1,2,...,N$ are the real time index and the roadway segment number respectively, and $\lceil x\rceil$ is the ``ceil'' function that returns the smallest integer not less than $x$. Here $T$ is the number of annual time periods, and $N$ is the number of roadway segments observed during each period. The change of indexing to auxiliary time $\tilde t$, given by equation~(\ref{EQ_general_freq_year_1}), is demonstrated in Figure~\ref{FIG_MARKOV} for the case when $T=5$ (in Section~\ref{RESULTS_FREQ_YEAR} we will consider a five-year accident frequency data). Separate roadway segments $n=1,2,..,N$ have different transition probabilities for their states of roadway safety [refer to equation~(\ref{EQ_MS_P})]. Therefore, in Equation~(\ref{EQ_general_freq_year_3}) the time interval number $r$ coincides with the roadway segment number $n$, that is $r=n$ and $R=N$. Equation~(\ref{EQ_general_freq_year_2}) follows from the fact that states $s_{\tilde t}$ are independent for different roadway segments $n=1,2,...,N$. Equation~(\ref{EQ_general_freq_year_4}) gives the conversion from the auxiliary time indexing back to the real time indexing.

        The observations are annual accident frequencies $A_{t,n}$ (refer to Sections~\ref{STANDARD_FREQUENCY} and~\ref{Markov_freq_year}). Therefore, we have $Y_{{\tilde t},{\tilde n}}=Y_{{\tilde t},1}=Y_{t,n}=A_{t,n}$, where $t$ and $n$ are calculated from ${\tilde t}$ by using equations~(\ref{EQ_general_freq_year_4}). Thus, according to equations~(\ref{EQ_MS_P_year}) and~(\ref{EQ_MS_NB_year}), the likelihood functions of a single observation $Y_{{\tilde t},{\tilde n}}=Y_{{\tilde t},1}=A_{t,n}$ in the states $0$ and $1$ are
        \begin{eqnarray}
        \begin{array}{l}
        f({Y_{\tilde t,\tilde n}|{\tilde\betabf_{(0)}}})=
        f({Y_{\tilde t,1}|{\tilde\betabf_{(0)}}})=
        {\cal I}(A_{t,n}),
        \\
        f({Y_{\tilde t,\tilde n}|{\tilde\betabf_{(1)}}})=
        f({Y_{\tilde t,1}|{\tilde\betabf_{(1)}}})=
        {\cal P}(A_{t,n}|{\tilde\betabf_{(1)}})
        \end{array}
        \label{EQ_MS_P_year_1}
        \end{eqnarray}
        for the MSP model of annual accident frequencies, and
        \begin{eqnarray}
        \begin{array}{l}
        f({Y_{\tilde t,\tilde n}|{\tilde\betabf_{(0)}}})=
        f({Y_{\tilde t,1}|{\tilde\betabf_{(0)}}})=
        {\cal I}(A_{t,n}),
        \\
        f({Y_{\tilde t,\tilde n}|{\tilde\betabf_{(1)}}})=
        f({Y_{\tilde t,1}|{\tilde\betabf_{(1)}}})=
        {\cal NB}(A_{t,n}|{\tilde\betabf_{(1)}})
        \end{array}
        \label{EQ_MS_NB_year_1}
        \end{eqnarray}
        for the MSNB model of annual accident frequencies. Here ${\tilde n}=1$, while $t$ and $n$ are calculated from ${\tilde t}$ by using equations~(\ref{EQ_general_freq_year_4}). Keep in mind that ${\tilde\betabf_{(1)}}$ is given by equation~(\ref{EQ_beta_tilde}).
    \item For Markov switching models of weekly accident frequencies, introduced in Section~\ref{Markov_freq_week}, we have
        \begin{eqnarray}
        &&{\tilde t}=t,\qquad{\tilde T}=T,\qquad
        {\tilde n}=n,\qquad{\tilde N_{\tilde t}}=N,
        \label{EQ_general_freq_week_1}\\
        &&{\cal T}_-=\{\emptyset\},\qquad {\cal T}=\{1,\:(T+1)\},
        \qquad r=1, \qquad R=1,
        \label{EQ_general_freq_week_2}
        \end{eqnarray}
        where $t$ and $n$ are the real time index and roadway segment number respectively, $T$ is the number of weekly time periods, and $N$ is the number of roadway segments observed (it is the same for all periods). Here the auxiliary time $\tilde t$ coincides with the real time $t$. The transition probabilities are constant over all time periods ${\tilde t}=t$ and are the same for all roadway segments $n=1,2,...,N$. Thus, $R=1$, set ${\cal T}$ consists of just two values, and set ${\cal T}_-$ is empty.

        The observations are weekly accident frequencies $A_{t,n}$ (refer to Section~\ref{Markov_freq_week}). Therefore, we have $Y_{{\tilde t},{\tilde n}}=Y_{t,n}=A_{t,n}$, where we use ${\tilde t}=t$ and ${\tilde n}=n$. Thus, according to equations~(\ref{EQ_MS_P_week}) and~(\ref{EQ_MS_NB_week}), the likelihood functions of a single observation $Y_{{\tilde t},{\tilde n}}=A_{t,n}$ in the states $0$ and $1$ are
        \begin{eqnarray}
        f({Y_{\tilde t,\tilde n}|{\tilde\betabf_{(0)}}})={\cal P}(A_{t,n}|{\tilde\betabf_{(0)}}),\quad\quad
        f({Y_{\tilde t,\tilde n}|{\tilde\betabf_{(1)}}})={\cal P}(A_{t,n}|{\tilde\betabf_{(1)}})
        \label{EQ_MS_P_week_1}
        \end{eqnarray}
        for the MSP model of weekly accident frequencies, and
        \begin{eqnarray}
        f({Y_{\tilde t,\tilde n}|{\tilde\betabf_{(0)}}})={\cal NB}(A_{t,n}|{\tilde\betabf_{(0)}}),\quad\quad
        f({Y_{\tilde t,\tilde n}|{\tilde\betabf_{(1)}}})={\cal NB}(A_{t,n}|{\tilde\betabf_{(1)}})
        \label{EQ_MS_NB_week_1}
        \end{eqnarray}
        for the MSNB model of weekly accident frequencies. Here $t={\tilde t}$ and $n={\tilde n}$. Note that ${\tilde\betabf_{(0)}}$ and ${\tilde\betabf_{(1)}}$ are given by equation~(\ref{EQ_beta_tilde}).
    \item For Markov switching models of accident severities, introduced in Section~\ref{Markov_sever}, we again consider weekly time periods and, therefore, have formulas very similar to equations~(\ref{EQ_general_freq_week_1})--(\ref{EQ_general_freq_week_2}),
        \begin{eqnarray}
        &&{\tilde t}=t,\qquad{\tilde T}=T,\qquad
        {\tilde n}=n,\qquad{\tilde N_{\tilde t}}=N_t,
        \label{EQ_general_sever_1}\\
        &&{\cal T}_-=\{\emptyset\},\qquad {\cal T}=\{1,\:(T+1)\},
        \qquad r=1, \qquad R=1.
        \label{EQ_general_sever_2}
        \end{eqnarray}
        Here, the auxiliary time $\tilde t$ again coincides with the real time $t$, scalar $T$ is the total number of weekly time periods, and $N_t$ is the number of accidents occurring during time period $t$.

        The observations are accident severity outcome dummies $\delta^{(i)}_{t,n}$ (refer to Section~\ref{Markov_sever}). Thus, we have $Y_{{\tilde t},{\tilde n}}=Y_{t,n}=\{\delta^{(i)}_{t,n}\}$, where $i=1,2,...,I$ and we use ${\tilde t}=t$ and ${\tilde n}=n$. According to equation~(\ref{EQ_MS_ML}), the likelihood functions of a single observation $Y_{{\tilde t},{\tilde n}}$ in the states $0$ and $1$ are
        \begin{eqnarray}
        f({Y_{\tilde t,\tilde n}|{\tilde\betabf_{(0)}}})=\prod\limits_{i=1}^I
        \left[{\cal ML}(i|{\tilde\betabf_{(0)}})\right]^{\delta^{(i)}_{t,n}},
        \nonumber\\
        f({Y_{\tilde t,\tilde n}|{\tilde\betabf_{(1)}}})=\prod\limits_{i=1}^I
        \left[{\cal ML}(i|{\tilde\betabf_{(1)}})\right]^{\delta^{(i)}_{t,n}},
        \label{EQ_MS_ML_1}
        \end{eqnarray}
        where $t={\tilde t}$ and $n={\tilde n}$. Note that ${\tilde\betabf_{(0)}}$ and ${\tilde\betabf_{(1)}}$ are given by equation~(\ref{EQ_beta_tilde}).
\end{itemize}

In the remaining sections of this chapter we use the above general representation of Markov switching models. For convenience and brevity of the presentation, we drop tildes~($\sim$) from all our notations. In other words, we use $t$, $T$, $n$, $N_t$ and $\betabf$ instead of $\tilde t$, $\tilde T$, $\tilde n$, $\tilde N_{\tilde t}$ and $\tilde\betabf$. We also call ``auxiliary time'' just ``time''. Thus, it is good to keep in mind that, in the rest of this chapter, time index/period/interval means auxiliary time index/period/interval.

\section{Choice of the prior probability distribution}
\label{PRIOR_CHOICE}
A full specification of Bayesian methodology and model estimation
requires a specification of the prior probability distribution. In this section we describe how we choose the prior distribution $\pi({\bf\Theta})$ of the vector ${\bf\Theta}$ of all parameters to be estimated. In our study, for the general representation given in the previous section, vector ${\bf\Theta}$ includes all unobservable state variables ($s_t$), model parameters ($\betabf_{(0)}$, $\betabf_{(1)}$) and transition probabilities for every $r^{\rm th}$ time interval ($p^{(r)}_{0\rightarrow1}$, $p^{(r)}_{1\rightarrow0}$, $r=1,2,...,R$). Thus,
\begin{eqnarray}
{\bf\Theta} & = & [\betabf_{(0)}',\betabf_{(1)}',p^{(1)}_{0\rightarrow1},...,
p^{(R)}_{0\rightarrow1},p^{(1)}_{1\rightarrow0},...,
p^{(R)}_{1\rightarrow0},{\bf S}']'.
\label{EQ_THETA_general}
\end{eqnarray}
Here, vectors $\betabf_{(0)}$ and $\betabf_{(1)}$ are the model parameter vectors for states $s=0$ and $s=1$, which are defined in equation~(\ref{EQ_beta_tilde}). Vector ${\bf S}=[s_1,s_2,...,s_T]'$ contains all state values and has length $T$, which is the total number of time periods.

The prior distribution is supposed to reflect our prior knowledge of the model parameters \citep[][]{SAS_06}. We choose the prior distributions of $\betabf_{(0)}$, $\betabf_{(1)}$, $p^{(r)}_{0\rightarrow1}$ and $p^{(r)}_{1\rightarrow0}$ ($r=1,2,...,R$) to be nearly flat and essentially non-informative.\footnote{equation~(\ref{EQ_POSTERIOR}) shows that for nearly flat prior distributions, when $\pi({\bf\Theta}|{\cal M})$ is approximately constant around the peak of the likelihood function, the posterior distribution only weakly depends on the exact choice of the prior. We have verified this result during our test MCMC runs.} The prior distribution of the state vector ${\bf S}$ must reflect the Markov switching property of the state variable $s_t$. The overall prior distribution of the vector ${\bf\Theta}$ of all parameters is chosen to be the product of the prior distributions of all its components [refer to equation~(\ref{EQ_THETA_general})]. Thus, our choice of the prior is as follows:
\begin{itemize}
    \item
    Prior probability distribution of model parameters vectors $\betabf_{(s)}$ is the product of prior distributions for the vector components in states $s=0$ and $s=1$,
    \begin{eqnarray}
    \pi(\betabf_{(0)},\betabf_{(1)})=\prod\limits_{s=0}^1
    \prod\limits_{k=1}^{K_{(s)}}\pi(\beta_{(s),k}),
    \label{EQ_PRIOR_beta_tilde}
    \end{eqnarray}
    where $\beta_{(s),k}$ is the $k^{\rm th}$ component of vector $\betabf_{(s)}$, and $K_{(s)}$ is the length of vector $\betabf_{(s)}$ (i.e. the number of model parameters in the state $s$ is equal to $K_{(s)}$, where $s=0,1$). For free parameters $\beta_{(s),k}$ (which are free to be estimated), the priors of $\beta_{(s),k}$ are chosen to be normal distributions: $\pi(\beta_{(s),k})={\cal N}(\beta_{(s),k}|\mu_k,\Sigma_k)$. [Keep in mind that for NB models $\ln(\alpha)$ is estimated instead of the over-dispersion parameter $\alpha$, and, thus, the prior distribution of $\alpha$ is log-normal.] Parameters that enter the prior distributions are called hyper-parameters. For these, the means $\mu_k$ are chosen to be equal to the maximum likelihood estimation (MLE) values of $\beta_k$ for the corresponding standard single-state models (Poisson, NB, ZIP, ZINB and multinomial logit models in this study). The variances $\Sigma_k$ are chosen to be ten times larger than the maximum between the MLE values of $\beta_k$ squared and the MLE variances of $\beta_k$ for the corresponding standard models (thus, variances $\Sigma_k$ are chosen to be relatively large in order to have wide prior distributions of $\beta_{(s),k}$).

    All $\beta$-parameters can be either free (which are free to be estimated) or restricted (which are not free to be estimated, but instead are set to some predetermined values). We choose normally-distributed priors only for free parameters. In this study, if a parameter is not free, then there are only three other possibilities: the non-free parameter is restricted to be equal to either zero, or $-\infty$, or a free parameter. Thus, in all these three cases we have prior knowledge about the value of the restricted parameter. For simplicity of presentation, in equation~(\ref{EQ_PRIOR_beta_tilde}) and below we do not explicitly show which $\beta$-parameters are free and which are restricted, and for presentation purposes only we portray all $\beta$-parameters as being free. However, it is important to remember that during numerical MCMC simulations we do not draw restricted parameters, but, instead, we set them to the appropriate values that they are restricted to.\footnote{A non-free parameter that is restricted to a free parameter is set immediately after the free parameter is drawn during the hybrid Gibbs sampler simulations. This is because these two parameters (the restricted ``child'' parameter and its ``parent'' free parameter) must always be the same. For example, if we have three beta-parameters $\beta_1$, $\beta_2$ and $\beta_3$, and if $\beta_3$ is restricted to $\beta_1$, then $\beta_3$ is set to the new value of $\beta_1$ immediately after this new value is drawn.}
    \item
    For weekly accident frequency and severity models, introduced in Sections~\ref{Markov_freq_week} and~\ref{Markov_sever}, the joint prior distribution for all transition probabilities $\{p^{(r)}_{0\rightarrow1},p^{(r)}_{1\rightarrow0}\}$, where $r=1,2,...,R$ (note that $R=1$ in case of basic weekly models), is
    \begin{eqnarray}
    \pi(\{p^{(r)}_{0\rightarrow1},p^{(r)}_{1\rightarrow0}\})\propto
    \prod\limits^R_{r=1}\pi(p^{(r)}_{0\rightarrow1})
    \pi(p^{(r)}_{1\rightarrow0})I(p^{(r)}_{0\rightarrow1}\le p^{(r)}_{1\rightarrow0}).
    \label{EQ_PRIOR_P}
    \end{eqnarray}
    Here $\pi(p^{(r)}_{0\rightarrow1})={\cal B}eta(p^{(r)}_{0\rightarrow1}|\upsilon_0,\nu_0)$ and $\pi(p^{(r)}_{1\rightarrow0})={\cal B}eta(p^{(r)}_{1\rightarrow0}|\upsilon_1,\nu_1)$ are chosen to be standard beta distributions. Function $I(p^{(r)}_{0\rightarrow1}\le p^{(r)}_{1\rightarrow0})$ is defined as equal to unity if restriction $p^{(r)}_{0\rightarrow1}\le p^{(r)}_{1\rightarrow0}$ is satisfied and to zero otherwise [refer to  equation~(\ref{EQ_MS_P_RESTRICT})]. For annual accident frequency models, introduced in Sections~\ref{Markov_freq_year}, the prior distribution for transition probabilities is given by equation~(\ref{EQ_PRIOR_P}) with functions $I(p^{(r)}_{0\rightarrow1}\le p^{(r)}_{1\rightarrow0})$ dropped out because there are no any restrictions for transition probabilities in this case [note that equation~(\ref{EQ_PRIOR_P}) becomes an equality in this case]. Thus, in the case of annual accident frequency models, functions $I(p^{(r)}_{0\rightarrow1}\le p^{(r)}_{1\rightarrow0})$ should be left out from all formulas in the rest of this chapter. The hyper-parameters in equation~(\ref{EQ_PRIOR_P}) are chosen to be $\upsilon_0=\nu_0=\upsilon_1=\nu_1=1$, in which case the beta distributions become the uniform distribution between zero and one. Similar to parameters $\beta_{(s),k}$, we draw only free transition probability parameters $p^{(r)}_{0\rightarrow1}$ and $p^{(r)}_{1\rightarrow0}$. All restricted transition probabilities are not drawn, but are set to the values that they are restricted to.
    \item
    The prior distribution for the state vector ${\bf S}=[s_1,s_2,...,s_T]'$ is equal to the likelihood function of $\bf S$ given the transitional probabilities $\{p^{(r)}_{0\rightarrow1},p^{(r)}_{1\rightarrow0}\}$,
    \begin{eqnarray}
    && f({\bf S}|\{p^{(r)}_{0\rightarrow1},p^{(r)}_{1\rightarrow0}\})=
    P(s_1)
    \prod\limits_{\left\{t:\:1\le t<T,\atop t\in {\cal T}_-\right\}}P(s_{t+1})
    \prod\limits_{\left\{t:\:1\leq t<T,\atop t\notin {\cal T}_-\right\}}P(s_{t+1}|s_t)
    \nonumber\\
    && \quad\quad\propto\prod\limits_{\left\{t:\:1\leq t<T,\atop
    t\notin{\cal T}_-\right\}}P(s_{t+1}|s_t)
    \nonumber\\
    && \quad\quad=\prod\limits^R_{r=1}
    \prod\limits_{\left\{t:\:{\cal T}(r)\leq t<{\cal T}(r+1),\atop
    t<T,\;\;t\notin{\cal T}_-\right\}}P(s_{t+1}|s_t)
    \nonumber\\
    && \quad\quad=\prod\limits^R_{r=1}\;
    [p^{(r)}_{0\rightarrow1}]^{m^{(r)}_{0\rightarrow1}}
    [1-p^{(r)}_{0\rightarrow1}]^{m^{(r)}_{0\rightarrow0}}
    [p^{(r)}_{1\rightarrow0}]^{m^{(r)}_{1\rightarrow0}}
    [1-p^{(r)}_{1\rightarrow0}]^{m^{(r)}_{1\rightarrow1}}.
    \label{EQ_PRIOR_S}
    \end{eqnarray}
    Here, index $r=1,2,...,R$ counts time intervals ${\cal T}(r)\leq t<{\cal T}(r+1)$ of constant transition probabilities $p^{(r)}_{0\rightarrow1}$ and $p^{(r)}_{1\rightarrow0}$ (see Section~\ref{GENERAL_MS_MODEL}). Number $m^{(r)}_{i\rightarrow j}$ is the total number of Markov switching state transitions from $s_t=i$ to $s_{t+1}=j$ inside time interval ${\cal T}(r)\leq t<{\cal T}(r+1)$ [here $\,i,j=\{0,1\}\,$ and history-independent transitions at $t\in{\cal T}_-$ are not counted]. In equation~(\ref{EQ_PRIOR_S}) we disregard probability distribution $P(s_1)$ and distributions $P(s_{t+1})$ for $t\in{\cal T}_-\,$. This is because their contribution is negligible when $T$ is large and the number of elements in set ${\cal T}_-$ is small relative to the value of $T$, which is true in this study.\footnote{Alternatively, we can assume that $P(s_1=0)=P(s_1=1)=1/2$~ and $P(s_{t+1}=0)=P(s_{t+1}=1)=1/2$ for all $t\in{\cal T}_-$. A more sophisticated alternative (not considered here) would be to treat these probabilities as free estimable parameters of the model.}

    It is important to note that formula~(\ref{EQ_PRIOR_S}) for distribution $f({\bf S}|\{p^{(r)}_{0\rightarrow1},p^{(r)}_{1\rightarrow0}\})$ follows from the Markov switching property of the state variable $s_t$ [compare to equation~(\ref{EQ_MS_P})]. In other words, we a priori specify that the state variable $s_t$ follows a Markov process in time, with transition probabilities $\{p^{(r)}_{0\rightarrow1},p^{(r)}_{1\rightarrow0}\}$, and this specification must be and is reflected in the prior distribution given by equation~(\ref{EQ_PRIOR_S}).
    \item
    Finally, the prior probability distribution $\pi({\bf\Theta})$ of parameter vector ${\bf\Theta}$, which is given by equation~(\ref{EQ_THETA_general}), is the product of the priors of all ${\bf\Theta}$'s components, given by equations~(\ref{EQ_PRIOR_beta_tilde}) - (\ref{EQ_PRIOR_S}),
    \begin{eqnarray}
    \pi({\bf\Theta}) & = & \pi({\bf S},\{p^{(r)}_{0\rightarrow1},p^{(r)}_{1\rightarrow0}\},
    \betabf_{(0)},\betabf_{(1)})
    \nonumber\\
    & = & f({\bf S}|\{p^{(r)}_{0\rightarrow1},p^{(r)}_{1\rightarrow0}\})
    \pi(\{p^{(r)}_{0\rightarrow1},p^{(r)}_{1\rightarrow0}\})
    \pi(\betabf_{(0)},
    \betabf_{(1)})
    \nonumber\\
    & \propto & \prod\limits_{\left\{t:\:1\leq t<T,\atop t\notin{\cal T}_-\right\}}P(s_{t+1}|s_t)
    \nonumber\\
    && \times\prod\limits^R_{r=1}{\cal B}eta(p^{(r)}_{0\rightarrow1}|\upsilon_0,\nu_0)
    {\cal B}eta(p^{(r)}_{1\rightarrow0}|\upsilon_1,\nu_1)
    I(p^{(r)}_{0\rightarrow1}\le p^{(r)}_{1\rightarrow0})
    \nonumber\\
    && \times\prod\limits_{s=0}^1\prod\limits_{k=1}^{K_{(s)}}
    {\cal N}(\beta_{(s),k}|\mu_k,\Sigma_k).
    \label{EQ_PRIOR}
    \end{eqnarray}
\end{itemize}

\section{MCMC simulations: step-by-step algorithm}
\label{MCMC_SIMUL_APPL}
In our research, for Bayesian inference on the parameter vector ${\bf\Theta}$, given by equation~(\ref{EQ_THETA_general}), we apply the hybrid Gibbs sampler and make draws of the components of vector ${\bf\Theta}$ from their conditional posterior distributions (refer to Section~\ref{GIBBS}). All conditional posterior distributions are proportional to the joint distribution $f({\bf Y},{\bf\Theta})$ [see equation~(\ref{EQ_CONDITIONAL_DISTRIBUTION})]. The joint distribution is $f({\bf Y},{\bf\Theta})=f({\bf Y}|{\bf\Theta})\pi({\bf\Theta})$, where the likelihood $f({\bf Y}|{\bf\Theta})$ is given by equation~(\ref{EQ_MS_L_1}) and the prior $\pi({\bf\Theta})$ is given by equation~(\ref{EQ_PRIOR}). Thus, for the joint distribution we have
\begin{eqnarray}
f({\bf Y},{\bf\Theta}) & = & f({\bf Y}|{\bf\Theta})\pi({\bf\Theta})
\nonumber\\
& \propto & \left[\prod\limits_{\{t:\:s_t=0\}}
\prod\limits_{n=1}^{N_t}f({Y_{t,n}|{\betabf_{(0)}}})\right]
\times\left[\prod\limits_{\{t:\:s_t=1\}}
\prod\limits_{n=1}^{N_t}f({Y_{t,n}|{\betabf_{(1)}}})\right]
\nonumber\\
&& \times \prod\limits_{\left\{t:\:1\leq t<T,\atop t\notin{\cal T}_-\right\}}P(s_{t+1}|s_t)
\nonumber\\
&& \times\prod\limits^R_{r=1}{\cal B}eta(p^{(r)}_{0\rightarrow1}|\upsilon_0,\nu_0)
{\cal B}eta(p^{(r)}_{1\rightarrow0}|\upsilon_1,\nu_1)
I(p^{(r)}_{0\rightarrow1}\le p^{(r)}_{1\rightarrow0})
\nonumber\\
&& \times\left[\prod\limits_k^{K_{(0)}}{\cal N}(\beta_{(0),k}|\mu_k,\Sigma_k)\right]
\times\left[\prod\limits_k^{K_{(1)}}{\cal N}(\beta_{(1),k}|\mu_k,\Sigma_k)\right].
\label{EQ_JOINT_L_PRIOR}
\end{eqnarray}
As a result, the conditional posterior distributions of all components of vector ${\bf\Theta}$, which are proportional to the joint distribution, are as follows:
\begin{itemize}
    \item
    The conditional posterior distribution of the $k^{\rm th}$ component of vector $\betabf_{(0)}$ is
    \begin{eqnarray}
    && f(\beta_{(0),k}|{\bf Y},{\bf\Theta}\backslash
    \beta_{(0),k})=\frac{f(\beta_{(0),k},{\bf Y},{\bf\Theta}
    \backslash\beta_{(0),k})}{f({\bf Y},{\bf\Theta}
    \backslash\beta_{(0),k})}\propto f({\bf Y},{\bf\Theta})
    \nonumber\\
    && \quad\quad\propto\left[\prod\limits_{\{t:\:s_t=0\}}
    \prod\limits_{n=1}^{N_t}f({Y_{t,n}|{\betabf_{(0)}}})\right]
    \times{\cal N}(\beta_{(0),k}|\mu_k,\Sigma_k)
    \nonumber\\
    && \quad\quad=\left[\prod\limits_{\{t:\:s_t=0\}}
    \prod\limits_{n=1}^{N_t}f({Y_{t,n}|{\betabf_{(0)}}})\right]
    \times\frac{1}{\sqrt{2\pi\Sigma_k}}
    e^{\left.-[\beta_{(0),k}-\mu_k]^2\right/2\Sigma_k}
    \nonumber\\
    && \quad\quad\propto\left[\prod\limits_{\{t:\:s_t=0\}}
    \prod\limits_{n=1}^{N_t}f({Y_{t,n}|{\betabf_{(0)}}})\right]
    \times e^{\left.-[\beta_{(0),k}-\mu_k]^2\right/2\Sigma_k},
    \label{EQ_COND_BETA0_DRAWS}
    \end{eqnarray}
    where ${\bf\Theta}\backslash\beta_{(0),k}$ means all components of ${\bf\Theta}$ except $\beta_{(0),k}$, and we keep only those multipliers that depend on $\beta_{(0),k}$. In equation~(\ref{EQ_COND_BETA0_DRAWS}) the conditional posterior distribution of $\beta_{(0),k}$ is known up to an unknown normalization constant. Therefore, we draw free parameters $\beta_{(0),k}$ by using the Metropolis-Hasting algorithm described in Section~\ref{GIBBS}. Note that $k=1,2,...,K_{(0)}$, where $K_{(0)}$ is the number of model's $\beta$-parameters in state $0$.
    \item
    The conditional posterior distribution of the $k^{\rm th}$ component of vector $\betabf_{(1)}$, is derived similarly to the conditional posterior distribution of $\beta_{(0),k}$ in equation~(\ref{EQ_COND_BETA0_DRAWS}),
    \begin{eqnarray}
    && f(\beta_{(1),k}|{\bf Y},{\bf\Theta}\backslash
    \beta_{(1),k})\propto f({\bf Y},{\bf\Theta})
    \nonumber\\
    && \quad\quad\propto\left[\prod\limits_{\{t:\:s_t=1\}}
    \prod\limits_{n=1}^{N_t}f({Y_{t,n}|{\betabf_{(1)}}})\right]
    \times e^{\left.-[\beta_{(1),k}-\mu_k]^2\right/2\Sigma_k}.
    \label{EQ_COND_BETA1_DRAWS}
    \end{eqnarray}
    Free parameters $\beta_{(1),k}$, where $k=1,2,...,K_{(1)}$, are also drawn by using the Metropolis-Hasting algorithm.
    \item
    The conditional posterior distribution of the transition probability $p^{(r)}_{0\rightarrow1}$ is
    \begin{eqnarray}
    && f(p^{(r)}_{0\rightarrow1}|{\bf Y},{\bf\Theta}\backslash p^{(r)}_{0\rightarrow1})=
    \frac{f(p^{(r)}_{0\rightarrow1},{\bf Y},{\bf\Theta}\backslash p^{(r)}_{0\rightarrow1})}
    {f({\bf Y},{\bf\Theta}\backslash p^{(r)}_{0\rightarrow1})}\propto
    f({\bf Y},{\bf\Theta})
    \nonumber\\
    %--------------
    && \quad\quad\propto\prod\limits_{\left\{t:\:1\leq t<T,\atop t\notin{\cal T}_-\right\}}P(s_{t+1}|s_t)
    \nonumber\\
    && \quad\quad\quad\times\:{\cal B}eta(p^{(r)}_{0\rightarrow1}|\upsilon_0,\nu_0)
    I(p^{(r)}_{0\rightarrow1}\le p^{(r)}_{1\rightarrow0})
    \nonumber\\
    %--------------
    && \quad\quad=\prod\limits^R_{{\hat r}=1}\;
    [p^{({\hat r})}_{0\rightarrow1}]^{m^{({\hat r})}_{0\rightarrow1}}
    [1-p^{({\hat r})}_{0\rightarrow1}]^{m^{({\hat r})}_{0\rightarrow0}}
    [p^{({\hat r})}_{1\rightarrow0}]^{m^{({\hat r})}_{1\rightarrow0}}
    [1-p^{({\hat r})}_{1\rightarrow0}]^{m^{({\hat r})}_{1\rightarrow1}}
    \nonumber\\
    && \quad\quad\quad\times\frac{\Gamma(\upsilon_0+\nu_0)}
    {\Gamma(\upsilon_0)\Gamma(\nu_0)}
    [p^{(r)}_{0\rightarrow1}]^{\upsilon_0-1}
    [1-p^{(r)}_{0\rightarrow1}]^{\nu_0-1}
    I(p^{(r)}_{0\rightarrow1}\le p^{(r)}_{1\rightarrow0})
    \nonumber\\
    %--------------
    && \quad\quad\propto
    [p^{(r)}_{0\rightarrow1}]^{(m^{(r)}_{0\rightarrow1}+{\upsilon_0})-1}
    [1-p^{(r)}_{0\rightarrow1}]^{(m^{(r)}_{0\rightarrow0}+{\nu_0})-1}
    I(p^{(r)}_{0\rightarrow1}\le p^{(r)}_{1\rightarrow0})
    \nonumber\\
    %--------------
    && \quad\quad\propto
    {\cal B}eta(p^{(r)}_{0\rightarrow1}|m^{(r)}_{0\rightarrow1}+{\upsilon_0},
    m^{(r)}_{0\rightarrow0}+{\nu_0})
    I(p^{(r)}_{0\rightarrow1}\le p^{(r)}_{1\rightarrow0}),
    \label{EQ_COND_P01_DRAWS}
    \end{eqnarray}
    where $\Gamma(\ldots)$ is the Gamma function, ${\bf\Theta}\backslash p^{(r)}_{0\rightarrow1}$ means all components of ${\bf\Theta}$ except $p^{(r)}_{0\rightarrow1}$, and we keep only those multipliers that depend on $p^{(r)}_{0\rightarrow1}$. We use formula~(\ref{EQ_PRIOR_S}) to obtain the fourth line in equation~(\ref{EQ_COND_P01_DRAWS}), and number $m^{(r)}_{i\rightarrow j}$ is the total number of Markov switching state transitions from $s_t=i$ to $s_{t+1}=j$ inside time interval ${\cal T}(r)\leq t<{\cal T}(r+1)$ [not counting history-independent transitions at $t\in{\cal T}_-$]. In equation~(\ref{EQ_COND_P01_DRAWS}) the conditional posterior distribution of $p^{(r)}_{0\rightarrow1}$ is a known truncated beta distribution. Therefore, we draw $p^{(r)}_{0\rightarrow1}$ directly from this distribution by using Gibbs sampling described in Section~\ref{GIBBS}, and by the rejection sampling (``accept-reject'') algorithm described in the next section. Note that index $r=1,2,...,R$, where $R$ is the total number of time intervals of constant transition probabilities.
    \item
    The conditional posterior distribution of the transition probability $p^{(r)}_{1\rightarrow0}$ is given by equation~(\ref{EQ_COND_P01_DRAWS}) with states $0$ and $1$ interchanged everywhere, except in function $I(p^{(r)}_{0\rightarrow1}\le p^{(r)}_{1\rightarrow0})$,
    \begin{eqnarray}
    && f(p^{(r)}_{1\rightarrow0}|{\bf Y},{\bf\Theta}\backslash p^{(r)}_{1\rightarrow0})
    \propto f({\bf Y},{\bf\Theta})
    \nonumber\\
    && \quad\quad\propto
    {\cal B}eta(p^{(r)}_{1\rightarrow0}|m^{(r)}_{1\rightarrow0}+{\upsilon_1},
    m^{(r)}_{1\rightarrow1}+{\nu_1})
    I(p^{(r)}_{0\rightarrow1}\le p^{(r)}_{1\rightarrow0}).
    \label{EQ_COND_P10_DRAWS}
    \end{eqnarray}
    We also draw $p^{(r)}_{1\rightarrow0}$ directly from its conditional posterior distribution by using Gibbs sampling.
    \item
    To speed up MCMC convergence for posterior draws of vector ${\bf S}=[s_1,s_2,...,s_T]'$, we draw subsections ${\bf S}_{t,\tau}=[s_t,s_{t+1},...,s_{t+\tau-1}]'$ of ${\bf S}$ at a time, instead of drawing vector ${\bf S}$ component-by-component~\citep[][]{T_02}. The conditional posterior distribution of ${\bf S}_{t,\tau}$ is
    \begin{eqnarray}
    && f({\bf S}_{t,\tau}|{\bf Y},{\bf\Theta}\backslash{\bf S}_{t,\tau})=
    \frac{f({\bf S}_{t,\tau},{\bf Y},{\bf\Theta}\backslash{\bf S}_{t,\tau})}
    {f({\bf Y},{\bf\Theta}\backslash{\bf S}_{t,\tau})}\propto
    f({\bf Y},{\bf\Theta})
    \nonumber\\
    %--------------
    && \quad\quad\propto\left[\prod\limits_{\{\hat t:\:s_{\hat t}=0\}}
    \prod\limits_{n=1}^{N_{\hat t}}
    f({Y_{\hat t,n}|{\betabf_{(0)}}})\right]
    \times\left[\prod\limits_{\{\hat t:\:s_{\hat t}=1\}}
    \prod\limits_{n=1}^{N_{\hat t}}
    f({Y_{\hat t,n}|{\betabf_{(1)}}})\right]
    \nonumber\\
    && \quad\quad\quad\times\prod\limits_{\left\{\hat t:\:1\leq {\hat t}<T,\atop
    \hat t\notin{\cal T}_-\right\}}P(s_{\hat t+1}|s_{\hat t})
    \nonumber\\
    %--------------
    && \quad\quad\propto\left[\prod\limits_{\left\{\hat{t}:\:s_{\hat{t}}=0,\atop t\leq\hat{t}\leq t+\tau-1\right\}}
    \prod\limits_{n=1}^{N_{\hat{t}}}f({Y_{\hat{t},n}|{\betabf_{(0)}}})\right]
    \times\left[\prod\limits_{\left\{\hat{t}:\:s_{\hat{t}}=1,\atop t\leq\hat{t}\leq t+\tau-1\right\}}
    \prod\limits_{n=1}^{N_{\hat{t}}}f({Y_{\hat{t},n}|{\betabf_{(1)}}})\right]
    \nonumber\\
    && \quad\quad\quad\times\prod\limits^R_{r=1}
    \prod\limits_{\left\{\hat{t}:\:{\cal T}(r)\leq\hat{t}<{\cal T}(r+1),\;\;{\hat t}<T,\atop
    t-1\leq\hat{t}\leq t+\tau-1,\;{\hat t}\notin{\cal T}_-\right\}}P(s_{\hat{t}+1}|s_{\hat{t}})
    \nonumber\\
    %--------------
    && \quad\quad=\prod\limits_{\hat{t}=t}^{t+\tau-1}\left[(1-s_{\hat{t}})
    \prod\limits_{n=1}^{N_{\hat{t}}}f({Y_{\hat{t},n}|{\betabf_{(0)}}})
    \;+\;s_{\hat{t}}
    \prod\limits_{n=1}^{N_{\hat{t}}}f({Y_{\hat{t},n}|{\betabf_{(1)}}})\right]
    \nonumber\\
    && \quad\quad\quad\times\prod\limits^R_{r=1}\;
    [p^{(r)}_{0\rightarrow1}]^{\hat{m}^{(r,t)}_{0\rightarrow1}}
    [1-p^{(r)}_{0\rightarrow1}]^{\hat{m}^{(r,t)}_{0\rightarrow0}}
    [p^{(r)}_{1\rightarrow0}]^{\hat{m}^{(r,t)}_{1\rightarrow0}}
    [1-p^{(r)}_{1\rightarrow0}]^{\hat{m}^{(r,t)}_{1\rightarrow1}}
    \nonumber\\
    %--------------
    && \quad\quad=\prod\limits_{\hat{t}=t}^{t+\tau-1}\left[(1-s_{\hat{t}})
    \prod\limits_{n=1}^{N_{\hat{t}}}f({Y_{\hat{t},n}|{\betabf_{(0)}}})
    \;+\;s_{\hat{t}}
    \prod\limits_{n=1}^{N_{\hat{t}}}f({Y_{\hat{t},n}|{\betabf_{(1)}}})\right]
    \nonumber\\
    && \quad\quad\quad\times \!\!\!\!\!\!\!\!\!\!\!\!\!\!
    \prod\limits_{\quad\quad\left\{r:\;\:\left[{\cal T}(r),{\cal T}(r+1)\vphantom{\int}\right)\bigcap\:
    \left[t-1,t+\tau-1\vphantom{\int}\right]\neq\{\emptyset\}\right\}
    \!\!\!\!\!\!\!\!\!\!\!\!\!\!\!\!\!\!\!\!\!\!\!\!\!\!\!\!\!\!\!\!\!
    \!\!\!\!\!\!\!\!\!\!\!\!\!\!\!\!\!\!\!\!\!\!\!\!\!\!\!\!\!\!\!\!\!
    \!\!\!\!\!\!}
    \!\!\!\!\!\!\!\!\!\!\!\!\!\!
    [p^{(r)}_{0\rightarrow1}]^{\hat{m}^{(r,t)}_{0\rightarrow1}}
    [1-p^{(r)}_{0\rightarrow1}]^{\hat{m}^{(r,t)}_{0\rightarrow0}}
    [p^{(r)}_{1\rightarrow0}]^{\hat{m}^{(r,t)}_{1\rightarrow0}}
    [1-p^{(r)}_{1\rightarrow0}]^{\hat{m}^{(r,t)}_{1\rightarrow1}},
    \label{EQ_COND_S_DRAWS}
    \end{eqnarray}
    where ${\bf\Theta}\backslash{\bf S}_{t,\tau}$ means all components of ${\bf\Theta}$ except for ${\bf S}_{t,\tau}$, and we keep only those multipliers that depend on ${\bf S}_{t,\tau}=[s_t,s_{t+1},...,s_{t+\tau-1}]'$. Number ${\hat m}^{(r,t)}_{i\rightarrow j}$ is the total number of Markov switching state transitions from $s_{\hat t}=i$ to $s_{\hat t+1}=j$ inside the intersection of time intervals ${\cal T}(r)\leq\hat t<{\cal T}(r+1)$ and $t-1\leq\hat{t}\leq t+\tau-1$ [here $\,i,j=\{0,1\}\,$ and history-independent transitions at $\hat t\in{\cal T}_-$ are not counted]. Number $\hat{m}_{i\rightarrow j}$ is zero for all $\,i,j=\{0,1\}\,$ if intervals ${\cal T}(r)\leq\hat t<{\cal T}(r+1)$ and $t-1\leq\hat{t}\leq t+\tau-1$ do not intersect, resulting in the final expression for the product over $r$ on the last line in equation~(\ref{EQ_COND_S_DRAWS}). Vector ${\bf S}_{t,\tau}$ has length $\tau$ and can assume $2^\tau$ possible values. By choosing $\tau$ small enough, we can compute the right-hand-side of equation~(\ref{EQ_COND_S_DRAWS}) for each of these values and find the normalization constant of $f({\bf S}_{t,\tau}|{\bf Y},{\bf\Theta}\backslash{\bf S}_{t,\tau})$. This allows us to make Gibbs sampling of ${\bf S}_{t,\tau}$. Our typical choice of $\tau$ is from 5 to 14.
\end{itemize}

All components of parameter vector ${\bf\Theta}$ are given by equation~(\ref{EQ_THETA_general}), and all conditional posterior distributions are given by equations~(\ref{EQ_COND_BETA0_DRAWS})--(\ref{EQ_COND_S_DRAWS}). We generate draws of ${\bf\Theta}^{(g)}$ from ${\bf\Theta}^{(g-1)}$ by using the hybrid Gibbs sampler explained in Section~\ref{GIBBS} as follows (for brevity, we drop $g$ indexing below):
\begin{enumerate}
    \item[(a)]
    We draw vector $\betabf_{(0)}$ component-by-component by using the Metropolis-Hasting ($\mbox{M-H}$) algorithm (note that we draw only those components that are free parameters). For each (free) component $\beta_{(0),k}$ of $\betabf_{(0)}$ we use a normal jumping distribution
    \begin{eqnarray}
    J(\hat{\beta}_{(0),k}|\beta_{(0),k})=
    {\cal N}(\hat{\beta}_{(0)}|\beta_{(0),k},\sigma^2_{(0),k})=
    \frac{1}{\sigma_{(0),k}\sqrt{2\pi}}e^{\left.-[\hat{\beta}_{(0),k}-
    \beta_{(0),k}]^2\right/2\sigma^2_{(0),k}}
    \label{EQ_NORMAL_JUMP}
    \end{eqnarray}
    Standard deviations $\sigma_{(0),k}$ are adjusted during the burn-in sampling (i.e.~during $g=1,2,...,G_{bi}$) to have approximately $30\%$ averaged acceptance rate in equation~(\ref{EQ_MH_DRAW}). The adjustment algorithm is explained in the next section. We also tried Cauchy jumping distribution
    \begin{eqnarray}
    J(\hat{\beta}_{(0),k}|\beta_{(0),k}) &=&
    {\cal C}auchy(\hat{\beta}_{(0)}|\beta_{(0),k},\sigma_{(0),k})
    \vphantom{\int}
    \nonumber\\
    & = & \frac{1/(\pi\sigma_{(0),k})}{1+\left[(\hat{\beta}_{(0),k}-\beta_{(0),k})
    /\sigma_{(0),k}\right]^2},
    \label{EQ_CAUCHY_JUMP}
    \end{eqnarray}
    and obtained similar results. As already explained in Section~\ref{PRIOR_CHOICE}, we draw $\beta_{(0),k}$ from its conditional posterior distribution, given by equation~(\ref{EQ_COND_BETA0_DRAWS}), only if it is a free parameter. We do not draw $\beta_{(0),k}$ in the following three cases. First, $\beta_{(0),k}$ is restricted to zero (which is the case if it is found to be statistically insignificant). Second, $\beta_{(0),k}$ is restricted to $-\infty$ [which is the case if state $0$ is the zero-accident state, and, therefore, the intercept in state $0$ is $-\infty$, see equations~(\ref{EQ_MLE_LAMBDA}),~(\ref{EQ_MLE_LAMBDA_1}),~(\ref{EQ_MS_P_year}) and~(\ref{EQ_MS_NB_year})]. The third case is when $\beta_{(0),k}$ is restricted to another, free $\beta$-coefficient.
    \item[(b)]
    We use Metropolis-Hasting algorithm and draw all components of $\betabf_{(1)}$ (that are free parameters) from their conditional posterior distributions, given in equation~(\ref{EQ_COND_BETA1_DRAWS}), in exactly the same way as we draw the components of $\betabf_{(0)}$.
    \item[(c)]
    By using Gibbs sampling, for all $r=1,2,...,R$ time intervals we draw transition probabilities $p^{(r)}_{0\rightarrow1}$, first, and $p^{(r)}_{1\rightarrow0}$, second, from their conditional posterior distributions given in equations~(\ref{EQ_COND_P01_DRAWS}) and~(\ref{EQ_COND_P10_DRAWS}).\footnote{We do not make draws of $p^{(r)}_{0\rightarrow1}$ and $p^{(r)}_{1\rightarrow0}$ from their conditional posterior distributions if these parameters are not free, but are restricted to other transition probabilities. For example, in the next chapter we will consider a model of weekly accident frequencies in which we will assume that different seasons have different transition probabilities, but the transition probabilities for the same seasons in different years are restricted to be the same. In this case, only transition probabilities for time intervals that are inside the first year of data are free and are drawn.}
    \item[(d)]
    Finally, we draw subsections ${\bf S}_{t,\tau}=[s_t,s_{t+1},...,s_{t+\tau-1}]'$ of the state vector ${\bf S}=[s_1,s_2,...,s_T]'$. We use Gibbs sampling and draw subsections ${\bf S}_{t,\tau}$ one after another from their conditional posterior distributions given by equation~(\ref{EQ_COND_S_DRAWS}).
\end{enumerate}

\section{Computational issues and optimization}
\label{MCMC_COMP_ISSUES}
A special numerical code was written in the MATLAB programming language for the MCMC simulations used in the present research study. Our code was written from scratch, and no standard MCMC computer scripts and procedures were used. This programming approach provided us with great flexibility and control during model estimation. Our code uses the general representation introduced Section~\ref{GENERAL_MS_MODEL}, and as a result, the code is applicable to estimation of all accident frequency and severity models considered here.

Below, in this section, we briefly discuss several numerical issues, tips and optimizations that turned out to be important for numerically accurate, reliable and fast MCMC runs during model estimation process.
\begin{itemize}
\item
We tested our MCMC code on artificial accident data sets. The test procedure included a generation of artificial data with a known probabilistic model (e.g. a MSNB model or a MSML model). Then these data were used to estimate the underlying model by means of our MCMC simulation code. With this procedure we found that the probabilistic models, used to generate the artificial data, were reproduced successfully with our estimation code.
\item
In order to avoid numerical zero and numerical infinity, during MCMC simulations we always use and calculate the logarithms of all probability distributions instead of the distributions themselves (for example, we work with log-likelihood functions instead of likelihood functions).
\item
Standard deviations $\sigma_{(0),k}$ of the normal and Cauchy jump distributions, given by equations~(\ref{EQ_NORMAL_JUMP}) and~(\ref{EQ_CAUCHY_JUMP}), are adjusted during the burn-in sampling ($g=1,2,\ldots,G_{bi}$) to have approximately $30\%$ averaged acceptance rate in equation~(\ref{EQ_MH_DRAW}). For each $k=1,2,...K_{(0)}$ that corresponds to a free model parameter $\beta_{(0),k}$, drawn by the Metropolis-Hasting (M-H) algorithm, the adjustment is done as follows. We calculate the mean candidate acceptance rate in equation~(\ref{EQ_MH_DRAW}), averaged over the last $50$ consecutive M-H draws. If this mean rate is above/below the $30\%$ target rate, we multiply/divide the standard deviation $\sigma_{(0),k}$ by factor $1.25$. Then we calculate the mean acceptance rate, averaged over the next $50$ M-H draws, and again adjust $\sigma_{(0),k}$ by multiplying or dividing it by $1.25$, and so on. During the burn-in sampling we collect and save all standard deviations used for the M-H draws and the corresponding mean acceptance rates (averaged over groups of 50 consecutive draws). After all $G_{bi}$ burn-in draws are made, we fit a decreasing exponential function to the dependence of the mean acceptance rates on the $\sigma_{(0),k}$ values [for this fit we use the acceptance rate data collected over the last $(2/3)G_{bi}$ burn-in draws]. Finally, we use this exponential function to obtain the best guess about the value of $\sigma_{(0),k}$ that will result in the $30\%$ target averaged acceptance rate. This value of $\sigma_{(0),k}$ stays constant for all further draws $g=G_{bi}+1,...,G$, which are used for Bayesian inference. As a result of the adjustments descried above, in our MCMC simulations the actual mean acceptance rates (averaged over draws $g=G_{bi}+1,...,G$) always turned out to be within $1\%$ of the target $30\%$ rate.
\item
The Gibbs sampling draws from the truncated betas distributions in equations~(\ref{EQ_COND_P01_DRAWS}) and~(\ref{EQ_COND_P10_DRAWS}) are done by the rejection sampling technique, also known as the accept-reject algorithm \citep[][]{HLD_04}. This algorithm works as follows. Let us assume that we need to make draws of $x$ from a probability density function $f(x)$, which is not easily available. Then, we construct an envelope function $F(x)$ such that, first, $F(x)\ge f(x)$ is satisfied for all $x$, and, second, $x$ can be easily drawn from the probability density function $F(x)\left/\int F(x)\,dx\right.$. To obtain correct draws from $f(x)$, we repeatedly, first, generate draws $x_g$ from $F(x)\left/\int F(x)\,dx\right.$, and, second, accept $x_g$ with probability $f(x_g)/F(x_g)$ [here $g=1,2,3,...$]. For the algorithm to be efficient, the envelope function $F(x)$ should be sufficiently close to $f(x)$ [so that the acceptance probability $f(x_g)/F(x_g)$ is not very small]. Because the logarithm of a truncated beta distribution is concave, we construct and use a piece-exponential envelope function (i.e. the logarithm of the envelope function is piece-linear), see \citet[][]{HLD_04}.
\item
The Gibbs sampling of subsections ${\bf S}_{t,\tau}=[s_t,s_{t+1},...,s_{t+\tau-1}]'$ from the conditional posterior distribution given in equation~(\ref{EQ_COND_S_DRAWS}) can be optimized as follows. First, for each value of time $\hat{t}=t,t+1,...,t+\tau-1$ we calculate and save in the computer memory the values of products $\prod\limits_{n=1}^{N_{\hat{t}}}f({Y_{\hat{t},n}|{\betabf_{(0)}}})$ and $\prod\limits_{n=1}^{N_{\hat{t}}}f({Y_{\hat{t},n}|{\betabf_{(1)}}})$, refer to equation~(\ref{EQ_COND_S_DRAWS}). Then, we use these values to compute the probabilities of all $2^\tau$ possible combination values of the subsection vector ${\bf S}_{t,\tau}$ without need to recalculate the likelihood functions $f({Y_{\hat{t},n}|{\betabf_{(0)}}})$ and $f({Y_{\hat{t},n}|{\betabf_{(1)}}})$ each time. This optimization procedure considerably speeds up Gibbs sampling of ${\bf S}_{t,\tau}$.
\item
There is an important issue that arises during Bayesian-MCMC estimation of Markov switching models, which is the ``label switching problem''. This problem can be understood and solved as follows. Note that the likelihood functions for the MSP, MSNB and MSML models, given by equations~(\ref{EQ_MS_P_week}),~(\ref{EQ_MS_NB_week}) and~(\ref{EQ_MS_ML}), are completely symmetric under the interchange \mbox{``$0$''$\leftrightarrow$``$1$''} of the labels of the two states of roadway safety. This label interchange is just equivalent to renaming labels for the two states (using label names "1" and "0" as opposed to using label names "0" and "1" for the first and second states respectively). During a MCMC run the labels might interchange many times back and forth, in which case the MCMC chain would not converge. This is called the ``label switching problem''. To avoid this problem, we impose a restriction $p_{0\rightarrow1}\leq p_{1\rightarrow0}$ on the Markov transition probabilities, see equations~(\ref{EQ_MS_P_RESTRICT}) and~(\ref{EQ_MS_P_RESTRICT_1}). This restriction breaks the symmetry of the likelihood function and the posterior distribution under the interchange \mbox{``$0$''$\leftrightarrow$``$1$''} of the label notations.\footnote{Instead of the restriction imposed on the transitional probabilities, we also tried restrictions imposed on the intercept coefficients (the first components of $\beta$-s). We found that the later works no better and no worse that the former for controlling the label switching problem. It is convenient to use the restriction on the transitional probabilities because there are more than two intercepts in the MSML models and because of its easier interpretation [the interpretation of restriction $p_{0\rightarrow1}\leq p_{1\rightarrow0}$ is that, on average, the state $0$ is more frequent than the state $1$, refer to equation~(\ref{EQ_MS_P_BAR})].} In practice, the restriction imposed on the transitional probabilities does not completely solve the label switching problem because few MCMC chains still happen to converge to the incorrect label setting (with the two labels interchanged as compared to the correct label setting). To deal with this problem, we monitor the posterior average of the logarithm of the joint probability distribution $f({\bf Y},{\bf\Theta})$ [note that monitoring the joint distribution is equivalent to monitoring the posterior distribution because the later is proportional to the former]. When a MCMC chain converges to an incorrect label setting, this average is considerably smaller (typically, by $10$ to $50$) than its value for the MCMC chains that converge to the correct label setting.  To distinguish label settings, we define the correct label setting as the one that provides the maximal value of the average of the posterior probability and, therefore, the maximal value of the average of the joint probability (since the posterior is proportional to the joint). If we had an unlimited computational time, then eventually all MCMC chains would converge to the correct label setting. Since our computational time is limited, we have to eliminate those few chains that did not converge to the correct label settings.\footnote{This may introduce a model estimation bias. However, this bias is negligible because the incorrect label setting corresponds to posterior (or joint) probability values that are much smaller than those for the correct label setting (typically the difference factors range from $\approx e^{-50}$ to $\approx e^{-10}$).}
\item
The convergence of the MCMC chains used for Bayesian inference (these chains have converged to the correct label setting) is checked by the Brooks-Gelman-Rubin diagnostic \citep[][]{BG_98,SAS_06}. This diagnostic works as follows. Let us consider only the continuous model parameters ($\beta$-s, $p^{(r)}_{0\rightarrow1}$ and $p^{(r)}_{1\rightarrow0}$). Let the number of these continuous parameters be equal to $N$ and vector $\thetabf$ be composed of these parameters (in other words, $\thetabf$ include all components of ${\bf\Theta}$ except the state values ${\bf S}$). Let us have $M$ chains with $G$ posterior draws of $\thetabf$ in each chain, obtained as a result of the MCMC simulations.\footnote{We usually use eight MCMC chains of vector ${\bf\Theta}$ of all model parameters, which correspond to different choices of the initial parameter vector ${\bf\Theta}^{(0)}$, see Section~\ref{GIBBS}. Keep in mind that a certain number of first ``burn-in'' draws of ${\bf\Theta}$ are discarded in each chain because they can depend on the initial choices of ${\bf\Theta}^{(0)}$.} Thus, we have draws $\thetabf^{(g,m)}$, where $g=1,2,...,G$ and $m=1,2,...,M$. Then, the potential scale reduction factors (PSRFs) and the multivariate potential scale reduction factor (MPSRF) are given by the following equations:
\begin{eqnarray}
&&{\bar\thetabf}^{(m)}=\frac{1}{G}\sum\limits_{g=1}^G {\thetabf}^{(g,m)},
\qquad\qquad
{\bar\thetabf}=\frac{1}{M}\sum\limits_{m=1}^M {\bar\thetabf}^{(m)},
\nonumber
\\
&&B=\frac{1}{M-1}\sum\limits_{m=1}^M
({\bar\thetabf}^{(m)}-{\bar\thetabf})({\bar\thetabf}^{(m)}-{\bar\thetabf})',
\nonumber
\\
&&W=\frac{1}{M}\sum\limits_{m=1}^M\biggl[\frac{1}{G-1}\sum\limits_{g=1}^G
({\bar\thetabf}^{(g,m)}-{\bar\thetabf}^{(m)})({\bar\thetabf}^{(g,m)}-{\bar\thetabf}^{(m)})'
\biggr],
\nonumber
\\
&&V=\frac{G-1}{G}\,W+\frac{M+1}{M}\,B,
\nonumber
\\
&&{\rm PSRF}_n=\sqrt{\frac{{\rm diag}(V)_n}{{\rm diag}(W)_n}},
\qquad\mbox{where $n=1,2,...,N$},
\label{EQ_PSRF}
\\
&&{\rm MPSRF}=\sqrt{\frac{G-1}{G}+\frac{M+1}{M}\lambda_{max}},
\label{EQ_MPSRF}
\\
&&\qquad\qquad\quad
\mbox{where $\lambda_{max}=\max\left\{\mbox{eigenvalues of matrix $W^{-1}B$}\right\}$}.
\nonumber
\end{eqnarray}
Here vectors ${\thetabf}^{(g,m)}$, ${\bar\thetabf}^{(m)}$ and ${\bar\thetabf}$ have their lengths equal to $N$; matrices $B$, $W$ and $V$ have their sizes equal to $N\!\times\!N$; ${\rm PSRF}_n$ is the potential scale reduction factor for the $n^{th}$ component of $\thetabf$ and ${\rm diag}(V)_n$ is the $n^{th}$ element of the diagonal of matrix $V$ [${\rm diag}(W)_n$ is defined analogously]; $\lambda_{max}$ is the maximal eigenvalue of the symmetric positive definite matrix $W^{-1}B$. Matrices $B$ and $W$ represent variance/covariance of $\thetabf$ between the MCMC chains and inside the chains, respectively. For a well-converged MCMC simulation, the resulting PSRFs and MPSRF should be close to unity. Note that there exist an alternative definition of PSRF and MPSRF, in which the square roots are missing in Equations~(\ref{EQ_PSRF}) and~(\ref{EQ_MPSRF}).
\end{itemize}

%% file: ch_results_freq.tex
%
% This is the frequency model estimation results chapter
%

\chapter{Frequency model estimation results}
\label{RESULTS_FREQ}
In this chapter we present model estimation results for accident frequencies. The chapter consists of two sections. In the first section, we consider annual accident frequencies and estimate Markov switching Poisson (MSP), Markov switching negative binomial (MSNB), standard Poisson, standard negative binomial (NB), standard zero-inflated Poisson (ZIP) and standard zero-inflated negative binomial (ZINB) models. In the second section, we consider weekly accident frequencies and estimate MSP, MSNB, standard Poisson and standard NB models. We compare the performance of the models in fitting the data.

In the present study, for both annual and weekly accident frequency models, we use the data from 5769 accidents that were observed on 335 interstate highway segments in Indiana in 1995-1999.

\section{Model estimation results for annual frequency data}
\label{RESULTS_FREQ_YEAR}
In this section we use annual time periods, $t=1,2,3,4,T=5$ in total.\footnote{We also considered quarterly time periods and obtained qualitatively similar results (not reported here).} For each roadway segment $n=1,2,\ldots,N=335$ the state $s_{t,n}$ can change every year. Three types of annual accident frequency models are estimated:
\begin{enumerate}
    \item
    We estimate standard (single-state) Poisson and negative binomial (NB) models, specified by equations~(\ref{EQ_MLE_POISSON}) and~(\ref{EQ_MLE_NB}). We estimate these models, first, by the maximum likelihood estimation (MLE) and, second, by the Bayesian inference approach and MCMC simulations.\footnote{\label{MLE_LIMDEP}The maximum likelihood estimation was done by using LIMDEP software package. To obtain optimal parsimonious standard models, estimated by MLE, we choose the explanatory variables and their dummies by using the Akaike Information Criterion (AIC)~\citep[][]{T_02,WKM_03}. For details see~\citet[][]{M_06}.} As one expects, for our choice of a non-informative prior distribution, for both the Poisson and NB models, the estimated results obtained by MLE and by MCMC estimation techniques, turned out to be very similar.
    \item
    We estimate standard zero-inflated \mbox{ZIP-$\tau$}, \mbox{ZIP-$\gamma$}, \mbox{ZINB-$\tau$} and \mbox{ZINB-$\gamma$} models, specified by equations~(\ref{EQ_MLE_ZIP})--(\ref{EQ_Q_GAMMA}). First, we estimate these models by the MLE (see footnote~\ref{MLE_LIMDEP} on page~\pageref{MLE_LIMDEP}). Second, we estimate them by the Bayesian inference approach and MCMC simulations. As one expects, for our choice of a non-informative prior distribution, the Bayesian-MCMC estimation results again turned out to be similar to the MLE estimation results for the \mbox{ZIP-$\tau$} and \mbox{ZINB-$\tau$} models.
    \item
    We estimate the two-state Markov switching Poisson (MSP) and two-state Markov switching negative binomial (MSNB) models, given in equations~(\ref{EQ_MS_P_year}) and~(\ref{EQ_MS_NB_year}), by the Bayesian-MCMC methods. To choose the explanatory variables for the final MSP and MSNB models reported here, first, we start with using the variables that enter the standard Poisson and NB models.\footnote{\label{VARIABLES_CHOICE}This approach makes comparison of explanatory variable effects in different models straightforward. A formal Bayesian approach to model variable selection is based on evaluation of model's marginal likelihood and the Bayes factor~(\ref{EQ_BAYES_FACTOR}). Unfortunately, because MCMC simulations are computationally expensive, evaluation of marginal likelihoods for a large number of trial models is not feasible in our study.} Then, we consecutively construct and use $60\%$, $85\%$ and $95\%$ Bayesian credible intervals for evaluation of the statistical significance of each $\beta$-coefficient in the MSP and MSNB models. As a result, in the final MSP and MSNB models some components of $\betabf$ are restricted to zero.\footnote{\label{BETA_RESTRICT}A $\beta$-coefficient is restricted to zero if it is statistically insignificant. A $\,1-a\,$ credible interval is chosen in such a way that the posterior probabilities of being below and above it are both equal to $a/2$ (we use significance levels $a=40\%,15\%,5\%$).} For MSNB models, no restriction is imposed on the over-dispersion parameter $\alpha$, which turns out to be statistically significant anyway.
\end{enumerate}

The estimation results for the standard Poisson and NB models of annual accident frequencies are given in Table~\ref{T_year_P-NB}. The estimation results for the zero-inflated and Markov switching Poisson models are given in Table~\ref{T_year_ZIP-MSP}. The estimation results for the zero-inflated and Markov switching negative binomial models are given in Table~\ref{T_year_ZINB-MSNB}. In these tables, posterior (or MLE) estimates of all continuous model parameters, \mbox{$\beta$-s} and $\alpha$, are given together with their $95\%$ confidence intervals (if MLE) or $95\%$ credible intervals (if Bayesian-MCMC), refer to the superscript and subscript numbers adjacent to parameter posterior/MLE estimates.\footnote{\label{MLE_NORMALITY}Note that MLE assumes asymptotic normality of the estimates, resulting in confidence intervals being symmetric around the means (a $95\%$ confidence interval is $\pm1.96$ standard deviations around the mean). In contrast, Bayesian estimation does not require this assumption, and posterior distributions of parameters and Bayesian credible intervals are usually non-symmetric.} Table~\ref{T_freq_stat} gives summary statistics of all roadway segment characteristic variables ${\bf X}_{t,n}$ except the intercept.

Because estimation results for Poisson models are very similar to estimation results for negative binomial models, let us focus on and discuss only the estimation results for negative binomial models. Our major findings, discussed below for negative binomial models, hold for Poisson models as well (unless otherwise stated). The findings are as follows.

The estimation results show that two states of roadway safety exist, and that the two-state MSNB model is strongly favored by the empirical data, as compared to the standard \mbox{ZINB-$\tau$} and \mbox{ZINB-$\gamma$} models, which in turn are favored over the simple standard NB model. Indeed, from Tables~\ref{T_year_P-NB} and~\ref{T_year_ZINB-MSNB} we see that the values of the logarithm of the marginal likelihood of the data for NB, \mbox{ZINB-$\tau$}, \mbox{ZINB-$\gamma$} and MSNB models are $-2554.16$, $-2519.90$, $-2447.33$ and $-2184.21$ respectively. Thus, the MSNB model provides considerable, $369.95$, $335.69$ and $263.12$, improvements of the logarithm of the marginal likelihood as compared to the NB, \mbox{ZINB-$\tau$} and \mbox{ZINB-$\gamma$} models respectively. As a result, from equation~(\ref{EQ_BAYES_FACTOR}), we find that, given the accident data, the posterior probability of the MSNB model is larger than the probabilities of the NB, \mbox{ZINB-$\tau$} and
\mbox{ZINB-$\gamma$} models by $e^{369.95}$, $e^{335.69}$ and $e^{263.12}$ respectively.\footnote{In addition, we find DIC (deviance information criterion) values $5105.4$, $5037.3$, $4891.4$, $4261.5$ for the NB, \mbox{ZINB-$\tau$}, \mbox{ZINB-$\gamma$} and MSNB models respectively. We also find DIC values $5346.0$, $5292.9$, $5063.8$, $4317.0$  for the Poisson, \mbox{ZIP-$\tau$}, \mbox{ZIP-$\gamma$} and MSP models respectively. This means that the MSNB (MSP) model is favored over the standard NB (Poisson) and ZINB (ZIP) models. However, we prefer to rely on the Bayes factor approach instead of the DIC (see footnote~\ref{DIC} on page~\pageref{DIC}).} Note that we use the harmonic mean formula, given in equation~(\ref{EQ_L_MARGINAL}), to calculate the values and the $95\%$ confidence intervals of the log-marginal-likelihoods reported in Tables~\ref{T_year_P-NB}--\ref{T_year_ZINB-MSNB}. The confidence intervals are found by bootstrap simulations.\footnote{\label{BOOTSTRAP}During bootstrap simulations we repeatedly draw, with replacement, posterior values of ${\bf\Theta}$ to calculate the posterior expectation in equation~(\ref{EQ_L_MARGINAL}). In each of $10^5$ bootstrap draws that we make, the number of ${\bf\Theta}$ values drawn is 1/100 of the total number of all posterior ${\bf\Theta}$ values available from MCMC simulations. The bootstrap simulations show that equation~(\ref{EQ_L_MARGINAL}) gives sufficiently accurate answers, and that the expectation in this equation is not dominated too much by just few posterior values of $\bf\Theta$, at which the likelihood function happens to be extremely small.}

We can also use a classical statistics approach for model comparison, based on the MLE. Referring to Tables~\ref{T_year_P-NB} and~\ref{T_year_ZINB-MSNB}, the MLE gives the maximum log-likelihood values $-2533.81$, $-2502.67$ and $-2426.54$ for the NB, \mbox{ZINB-$\tau$} and \mbox{ZINB-$\gamma$} models respectively. The maximum log-likelihood value observed during our MCMC simulations for the MSNB model is equal to $-2049.45$. An imaginary MLE, at its convergence, would give MSNB log-likelihood value that would be even larger than this observed value. Therefore, if estimated by the MLE, the MSNB model would provide large, at least $484.36$, $453.22$ and $377.09$, improvements in the maximum log-likelihood value over the NB, \mbox{ZINB-$\tau$} and \mbox{ZINB-$\gamma$} models. These improvements would come with no increase or a decrease in the number of free continuous model parameters ($\beta$-s, $\alpha$, $\tau$, $\gamma$-s) that enter the likelihood function. Both the Akaike Information Criterion (AIC) and the Bayesian Information Criterion (BIC) would strongly favor the MSNB models over the NB model.\footnote{\label{AIC_BIC}Minimization of $AIC=2K-2LL$ and $BIC=K\ln(N)-2LL$ ensures an optimal choice of explanatory variables in a model and avoids overfitting \citep[][]{T_02,WKM_03}. Here $K$ is the number of free continuous model parameters that enter the likelihood function, $N$ is the number of observations and $LL$ is the log-likelihood. When $N\geq8$, BIC favors fewer free parameters than AIC does.}

\begin{landscape}
\begin{table}[p]
\centering
\caption{Estimation results for standard Poisson and negative binomial models of annual accident frequencies}
\label{T_year_P-NB}
\begin{scriptsize}
\input{frequency_year_Poisson-NB_table_a.tex}

\end{scriptsize}
\end{table}
%\addtocounter{table}{-1}
\begin{table}[p]
\centering
%\caption[]{(Continued)}
\centerline{Table \arabic{chapter}.\arabic{table}: (Continued)}\vspace{1.7ex}
\begin{scriptsize}
\input{frequency_year_Poisson-NB_table_b.tex}

\end{scriptsize}
\end{table}
\end{landscape}

\begin{landscape}
\begin{table}[p]
\centering
\caption{Estimation results for zero-inflated and Markov switching Poisson models of annual accident frequencies}
\label{T_year_ZIP-MSP}
\begin{scriptsize}
\input{frequency_year_ZIP-MSP_table_a.tex}

\end{scriptsize}
\end{table}
%\addtocounter{table}{-1}
\begin{table}[p]
\centering
%\caption{(Continued)}
\centerline{Table \arabic{chapter}.\arabic{table}: (Continued)}\vspace{1.7ex}
\begin{scriptsize}
\input{frequency_year_ZIP-MSP_table_b.tex}

\end{scriptsize}
\end{table}
\begin{table}[p]
\centering
\caption{Estimation results for zero-inflated and Markov switching negative binomial models of annual accident frequencies}
\label{T_year_ZINB-MSNB}
\begin{scriptsize}
\input{frequency_year_ZINB-MSNB_table_a.tex}

\end{scriptsize}
\end{table}
%\addtocounter{table}{-1}
\begin{table}[p]
\centering
%\caption{(Continued)}
\centerline{Table \arabic{chapter}.\arabic{table}: (Continued)}\vspace{1.7ex}
\begin{scriptsize}
\input{frequency_year_ZINB-MSNB_table_b.tex}

\end{scriptsize}
\end{table}
\end{landscape}

\begin{table}[p]
\centering
\caption{Summary statistics of explanatory variables that enter the models of annual and weekly accident frequencies}
\label{T_freq_stat}
\begin{scriptsize}
\input{frequency_statistics_table.tex}

\end{scriptsize}
\end{table}

\begin{figure}[t]
\vspace{7.0truecm}
\includegraphics{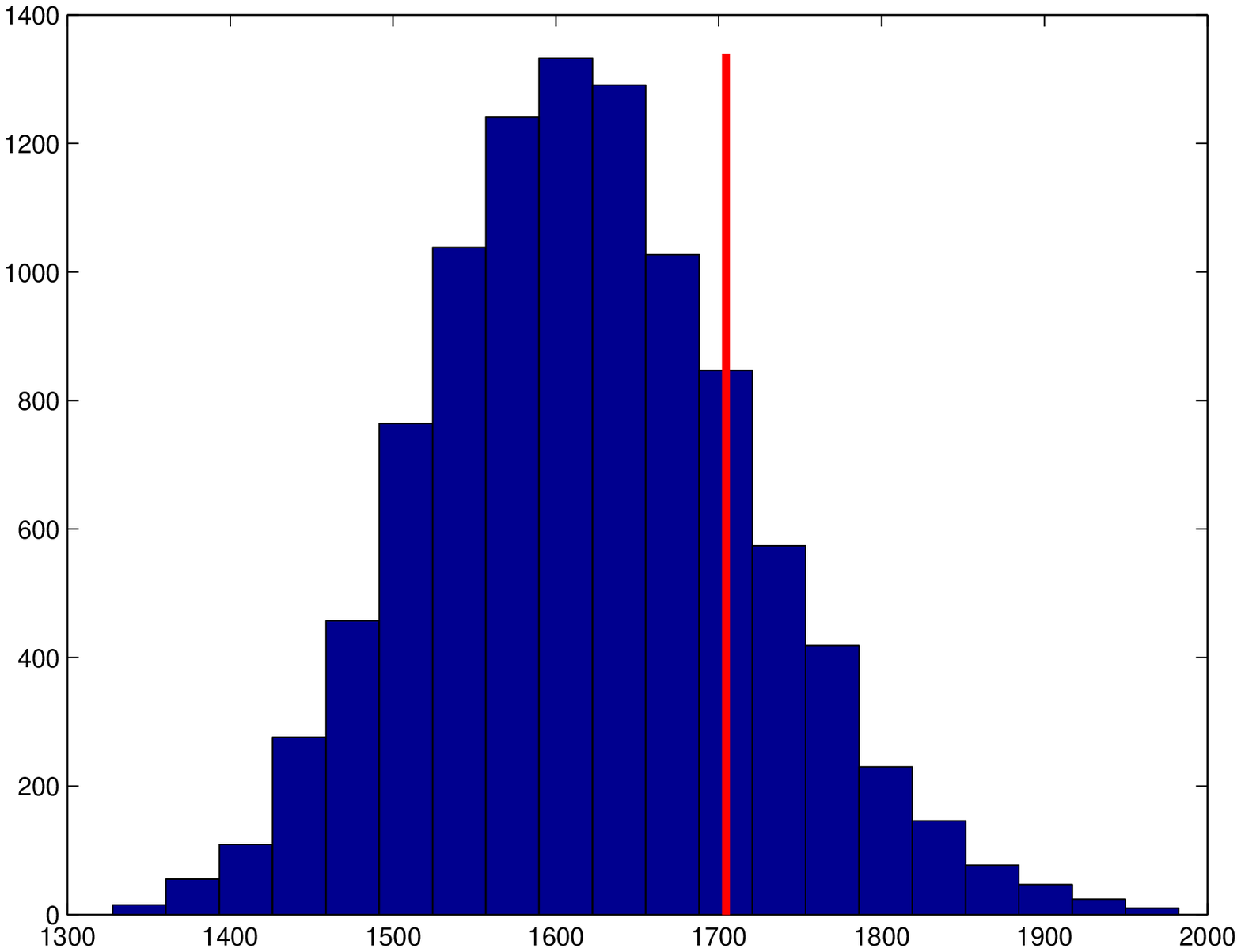}
\caption{The histogram of $10^4$ generated $\chi^2$ values for the MSNB model of annual accident frequencies. The vertical line shows the observed value of $\chi^2$.
\label{FIG_GOODNESS}
}
\end{figure}

To evaluate the goodness-of-fit for a model, we use the posterior (or MLE) estimates of all continuous model parameters ($\beta$-s, $\alpha$, $p^{(n)}_{0\rightarrow1}$, $p^{(n)}_{1\rightarrow0}$) and generate $10^4$ artificial data sets under the hypothesis that the model is true.\footnote{Note that the state values $\bf S$ are generated by using $p^{(n)}_{0\rightarrow1}$ and $p^{(n)}_{1\rightarrow0}$, where $n=1,2,...,N$.} We find the distribution of $\chi^2$, given by equation~(\ref{CHI_SQUARED}), and calculate the goodness-of-fit p-value for the observed value of $\chi^2$. As a demonstration, refer to Figure~\ref{FIG_GOODNESS}, where the histogram of the generated $\chi^2$ values is plotted for the MSNB model. The observed value of $\chi^2$ is shown by the vertical line in this figure, the goodness-of-fit p-value is equal to the ratio of the histogram area located to the right of the vertical line and the total histogram area. The resulting p-values for all negative binomial models are given in Tables~\ref{T_year_P-NB} and~\ref{T_year_ZINB-MSNB}. For the \mbox{ZINB-$\gamma$} and MSNB models the p-values are sufficiently large, around $20\%$, which indicates that these models fit the data reasonably well. At the same time, for the \mbox{ZINB-$\tau$} model the goodness-of-fit p-value is only around $0.5\%$
and for the standard NB model the p-value is only around $0.3\%$, which indicate much poorer fit. Note that all Poisson models (including the MSP model) provide relatively poor goodness-of-fit with p-value below $1\%$ (refer to Tables~\ref{T_year_P-NB} and~\ref{T_year_ZIP-MSP}), which can be explained by over-dispersion present in the annual frequency accident data.

The estimation results also show that the over-dispersion parameter $\alpha$ is higher for the \mbox{ZINB-$\tau$} and \mbox{ZINB-$\gamma$} models, as compared to the MSNB model (refer Table~\ref{T_year_ZINB-MSNB}). This suggests that over-dispersed volatility of accident frequencies, which is often observed in empirical data, could be in part due to the latent switching between the states of roadway safety.

\begin{figure}[t]
\vspace{9.4truecm}
\includegraphics{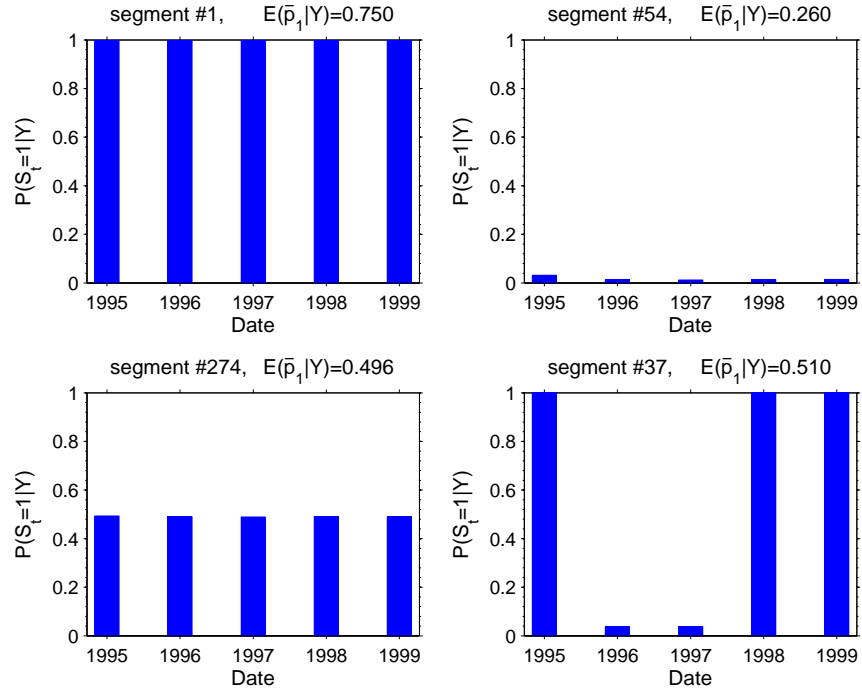}
\caption{Five-year time series of the posterior probabilities
$P(s_{t,n}=1|{\bf Y})$ of the unsafe state $s_{t,n}=1$
for four selected roadway segments ($t=1,2,3,4,5$). These plots are for the MSNB model of annual accident frequencies.
\label{FIGURE_freq_year_1}
}
\end{figure}

\begin{figure}[t]
\vspace{9.2truecm}
\includegraphics{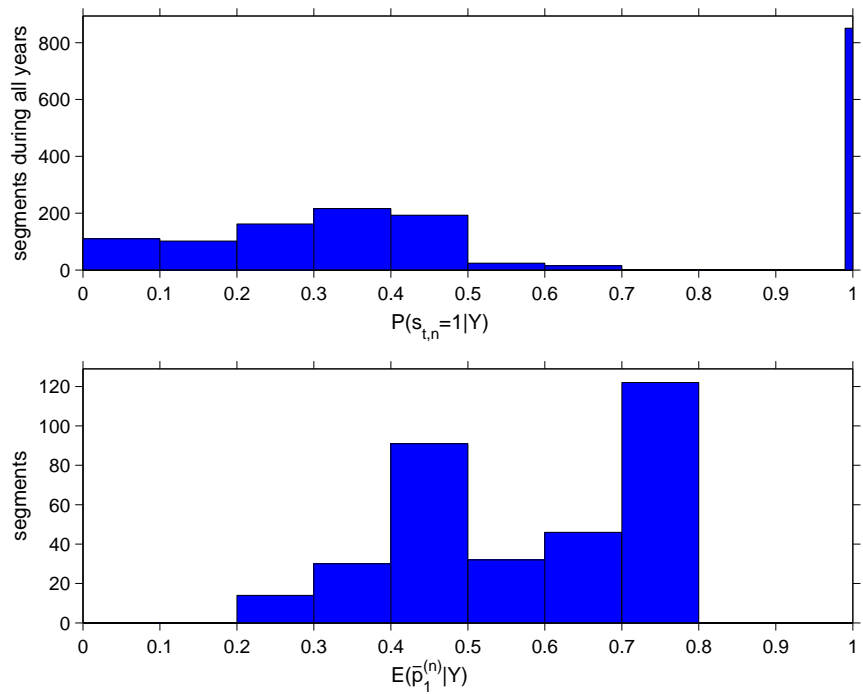}
\caption{Histograms of the posterior probabilities
$P(s_{t,n}=1|{\bf Y})$ (the top plot) and of the posterior
expectations $E[\bar p^{(n)}_1|{\bf Y}]$ (the bottom plot).
Here $t=1,2,3,4,5$ and $n=1,2,\ldots,335$. These histograms are for the MSNB model of annual accident frequencies.
\label{FIGURE_freq_year_2}
}
\end{figure}

Now, refer to Figure~\ref{FIGURE_freq_year_1}, created for the case of the MSNB model (note that the corresponding figure for the MSP model is similar and is not reported). The four plots in this figure show five-year time series of the posterior probabilities $P(s_{t,n}=1|{\bf Y})$ of the unsafe state for four selected roadway segments. These plots represent the following four categories of roadway segments:
\begin{itemize}
    \item
    For roadway segments from the first category we have $P(s_{t,n}=1|{\bf Y})=1$ for all $t=1,2,3,4,5$. Thus, we can say with absolute certainty that these segments were always in the unsafe state $s_{t,n}=1$ during the considered five-year time interval. A roadway segment belongs to this category if and only if it had at least one accident during each year ($t=1,2,3,4,5$). An example of such roadway segment is given in the top-left plot in Figure~\ref{FIGURE_freq_year_1}. For this segment the posterior expectation of the long-term unconditional probability ${\bar p}_1$ of being in the unsafe state is relatively large, $E({\bar p}_1|Y)=0.750$.
    \item
    For roadway segments from the second category $P(s_{t,n}=1|{\bf Y})\ll1$ for all $t=1,2,3,4,5$. Thus, we can say with high degree of certainty that these segments were always in the zero-accident state $s_{t,n}=0$ during the considered five-year time interval. A roadway segment $n$ belongs to this category if it had no any accidents observed over the five-year interval despite the accident rates given by equation~(\ref{EQ_MLE_LAMBDA_1}) were large, $\lambda_{t,n}\gg1$ for all $t=1,2,3,4,5$. Clearly this segment would be unlikely to have zero accidents observed, if it were not in the zero-accident state all the time.\footnote{Note that the zero-accident state may exist due to under-reporting of minor, low-severity accidents \citep[][]{SMM_97}.} An example of such roadway segment is given in the top-right plot in Figure~\ref{FIGURE_freq_year_1}. For this segment $E({\bar p}_1|Y)=0.260$ is relatively small.
    \item
    For roadway segments from the third category $P(s_{t,n}=1|{\bf Y})$ is neither one nor close to zero for all $t=1,2,3,4,5$.\footnote{If there were no Markov switching, which introduces time-dependence of states via equations~(\ref{EQ_MS_P}), then, assuming non-informative priors $\pi(s_{t,n}=0)=\pi(s_{t,n}=1)=1/2$ for states $s_{t,n}$, the posterior probabilities $P(s_{t,n}=1|{\bf Y})$ would be either exactly equal to $1$ (when $A_{t,n}>0$) or necessarily below $1/2$ (when $A_{t,n}=0$). In other words, we would have $P(s_{t,n}=1|{\bf Y})\notin[0.5,1)$ for any $t$ and $n$. Even with Markov switching existent, in this study we have never found any $P(s_{t,n}=1|{\bf Y})$ close but not equal to $1$, refer to the top plot in Figure~\ref{FIGURE_freq_year_2}.} For these segments we cannot determine with high certainty what states these segments were in during years $t=1,2,3,4,5$. A roadway segment $n$ belongs to this category if it had no any accidents observed over the considered five-year time interval and the accident rates were not large, $\lambda_{t,n}\lesssim 1$ for all $t=1,2,3,4,5$. In fact, when $\lambda_{t,n}\ll1$, the posterior probabilities of the two states are close to one-half, $P(s_{t,n}=1|{\bf Y})\approx P(s_{t,n}=0|{\bf Y})\approx0.5$, and no inference about the value of the state variable $s_{t,n}$ can be made. In this case of small accident rates, the observation of zero accidents is perfectly consistent with both states $s_{t,n}=0$ and $s_{t,n}=1$. An example of a roadway segment from the third category is given in the bottom-left plot in Figure~\ref{FIGURE_freq_year_1}. For this segment $E({\bar p}_1|Y)=0.496$ is about one-half.
    \item
    Finally, the fourth category is a mixture of the three categories described above. Roadway segments from this fourth category have posterior probabilities $P(s_{t,n}=1|{\bf Y})$ that change in time between the three possibilities given above. In particular, for some roadway segments we can say with high certainty that they changed their states in time from the zero-accident state $s_{t,n}=0$ to the unsafe state $s_{t,n}=1$ or vice versa. An example of a roadway segment from the fourth category is given in the bottom-right plot in Figure~\ref{FIGURE_freq_year_1}. For this segment $E({\bar p}_1|Y)=0.510$ is about one-half. Thus we find a direct empirical evidence that some roadway segments do change their states over time.
\end{itemize}

Next, it is useful to consider roadway segment statistics by state of roadway safety. Refer to Figure~\ref{FIGURE_freq_year_2}, made for the case of the MSNB model (note that the corresponding figure for the MSP model is similar and is not reported). The top plot in this figure shows the histogram of the posterior probabilities $P(s_{t,n}=1|{\bf Y})$ for all $N=335$ roadway segments during all $T=5$ years ($1675$ values of $s_{t,n}$ in total). For example, we find that during five years roadway segments had $P(s_{t,n}=1|{\bf Y})=1$ and were unsafe in $851$ cases, and they had $P(s_{t,n}=1|{\bf Y})<0.2$ and were likely to be safe in $212$ cases. The bottom plot in Figure~\ref{FIGURE_freq_year_2} shows the histogram of the posterior expectations $E[\bar p^{(n)}_1|{\bf Y}]$, where $\bar p^{(n)}_1=p^{(n)}_{0\rightarrow1}/(p^{(n)}_{0\rightarrow1} +p^{(n)}_{1\rightarrow0})$ are the stationary unconditional probabilities of the unsafe state (see Section~\ref{Markov}). We find that $0.2\le E[\bar p^{(n)}_1|{\bf Y}]\le0.8$ for all segments $n=1,2,\ldots,335$. This means that in the long run, all roadway segments have significant probabilities of visiting both the safe and the unsafe states.

\section{Model estimation results for weekly frequency data}
\label{RESULTS_FREQ_WEEK}
In this section we use weekly time periods, $t=1,2,3,\ldots,T=260$ in total.\footnote{A week is from Sunday to Saturday, there are 260 full weeks in the 1995-1999 time interval. We also considered daily time periods and obtained qualitatively similar results (not reported here).} The state $s_t$ is the same for all roadway segments and can change every week. Four types of weekly accident frequency models are estimated:
\begin{itemize}
    \item
    First, we estimate the standard (single-state) Poisson and negative binomial (NB) models, specified by equations~(\ref{EQ_MLE_POISSON}) and~(\ref{EQ_MLE_NB}). We estimate these models, first, by the maximum likelihood estimation (MLE) and, second, by the Bayesian inference approach and MCMC simulations (see footnote~\ref{MLE_LIMDEP} on page~\pageref{MLE_LIMDEP}). We refer to these models as ``P-by-MLE'' (for the Poisson model estimated by MLE), ``NB-by-MLE'' (for NB by MLE), ``P-by-MCMC'' (for Poisson by MCMC) and ``NB-by-MCMC'' (for NB by MCMC). As one expects, for our choice of a non-informative prior distribution, the estimated P-by-MCMC and NB-by-MCMC models turned out to be very similar to the P-by-MLE and NB-by-MLE models respectively.
    \item
    Second, we estimate a restricted two-state Markov switching Poisson model and a restricted two-state Markov switching negative binomial (MSNB) model. In these restricted switching models only the intercept in the model parameters vector $\betabf$ and the over-dispersion parameter $\alpha$ are allowed to switch between the two states of roadway safety. In other words, in equations~(\ref{EQ_MS_P_week}) and~(\ref{EQ_MS_NB_week}) only the first components of vectors $\betabf_{(0)}$ and $\betabf_{(1)}$ may differ, while the remaining components are restricted to be the same. In this case, the two states can have different average accident rates, given by equation~(\ref{EQ_MLE_LAMBDA}), but the rates have the same dependence on the explanatory variables. We refer to these models as ``restricted MSP'' and ``restricted MSNB''; they are estimated by the Bayesian-MCMC methods.
    \item
    Third, we estimate a full two-state Markov switching Poisson (MSP) model and a full two-state Markov switching negative binomial (MSNB) model, specified by equations~(\ref{EQ_MS_P_week}) and~(\ref{EQ_MS_NB_week}). In these models all estimable model parameters ($\beta$-s and $\alpha$) are allowed to switch between the two states of roadway safety. To choose the explanatory variables for the final restricted and full MSP and MSNB models reported here, we start with using the variables that enter the standard Poisson and NB models (see footnote~\ref{VARIABLES_CHOICE} on page~\pageref{VARIABLES_CHOICE}). Then we consecutively construct and use $60\%$, $85\%$ and $95\%$ Bayesian credible intervals for evaluation of the statistical significance of each $\beta$-parameter. As a result, in the final models some components of $\betabf_{(0)}$ and $\betabf_{(1)}$ are restricted to zero or restricted to be the same in the two states.\footnote{Of course, in the restricted models only the intercept is not restricted to be the same in the two states. For restrictions on other model coefficients, see footnote~\ref{BETA_RESTRICT} on page~\pageref{BETA_RESTRICT}.} We do not impose any restrictions on over-dispersion parameters ($\alpha$-s). We refer to the final full MSP and MSNB models as ``full MSP'' and ``full MSNB''; they are estimated by the Bayesian-MCMC methods.
\end{itemize}
Note that the two states, and thus the MSP and MSNB models, do not have to exist. For example, they will not exist if all estimated model parameters turn out to be statistically the same in the two states, $\betabf_{(0)}=\betabf_{(1)}$, (which suggests the two states are identical and the MSP and MSNB models reduce to the standard non-switching Poisson and NB models respectively). Also, the two states will not exist if all estimated state variables $s_t$ turn out to be close to zero, resulting in $p_{0\rightarrow1}\ll p_{1\rightarrow0}$ [compare to equation~(\ref{EQ_MS_P_RESTRICT})], then the less frequent state $s_t=1$ is not realized and the process always stays in state $s_t=0$.

The estimation results for all Poisson and NB models of weekly accident frequencies are given in Tables~\ref{T_week_P} and~\ref{T_week_NB} respectively. Posterior (or MLE) estimates of all continuous model parameters ($\beta$-s, $\alpha$, $p_{0\rightarrow1}$ and $p_{1\rightarrow0}$) are given together with their $95\%$ confidence intervals for MLE models and $95\%$ credible intervals for Bayesian-MCMC models (refer to the superscript and subscript numbers adjacent to parameter posterior/MLE estimates in Tables~\ref{T_week_P} and~\ref{T_week_NB}, and see footnote~\ref{MLE_NORMALITY} on page~\pageref{MLE_NORMALITY}). Table~\ref{T_freq_stat} on page~\pageref{T_freq_stat} gives summary statistics of all roadway segment characteristic variables ${\bf X}_{t,n}$ (except the intercept).

\begin{figure}[t]
\vspace{9.5truecm}
\includegraphics{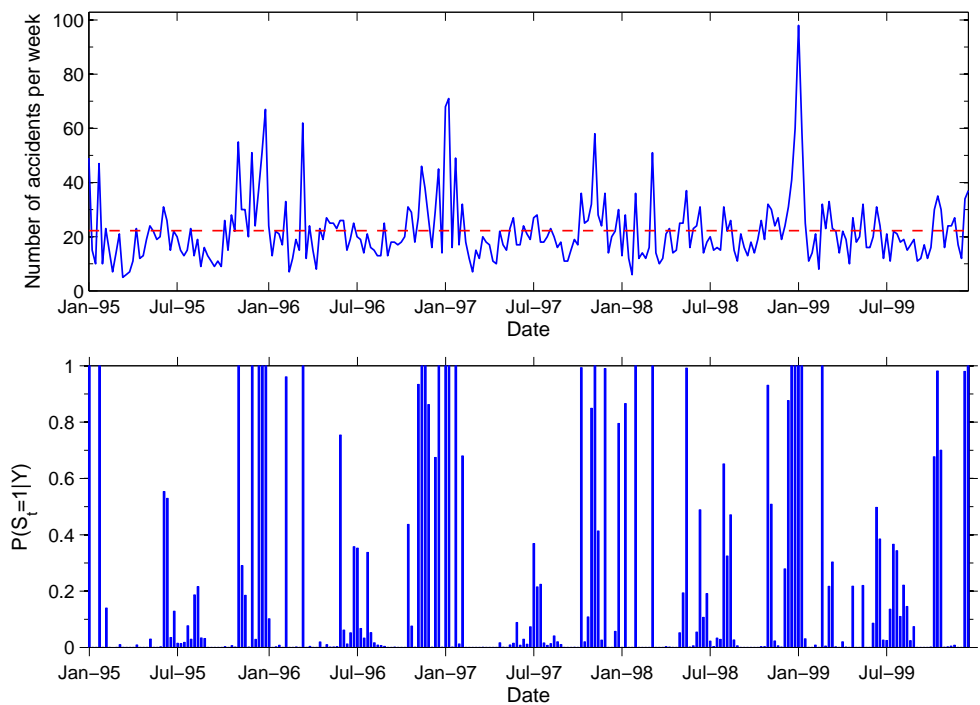}
\caption{The top plot shows the weekly accident frequencies in Indiana. The bottom plot shows weekly posterior probabilities \mbox{$P(s_t=1|{\bf Y})$} for the full MSNB model of weekly accident frequencies.
\label{FIGURE_freq_week}
}
\end{figure}

To visually see how the model tracks the data, consider Figure~\ref{FIGURE_freq_week}. The top plot in Figure~\ref{FIGURE_freq_week} shows the weekly time series of the number of accidents on selected Indiana interstate segments during the 1995-1999 time interval (the horizontal dashed line shows the average value). This plot shows that the number of accidents per week fluctuates strongly over time. Thus, under different conditions, roads can become considerably more or less safe. As a result, it is reasonable to assume that there exist two or more states of roadway safety. These states can help account for the existence of numerous unidentified and/or unobserved factors that influence roadway safety (unobserved heterogeneity). The bottom plot in Figure~\ref{FIGURE_freq_week} shows corresponding weekly posterior probabilities $P(s_t=1|{\bf Y})$ of the less frequent state $s_t=1$ for the full MSNB model. These probabilities are equal to the posterior expectations of $s_t$, $P(s_t=1|{\bf Y})=1\times P(s_t=1|{\bf Y})+0\times P(s_t=0|{\bf Y})=E(s_t|{\bf Y})$. Weekly values of $P(s_t=1|{\bf Y})$ for the restricted MSNB model and for the MSP models are very similar to those given on the bottom plot in Figure~\ref{FIGURE_freq_week}, and, as a result, are not shown on separate plots. Indeed, for example, the time-correlation\footnote{\label{WEIGHTED_CORR}Here and below we calculate weighted correlation coefficients. For variable $P(s_t=1|{\bf Y})\equiv E(s_t|{\bf Y})$ we use weights $w_t$ inversely proportional to the posterior standard deviations of $s_t$. That is $w_t\propto \min\left\{1/{\rm std}(s_t|{\bf Y}),{\rm median}[1/{\rm std}(s_t|{\bf Y})]\right\}$.} between $P(s_t=1|{\bf Y})$ for the two MSNB models (restricted and full) is about $99.5\%$.

Let us now turn to model estimation results. Because estimation results for Poisson models are very similar to estimation results for negative binomial models, let us focus on and discuss only the estimation results for negative binomial models. Our major findings, discussed below for negative binomial models, hold for Poisson models as well (unless otherwise stated). The findings are as follows.

\begin{landscape}
\begin{table}[p]
\centering
\caption{Estimation results for Poisson models of weekly accident frequencies}
\label{T_week_P}
\begin{scriptsize}
\input{frequency_week_Poisson_table_a.tex}

\end{scriptsize}
\end{table}
%\addtocounter{table}{-1}
\begin{table}[p]
\centering
%\caption{(Continued)}
\centerline{Table \arabic{chapter}.\arabic{table}: (Continued)}\vspace{1.7ex}
\begin{scriptsize}
\input{frequency_week_Poisson_table_b.tex}

\end{scriptsize}
\end{table}
\begin{table}[p]
\centering
\caption{Estimation results for negative binomial models of weekly accident frequencies}
\label{T_week_NB}
\begin{scriptsize}
\input{frequency_week_NB_table_a.tex}

\end{scriptsize}
\end{table}
%\addtocounter{table}{-1}
\begin{table}[p]
\centering
%\caption{(Continued)}
\centerline{Table \arabic{chapter}.\arabic{table}: (Continued)}\vspace{1.7ex}
\begin{scriptsize}
\input{frequency_week_NB_table_b.tex}

\end{scriptsize}
\end{table}
\end{landscape}

The findings show that two states exist and Markov switching models are non-trivial (in the sense that they do not reduce to the standard single-state models). In particular, we found that in the restricted MSNB model we over $99.9\%$ confident that the difference in values of $\beta$-intercept in the two states is non-zero.\footnote{\label{INTERCEPTS}The difference of the intercept values is statistically non-zero despite the fact that the $95\%$ credible intervals for these values overlap (see the ``Intercept'' line and the ``Restricted MSNB'' columns in Table~\ref{T_week_NB}). The reason is that the posterior draws of the intercepts are correlated. The statistical test of whether the intercept values differ, must be based on evaluation of their difference.} In addition, Markov switching models (restricted and full) are strongly favored by the empirical data as compared to the corresponding standard models. To compare the former with the later, we calculate and use Bayes factors given by equation~(\ref{EQ_BAYES_FACTOR}). From Table~\ref{T_week_NB} we see that the values of the logarithm of the marginal likelihood of the data for the standard NB, restricted MSNB and full MSNB models are $-16108.6$, $-15850.2$ and $-15809.4$ respectively. Thus, the restricted and full MSNB models provide considerable, $258.4$ and $299.2$, improvements of the logarithm of the marginal likelihood as compared to the standard non-switching NB model. As a result, given the accident data, the posterior probabilities of the restricted and full MSNB models are larger than the probability of the standard NB model by $e^{258.4}$ and $e^{299.2}$ respectively.\footnote{In addition, we find DIC (deviance information criterion) values $32219$, $31662$, $31577$ for the NB, restricted MSNB and full MSNB models respectively. We also find DIC values $32771$, $32086$, $31946$ or the Poisson, restricted MSP and full MSP models respectively. This means that the MSNB (MSP) models are favored over the standard NB (Poisson) model [the full MSNB (MSP) is favored most]. However, we prefer to rely on the Bayes factor approach instead of the DIC (see footnote~\ref{DIC} on page~\pageref{DIC}).} Note that we use equation~(\ref{EQ_L_MARGINAL}) for calculation of the values and the $95\%$ confidence intervals of the logarithms of the marginal likelihoods reported in Tables~\ref{T_week_P} and~\ref{T_week_NB}. The confidence intervals are found by bootstrap simulations (see footnote~\ref{BOOTSTRAP} on page~\pageref{BOOTSTRAP}).

We can also use a classical statistics approach for model comparison, based on the maximum likelihood estimation (MLE). Referring to Table~\ref{T_week_NB}, the MLE gives the maximum log-likelihood value $-16081.2$ for the standard NB model. The maximum log-likelihood values observed during our MCMC simulations for the restricted and full MSNB models are $-15786.6$ and $-15744.8$ respectively. An imaginary MLE, at its convergence, would give MSNB log-likelihood values that would be even larger than these observed values. Therefore, if estimated by the MLE, the MSNB models would provide very large (at least $294.6$ and $336.4$) improvements in the maximum log-likelihood value over the standard NB model. These improvements would come with only modest increases in the number of free continuous model parameters ($\beta$-s and $\alpha$-s) that enter the likelihood function. Both the Akaike Information Criterion (AIC) and the Bayesian Information Criterion (BIC) would strongly favor the MSNB models over the NB model (see footnote~\ref{AIC_BIC} on page~\pageref{AIC_BIC}).

To evaluate the goodness-of-fit for a model, we use the posterior (or MLE) estimates of all continuous model parameters ($\beta$-s, $\alpha$, $p_{0\rightarrow1}$, $p_{1\rightarrow0}$) and generate $10^4$ artificial data sets under the hypothesis that the model is true\footnote{\label{S_GENERATION}Note that the state values $\bf S$ are generated by using $p_{0\rightarrow1}$ and $p_{1\rightarrow0}$.}. We find the distribution of $\chi^2$, given by equation~(\ref{CHI_SQUARED}), and calculate the goodness-of-fit p-value for the observed value of $\chi^2$. The resulting p-values for the NB models are given in Table~\ref{T_week_NB}. These p-values are around \mbox{$65$--$70\%$}. Therefore, all models fit the data well.

Focusing on the full MSNB model, which is statistically superior because it has the maximal marginal likelihood of the data, its estimation results show that the less frequent state $s_t=1$ is about four times as rare as the more frequent state $s_t=0$ [refer to the estimated values of the unconditional probabilities ${\bar p}_0$ and ${\bar p}_1$ of the states $0$ and $1$, which are given by equation~(\ref{EQ_MS_P_BAR}) and reported in the ``Full MSNB'' columns in Table~\ref{T_week_NB}].

Also, the findings show that the less frequent state $s_t=1$ is considerably less safe than the more frequent state $s_t=0$. This result follows from the values of the mean weekly accident rate $\lambda_{t,n}$ [given by equation~(\ref{EQ_MLE_LAMBDA_1}) with model parameters $\beta$-s set to their posterior means in the two states], averaged over all values of the explanatory variables ${\bf X}_{t,n}$ observed in the data sample (see ``mean accident rate'' in Table~\ref{T_week_NB}). For the full MSNB model, on average, state $s_t=1$ has about two times more accidents per week than state $s_t=0$ has.\footnote{Note that accident frequency rates can easily be converted from one time period to another (for example, weekly rates can be converted to annual rates). Because accident events are independent, the conversion is done by a summation of moment-generating (or characteristic) functions. The sum of Poisson variates is Poisson. The sum of NB variates is also NB if all explanatory variables do not depend on time (${\bf X}_{t,n}={\bf X}_n$).} Therefore, it is not a surprise, that in Figure~\ref{FIGURE_freq_week} the weekly number of accidents (shown on the top plot) is larger when the posterior probability $P(s_t=1|{\bf Y})$ of the state $s_t=1$ (shown on the bottom plot) is higher.

Note that the long-term unconditional expectation of accident frequency $A_{t,n}$ is $E(A_{t,n})={\bar p}_0\big\langle\lambda^{(0)}_{t,n}\big\rangle_t+{\bar p}_1\langle\lambda^{(1)}_{t,n}\big\rangle_t$, where $\lambda^{(0)}_{t,n}=\exp(\betabf_{(0)}'{\bf X}_{t,n})$ and $\lambda^{(1)}_{t,n}=\exp(\betabf_{(1)}'{\bf X}_{t,n})$ are the mean accident rates in the states $s_t=0$ and $s_t=1$ respectively [see equation~(\ref{EQ_MLE_LAMBDA_1})], and $\langle\ldots\rangle_t$ means averaging over time. The unconditional expectation $E(A_{t,n})$ should be used in all predictions of long-term averaged accident rates on the $n^{\rm th}$ roadway segment. In the formula for this expectation, the mean accident rate $\lambda_{t,n}$ is averaged over the two states by using the stationary unconditional probabilities $\bar p_{0}$ and $\bar p_{1}$ (see the ``unconditional probabilities of states 0 and 1'' in Table~\ref{T_week_NB}).

It is also noteworthy that the number of accidents is more volatile in the less frequent and less-safe state ($s_t=1$). This is reflected in the fact that the standard deviation of the accident rate (${\rm std}_{t,n}=\sqrt{\lambda_{t,n}(1+\alpha\lambda_{t,n})}$ for NB distribution), averaged over all values of explanatory variables ${\bf X}_{t,n}$, is higher in state $s_t=1$ than in state $s_t=0$ (refer to Table~\ref{T_week_NB}). Moreover, for the full MSNB model the over-dispersion parameter $\alpha$ is higher in state $s_t=1$ ($\alpha=0.443$ in state $s_t=0$ and $\alpha=1.16$ in state $s_t=1$). Because state $s_t=1$ is relatively rare, this suggests that over-dispersed volatility of accident frequencies, which is often observed in empirical data, could be in part due to the latent switching between the states, and in part due to high accident volatility in the less frequent and less safe state $s_t=1$.

To study the effect of weather (which is usually unobserved heterogeneity in most data bases) on states, Table~\ref{T_week_corr} gives time-correlation coefficients between posterior probabilities $P(s_t=1|{\bf Y})$ for the full MSNB model and weather-condition variables. These correlations were found by using daily and hourly historical weather data in Indiana, available at the Indiana State Climate Office at Purdue University (www.agry.purdue.edu/climate). For these correlations, the precipitation and snowfall amounts are daily amounts in inches averaged over the week and across several weather observation stations that are located close to the roadway segments.\footnote{\label{SNOWFALL}Snowfall and precipitation amounts are weakly related with each other because snow density $(g/cm^3)$ can vary by more than a factor of ten.} The temperature variable is the mean daily air temperature $(^oF)$ averaged over the week and across the weather stations. The effect of fog/frost is captured by a dummy variable that is equal to one if and only if the difference between air and dewpoint temperatures does not exceed $5^oF$ (in this case frost can form if the dewpoint is below the freezing point $32^oF$, and fog can form otherwise). The fog/frost dummies are calculated for every hour and are averaged over the week and across the weather stations. Finally, visibility distance variable is the harmonic mean of hourly visibility distances, which are measured in miles every hour and are averaged over the week and across the weather stations.\footnote{\label{HARMONIC_MEAN}The harmonic mean $\bar d$ of distances $d_n$ is calculated as $\bar d^{-1}=(1/N)\sum^N_{n=1}d^{-1}_n$, assuming $d_n=0.25$ miles if $d_n\leq0.25$ miles.}

\begin{table}[t]
\centering
\caption{Correlations of the posterior probabilities $P(s_t=1|{\bf Y})$
with weather-condition variables for the full MSNB model}
\label{T_week_corr}
\begin{footnotesize}
\input{frequency_week_corr_table.tex}

\end{footnotesize}
\end{table}

Table~\ref{T_week_corr} shows that the less frequent and less safe state $s_t=1$ is positively correlated with extreme temperatures (low during winter and high during summer), rain precipitations and snowfalls, fogs and frosts, low visibility distances. It is reasonable to expect that during bad weather, roads can become significantly less safe, resulting in a change of the state of roadway safety. As a useful test of the switching between the two states, all weather variables, listed in Table~\ref{T_week_corr}, were added into our full MSNB model. However, when doing this, the two states did not disappear and the posterior probabilities $P(s_t=1|{\bf Y})$ did not changed substantially (the correlation between the new and the old probabilities was around $90\%$). As another test, we modified the standard single-state NB model by adding the weather variables into it. As a result, the marginal likelihood for this model improved noticeably, but the modified single-state NB model was still strongly disfavored by the data as compared to the restricted and full MSNB models. This result emphasizes the importance of the two-state approach.

Let us give a brief summary of the effects of explanatory variables on accident rates. We will focus on those variables that are significantly different between the two states in the full MSNB model. Table~\ref{T_week_NB} shows that parameter estimates for pavement quality index, total number of ramps on the road viewing and opposite sides, average annual daily traffic (AADT), number of bridges per mile, percentage of single unit trucks, and season dummy variables are all significantly different between the two states. All these differences are reasonable and could be explained by adverse weather/pavement conditions in the less-safe state $s_t=1$, and by the resulting lighter-than-usual traffic and more alert/defensive driving in this state. In particular, as compared to variable effects in the safe state $s_t=0$, in the less safe state $s_t=1$ an improvement of pavement quality leads to a smaller reduction of the accident rate, an increase in percentage of single unit trucks results in a larger increase of the accident rate, and an increase in AADT leads to a
smaller increase of the accident rate (note that the effects of AADT and its logarithm should be considered simultaneously). An increase in number of ramps and bridges, and the summer season indicator significantly reduce the accident rate only in the less-safe state $s_t=1$. The winter season indicator reduces the accident rate only in the safe state $s_t=0$ (this result, which might look counter-intuitive, could be explained by an increase in cases of over-confident, reckless driving during good weather/pavement conditions, unless there is a winter).

Finally, because the time series in Figure~\ref{FIGURE_freq_week} seem to exhibit a seasonal pattern [roads appear to be less safe and $P(s_t=1|{\bf Y})$ appears to be higher during winters], we estimated MSNB and MSP models in which the transition probabilities $p_{0\rightarrow1}$ and $p_{1\rightarrow0}$ are not constant (allowing each of them to assume two different values: one during winters and the other during non-winter seasons).\footnote{Let us briefly describe how these models can be specified by using the general representation of Markov switching models, given in Section~\ref{GENERAL_MS_MODEL}. We define the winter seasons to be from November to March. The non-winter seasons are from April to October. For relations between the real time indexing and the auxiliary time indexing we have ${\tilde t}=t$, ${\tilde T}=T$, ${\tilde n}=n$, ${\tilde N_{\tilde t}}=N$, ${\cal T\!\_}=\{\}$. The elements of set ${\cal T}=\{1,14,45,67,97,119,149,171,201,223,254,261\}$ are in weekly time units and contain the left boundaries of the winter and non-winter time intervals for the years 1995-1999. The total number of time intervals is $R=11$. Transition probabilities $p^{(1)}_{0\rightarrow1}$, $p^{(1)}_{1\rightarrow0}$, $p^{(2)}_{0\rightarrow1}$ and $p^{(2)}_{1\rightarrow0}$, which are for the first winter and first non-winter intervals are free parameters. All other transition probabilities are not free: for the remaining winter intervals they are restricted to $p^{(1)}_{0\rightarrow1}$ and $p^{(1)}_{1\rightarrow0}$, and for the remaining non-winter intervals they are restricted to $p^{(2)}_{0\rightarrow1}$ and $p^{(2)}_{1\rightarrow0}$.} However, these models did not perform as well as the MSNB and MSP models with constant transition probabilities [as judged by the Bayes factors, see equation~(\ref{EQ_BAYES_FACTOR})].\footnote{We have only six (five full) winter periods in our five-year data. MSNB and MSP with seasonally changing transition probabilities could perform better for an accident data that covers a longer time period.}

%% file: frequency_year_Poisson-NB_table_a.tex
\begin{tabular}{|l|c|c|c|c|}
\hline
${}_{{}_{{}\atop{\displaystyle\mbox{\bf Variable}}}}$
& \multicolumn{2}{|c|}{\bf Poisson} & \multicolumn{2}{|c|}{\bf NB}\\
\cline{2-5}
%${}^{{}^{\bf\displaystyle Variable}}$
& {\bf by MLE$^{\,\rm a}$} & {\bf by MCMC$^{\,\rm b}$} &
{\bf by MLE$^{\,\rm c}$} & {\bf by MCMC$^{\,\rm d}$}\\
\hline
Intercept (constant term) &
$-15.7^{-13.6}_{-17.8}$ &
$-15.7^{-13.6}_{-17.8}$ &
$-20.0^{-16.8}_{-23.2}$ &
$-20.3^{-16.9}_{-23.8}$\\
\hline
Accident occurring on interstates I-70 or I-164 (dummy) &
$-.689^{-.599}_{-.778}$ &
$-.689^{-.600}_{-.778}$ &
$-.756^{-.608}_{-.905}$ &
$-.760^{-.623}_{-.898}$\\
\hline
Pavement quality index (PQI) average$^{\,\rm e}$ &
$-.0184^{-.0133}_{-.0235}$ &
$-.0184^{-.0134}_{-.0234}$ &
$-.0150^{-.00646}_{-.0235}$ &
$-.0149^{-.00668}_{-.0231}$\\
\hline
Road segment length (in miles) &
$.0506^{.0761}_{.0251}$ &
$.0504^{.0756}_{.0250}$ &
-- &
--\\
\hline
Logarithm of road segment length (in miles) &
$.924^{.979}_{.869}$ &
$.925^{.979}_{.871}$ &
$.989^{1.05}_{.930}$ &
$.990^{1.04}_{.938}$\\
\hline
Number of ramps on the viewing side &
$-.0397^{-.0142}_{-.0651}$ &
$-.0396^{-.0144}_{-.0649}$ &
-- &
--\\
\hline
Number of ramps on the viewing side per lane per mile &
$.414^{.493}_{.335}$ &
$.414^{.492}_{.335}$ &
$.407^{.501}_{.312}$ &
$.410^{.510}_{.312}$\\
\hline
Number of lanes on a roadway &
-- &
-- &
$.513^{.117}_{.910}$ &
$.553^{1.23}_{-}$\\
\hline
Median configuration is depressed (dummy) &
$.177^{.277}_{.0775}$ &
$.178^{.278}_{.0788}$ &
$.187^{.317}_{.0558}$ &
$.186^{.319}_{.0545}$\\
\hline
Median barrier presence (dummy) &
$-3.01^{-2.39}_{-3.63}$ &
$-3.06^{-2.47}_{-3.71}$ &
$-2.41^{-2.00}_{-2.82}$ &
$-2.44^{-1.90}_{-3.02}$\\
\hline
Interior shoulder presence (dummy) &
$-1.09^{-.428}_{-1.75}$ &
$-1.12^{-.493}_{-1.81}$ &
-- &
--\\
\hline
Width of the interior shoulder is less that 5 feet (dummy) &
$.358^{.456}_{.259}$ &
$.358^{.457}_{.261}$ &
$.358^{.509}_{.207}$ &
$.358^{.503}_{.214}$\\
\hline
Outside shoulder width (in feet) &
$-.0612^{-.0377}_{-.0847}$ &
$-.0614^{-.0380}_{-.0849}$ &
$-.0632^{-.0281}_{-.0982}$ &
$-.0633^{-.0289}_{-.0979}$\\
\hline
Outside barrier absence (dummy) &
$-.244^{-.136}_{-.353}$ &
$-.244^{-.135}_{-.351}$ &
$-.251^{-.111}_{-.391}$ &
$-.252^{-.0984}_{-.406}$\\
\hline
Average annual daily traffic (AADT) &
$\displaystyle{-3.98^{-3.15\vphantom{O^o}}_{-4.81}}\atop\displaystyle{{}\times10^{-5}}$ &
$\displaystyle{-3.99^{-3.17\vphantom{O^o}}_{-4.85}}\atop\displaystyle{{}\times10^{-5}}$ &
$\displaystyle{-4.83^{-3.72\vphantom{O^o}}_{-5.95}}\atop\displaystyle{{}\times10^{-5}}$ &
$\displaystyle{-4.93^{-3.84\vphantom{O^o}}_{-6.08}}\atop\displaystyle{{}\times10^{-5}}$\\
\hline
Logarithm of average annual daily traffic &
$2.05^{2.28}_{1.82}$ &
$2.05^{2.29}_{1.82}$ &
$2.25^{2.61}_{1.89}$ &
$2.28^{2.61}_{1.96}$\\
\hline
Posted speed limit (in mph) &
$.0121^{.0205}_{.00370}$ &
$.0121^{.0204}_{.00379}$ &
$.0145^{.0282}_{.000762}$ &
$.0146^{.0280}_{.00128}$\\
\hline
Number of bridges per mile &
$-.0257^{-.00860}_{-.0428}$ &
$-.0262^{-.00966}_{-.0435}$ &
$-.0261^{-.00425}_{-.0479}$ &
$-.0270^{-.00652}_{-.0488}$\\
\hline
Maximum of reciprocal values of horizontal curve radii (in $1/{\rm mile}$) &
$-.164^{-.107}_{-.222}$ &
$-.165^{-.107}_{-.222}$ &
$-.194^{-.108}_{-.281}$ &
$-.196^{-.112}_{-.282}$\\
\hline
Maximum absolute value of change in grade of a vertical curve &
$.0456^{.0226}_{.0686}$ &
$.0456^{.0684}_{.0226}$ &
-- &
--\\
\hline
Number of vertical curves per roadway section &
$-.158^{-.0599}_{-.255}$ &
$-.158^{-.0621}_{-.255}$ &
-- &
--\\
\hline
Percentage of single unit trucks (daily average) &
$1.40^{1.86}_{.939}$ &
$1.40^{1.86}_{.942}$ &
$1.66^{2.51}_{.810}$ &
$1.66^{2.42}_{.912}$\\
\hline
Number of changes per vertical profile along a roadway segment &
$.0616^{.109}_{.0140}$ &
$.0619^{.109}_{.0151}$ &
$.0597^{.107}_{.0119}$ &
$.0614^{.105}_{.0191}$\\
\hline
\end{tabular}

%% file: frequency_year_Poisson-NB_table_b.tex
\begin{tabular}{|l|c|c|c|c|}
\hline
${}_{{}_{{}\atop{\displaystyle\mbox{\bf Variable}}}}$
& \multicolumn{2}{|c|}{\bf Poisson} & \multicolumn{2}{|c|}{\bf NB}\\
\cline{2-5}
%${}^{{}^{\bf\displaystyle Variable}}$
& {\bf by MLE$^{\,\rm a}$} & {\bf by MCMC$^{\,\rm b}$} &
{\bf by MLE$^{\,\rm c}$} & {\bf by MCMC$^{\,\rm d}$}\\
\hline
Over-dispersion parameter $\alpha$ in NB models & -- &
-- &
$.227^{.278}_{.175}$ &
$.240^{.301}_{.187}$\\
\hline
\hline
Mean accident rate ($\lambda_{t,n}$ for Poisson and NB),
averaged over all values of ${\bf X}_{t,n}$ &
-- &
$3.45$ &
-- &
$3.54$\\
\hline
Standard deviation of accident rate
($\lambda_{t,n}$ for Poisson; $\sqrt{\lambda_{t,n}(1+\alpha\lambda_{t,n})}\,$ for NB), & & & &\\
averaged over all values of explanatory variables ${\bf X}_{t,n}$ &
-- &
$1.38$ &
-- &
$2.33$\\
\hline
Total number of free model parameters &
$22$ &
$22$ &
$19$ &
$19$\\
\hline
Posterior average of the log-likelihood (LL) &
-- &
$-2662.09^{-2656.61}_{-2669.48}$ &
-- &
$-2543.32^{-2538.27}_{-2550.24}$\\
\hline
Max$(LL)$:\quad true maximum value of log-likelihood (LL) for MLE; maximum & & & &\\
observed value of LL for Bayesian-MCMC &
$\displaystyle{-2651.16}\atop{\rm(true)}$ &
$\displaystyle{-2652.37}\atop{\rm(observed)}$ &
$\displaystyle{-2533.81}\atop{\rm(true)}$ &
$\displaystyle{-2534.52}\atop{\rm(observed)}$\\
\hline
Logarithm of marginal likelihood of data ($\ln[f({\bf Y}|{\cal M})]$) &
-- &
$-2672.27^{-2669.92}_{-2674.02}$ &
-- &
$-2554.16^{-2550.49}_{-2556.52}$\\
\hline
Goodness-of-fit p-value &
-- &
$0.000$ &
-- &
$0.003$\\
\hline
Maximum of the potential scale reduction factors (PSRF)$^{\,\rm f}$ &
-- &
$1.02304$ &
-- &
$1.01813$\\
\hline
Multivariate potential scale reduction factor (MPSRF)$^{\,\rm f}$ &
-- &
$1.02434$ &
-- &
$1.01938$\\
%\hline
%Average candidate acceptance rate in Metropolis-Hasting sampling &
%-- &
%$\!.296\leq {\rm rate}\leq.297\!$ &
%-- &
%$\!.296\leq {\rm rate}\leq.297\!$\\
\hline
\multicolumn{5}{l}{$^{\rm a,\:c}$ Standard (conventional) Poisson and negative binomial
correspondingly estimated by maximum likelihood estimation (MLE).}\\
\multicolumn{5}{l}{$^{\rm b,\:d}$ Standard Poisson and negative binomial
correspondingly estimated by Markov Chain Monte Carlo (MCMC) simulations.}\\
\multicolumn{5}{l}{$^{\rm e}$ The pavement quality index (PQI) is a composite
measure of overall pavement quality evaluated on a 0 to 100 scale.}\\
\multicolumn{5}{l}{$^{\rm f}$ PSRF/MPSRF are calculated separately/jointly
for all continuous model parameters. PSRF and MPSRF are close to 1 for
converged MCMC chains.}\\
\end{tabular}

%% file: frequency_year_ZIP-MSP_table_a.tex
\begin{tabular}{|l|c|c|c|c|c|}
\hline
${}_{{}_{{}\atop{\displaystyle\mbox{\bf Variable}}}}$
 & \multicolumn{2}{|c|}{\bf\boldmath ZIP-$\tau^{\,\rm a}$} &
   \multicolumn{2}{|c|}{\bf\boldmath ZIP-$\gamma^{\,\rm b}$} & {\bf MSP$^{\,\rm c}$}
\\
\cline{2-3}\cline{4-5}
 & {\bf by MLE} & {\bf by MCMC} & {\bf by MLE} & {\bf by MCMC}
 & {\bf by MCMC}
\\
\hline
\multicolumn{6}{|c|}{$\beta$-coefficients
in Equation~(\ref{EQ_MLE_LAMBDA})}\\
\hline
Intercept (constant term) &
$-6.66^{-5.90}_{-7.41}$ &
$-6.65^{-5.75}_{-7.57}$ &
$-7.82^{-7.01}_{-8.64}$ &
$-7.85^{-6.85}_{-8.87}$ &
$-13.4^{-10.7}_{-16.1}$\\
\hline
Accident occurring on interstates I-70 or I-164 (dummy) &
$-.525^{-.455}_{-.595}$ &
$-.526^{-.441}_{-.611}$ &
$-.594^{-.523}_{-.665}$ &
$-.596^{-.505}_{-.686}$ &
$-.631^{-.544}_{-.718}$\\
\hline
Pavement quality index (PQI) average$^{\,\rm d}$ &
$-.00859^{-.00489}_{-.123}$ &
$-.00860^{-.00389}_{-.0133}$ &
$-.0101^{-.00622}_{-.0139}$ &
$-.0101^{-.00493}_{-.0152}$ &
$-.015^{-.00969}_{-.0202}$\\
\hline
Road segment length (in miles) &
$.0803^{.0982}_{.0624}$ &
$.0801^{.0561}_{.104}$ &
$.0674^{.0871}_{.0477}$ &
$.0667^{.0929}_{.0402}$ &
$.092^{.117}_{.0668}$\\
\hline
Logarithm of road segment length (in miles) &
$.741^{.781}_{.702}$ &
$.742^{.789}_{.694}$ &
$.804^{.853}_{.756}$ &
$.808^{.875}_{.742}$ &
$.714^{.776}_{.652}$\\
\hline
Number of ramps on the viewing side &
$-.0301^{-.00906}_{-.0512}$ &
$-.0301^{-.00559}_{-.0545}$ &
$-.0247^{-.00328}_{-.0461}$ &
$-.0247^{.00000181}_{-.0494}$ &
$-.0332^{-.00825}_{-.0581}$\\
\hline
Number of ramps on the viewing side per lane per mile &
$.309^{.369}_{.249}$ &
$.308^{.381}_{.234}$ &
$.301^{.369}_{.233}$ &
$.302^{.386}_{.218}$ &
$.303^{.382}_{.223}$\\
\hline
Median configuration is depressed (dummy) &
$.144^{.220}_{.0679}$ &
$.144^{.233}_{.0554}$ &
$.149^{.232}_{.0655}$ &
$.150^{.257}_{.0443}$ &
$.126^{.231}_{.0224}$\\
\hline
Median barrier presence (dummy) &
$-2.50^{-2.35}_{-2.65}$ &
$-2.50^{-2.24}_{-2.76}$ &
$-.0821^{-.598}_{-1.04}$ &
$-.828^{-.525}_{-1.14}$ &
$-2.30^{-1.57}_{-3.10}$\\
\hline
Interior shoulder presence (dummy) &
-- &
-- &
-- &
-- &
$-2.02^{-1.22}_{-2.87}$\\
\hline
Width of the interior shoulder is less that 5 feet (dummy) &
$.341^{.414}_{.269}$ &
$.342^{.433}_{.251}$ &
$.324^{.403}_{.245}$ &
$.325^{.428}_{.223}$ &
$.304^{.405}_{.204}$\\
\hline
Outside shoulder width (in feet) &
$-.0575^{-.0394}_{-.0756}$ &
$-.0578^{-.0358}_{-.0799}$ &
$-.0635^{-.0439}_{-.0832}$ &
$-.0639^{-.0393}_{-.0887}$ &
$-.0419^{-.0172}_{-.0667}$\\
\hline
Outside barrier absence (dummy) & $-.179^{-.106}_{-.252}$ &
$-.178^{-.0765}_{-.279}$ &
$-.253^{-.175}_{-.331}$ &
$-.253^{-.140}_{-.365}$ &
$-.233^{-.121}_{-.343}$\\
\hline
Average annual daily traffic (AADT) & -- &
-- &
-- &
-- &
$\displaystyle{-3.60^{-2.49\vphantom{O^o}}_{-4.69}}\atop\displaystyle{{}\times10^{-5}}$\\
\hline
Logarithm of average annual daily traffic &
$.841^{.903}_{.779}$ &
$.841^{.914}_{.769}$ &
$1.03^{1.10}_{.959}$ &
$1.03^{1.11}_{.947}$ &
$1.89^{2.19}_{1.60}$\\
\hline
Posted speed limit (in mph) &
$.0163^{.0227}_{.00993}$ &
$.0164^{.0241}_{.00860}$ &
$.00825^{.0149}_{.00164}$ &
$.00837^{.0169}_{-.0000724}$ &
$.00899^{.0175}_{.000528}$\\
\hline
Number of bridges per mile &
$-.0366^{-.0254}_{-.0478}$ &
$-.0368^{-.0222}_{-.0518}$ &
$-.0246^{-.00925}_{-.0400}$ &
$-.0249^{-.00644}_{-.0440}$ &
$-.0223^{-.00574}_{-.0401}$\\
\hline
Maximum of reciprocal values of horizontal curve radii (in $1/{\rm mile}$) &
$-.148^{-.105}_{-.190}$ &
$-.148^{-.0944}_{-.201}$ &
$-.106^{-.0594}_{-.152}$ &
$-.107^{-.0455}_{-.168}$ &
$-.127^{-.0684}_{-.186}$\\
\hline
Maximum absolute value of change in grade of a vertical curve &
$.0328^{.0499}_{.0157}$ &
$.0328^{.0551}_{.0104}$ &
$.0308^{.0494}_{.0122}$ &
$.0309^{.0539}_{.00760}$ &
$.0208^{.0380}_{.00341}$\\
\hline
Number of vertical curves per roadway section &
$-.1498^{-.0765}_{-.2231}$ &
$-.151^{-.0609}_{-.241}$ &
$-.1157^{-.0362}_{-.1952}$ &
$-.117^{-.0200}_{-.214}$ &
--\\
\hline
Percentage of single unit trucks (daily average) &
$.614^{.941}_{.287}$ &
$.616^{1.05}_{.184}$ &
$.8143^{1.15}_{.478}$ &
$.821^{1.28}_{.363}$ &
$1.00^{1.45}_{.548}$\\
\hline
Number of changes per vertical profile along a roadway segment &
$.0681^{.104}_{.0320}$ &
$.0684^{.112}_{.0248}$ &
$.0398^{.0784}_{.00123}$ &
$.0402^{.0872}_{-.00665}$ &
--\\
\hline
\end{tabular}

%% file: frequency_year_ZIP-MSP_table_b.tex
\begin{tabular}{|l|c|c|c|c|c|}
\hline
${}_{{}_{{}\atop{\displaystyle\mbox{\bf Variable}}}}$
 & \multicolumn{2}{|c|}{\bf\boldmath ZIP-$\tau^{\,\rm a}$} &
   \multicolumn{2}{|c|}{\bf\boldmath ZIP-$\gamma^{\,\rm b}$} & {\bf MSP$^{\,\rm c}$}
\\
\cline{2-3}\cline{4-5}
 & {\bf by MLE} & {\bf by MCMC} & {\bf by MLE} & {\bf by MCMC}
 & {\bf by MCMC}
\\
\hline
\multicolumn{6}{|c|}{$\tau$- and $\gamma$-coefficients
in Equations~(\ref{EQ_Q_TAU}) and~(\ref{EQ_Q_GAMMA})}\\
\hline
The model parameter $\tau$ in Equation~(\ref{EQ_Q_TAU}) &
$-1.42^{-1.22}_{-1.62}$ &
$-1.42^{-1.24}_{-1.61}$ &
-- &
-- &
--\\
\hline
Intercept (constant term) &
-- &
-- &
-- &
-- &
--\\
\hline
Logarithm of road segment length (in miles) &
-- &
-- &
$-1.40^{-1.09}_{-1.71}$ &
$-1.42^{-1.14}_{-1.71}$ &
--\\
\hline
Median barrier presence (dummy) &
-- &
-- &
$.157^{.961}_{-.647}$ &
$4.16^{4.98}_{3.41}$ &
--\\
\hline
Width of the interior shoulder is less that 5 feet (dummy) &
-- &
-- &
$-.921^{-.370}_{-1.47}$ &
$-.937^{-.425}_{-1.45}$ &
--\\
\hline
Outside shoulder width (in feet) &
-- &
-- &
$-.222^{-.159}_{-.285}$ &
$-.288^{-.176}_{-.286}$ &
--\\
\hline
Maximum of reciprocal values of horizontal curve radii (in $1/{\rm mile}$) &
-- &
-- &
$.573^{.944}_{.201}$ &
$.581^{.952}_{.215}$ &
--\\
\hline
\hline
Mean accident rate ($\lambda_{t,n}$), averaged over all values of ${\bf X}_{t,n}$ &
-- &
$3.41$ &
-- &
$3.42$ &
$3.94$\\
\hline
Standard deviation of accident rate
($\lambda_{t,n}$), averaged over all & & & & &\\
values of explanatory variables ${\bf X}_{t,n}$ &
-- &
$1.62$ &
-- &
$1.67$ &
$1.60$\\
\hline
Total number of free model parameters ($\beta$-s, $\gamma$-s, $\alpha$ and $\tau$) &
$22$ &
$21$ &
$25$ &
$25$ &
$20$\\
\hline
Posterior average of the log-likelihood (LL) &
-- &
$-2636.01^{-2630.69}_{-2643.18}$ &
-- &
$-2519.54^{-2513.62}_{-2527.33}$ &
$-2149.82^{-2122.28}_{-2178.53}$\\
\hline
Max$(LL)$:\quad true maximum value of log-likelihood (LL) for MLE; & & & & &\\
maximum observed value of LL for Bayesian-MCMC &
$\displaystyle{-2625.58}\atop{\rm(true)}$ &
$\displaystyle{-2626.63}\atop{\rm(observed)}$ &
$\displaystyle{-2507.07}\atop{\rm(true)}$ &
$\displaystyle{-2508.61}\atop{\rm(observed)}$ &
$\displaystyle{-2080.32}\atop{\rm(observed)}$\\
\hline
Logarithm of marginal likelihood of data ($\ln[f({\bf Y}|{\cal M})]$) &
-- &
$-2648.41^{-2643.46}_{-2651.00}$ &
-- &
$-2532.21^{-2527.95}_{-2534.02}$ &
$-2229.27^{-2194.42}_{-2214.49}$\\
\hline
Goodness-of-fit p-value &
-- &
$0.000$ &
-- &
$0.009$ &
$0.009$\\
\hline
Maximum of the potential scale reduction factors (PSRF)$^{\,\rm e}$ &
-- &
$1.00163$ &
-- &
$1.00252$ &
$1.02803$\\
\hline
Multivariate potential scale reduction factor (MPSRF)$^{\,\rm e}$ &
-- &
$1.00171$ &
-- &
$1.00255$ &
$1.02852$\\
%\hline
%Average candidate acceptance rate in Metropolis-Hasting sampling &
%-- &
%$\!.296\leq {\rm rate}\leq.297\!$ &
%-- &
%$\!.295\leq {\rm rate}\leq.297\!$ &
%$\!.296\leq {\rm rate}\leq.297\!$\\
\hline
\multicolumn{6}{l}{$^{\rm a}$ Standard (conventional) \mbox{ZIP-$\tau$}
model estimated by maximum likelihood estimation (MLE) and Markov Chain Monte Carlo (MCMC) simulations.}
\\
\multicolumn{6}{l}{$^{\rm b}$ Standard \mbox{ZIP-$\gamma$}
model estimated by maximum likelihood estimation (MLE) and Markov Chain Monte Carlo (MCMC) simulations.}
\\
\multicolumn{6}{l}{$^{\rm c}$ Two-state Markov switching Poisson (MSP) model where all reported parameters are for the unsafe state $s=1$.}
\\
\multicolumn{6}{l}{$^{\rm d}$ The pavement quality index (PQI) is a composite
measure of overall pavement quality evaluated on a 0 to 100 scale.}
\\
\multicolumn{6}{l}{$^{\rm e}$ PSRF/MPSRF are calculated separately/jointly
for all continuous model parameters. PSRF and MPSRF are close to 1 for
converged MCMC chains.}
\end{tabular}

%% file: frequency_year_ZINB-MSNB_table_a.tex
\begin{tabular}{|l|c|c|c|c|c|}
\hline
${}_{{}_{{}\atop{\displaystyle\mbox{\bf Variable}}}}$
 & \multicolumn{2}{|c|}{\bf\boldmath ZINB-$\tau^{\,\rm a}$} &
   \multicolumn{2}{|c|}{\bf\boldmath ZINB-$\gamma^{\,\rm b}$} & {\bf MSNB$^{\,\rm c}$}
\\
\cline{2-3}\cline{4-5}
 & {\bf by MLE} & {\bf by MCMC} & {\bf by MLE} & {\bf by MCMC}
 & {\bf by MCMC}
\\
\hline
\multicolumn{6}{|c|}{$\beta$- and $\alpha$-parameters in
Equation~(\ref{EQ_MLE_LAMBDA_1})}\\
\hline
Intercept (constant term) &
$-15.0^{-12.5}_{-17.5}$ &
$-15.2^{-13.0}_{-17.4}$ &
$-11.6^{-8.32}_{-14.8}$ &
$-11.6^{-8.29}_{-14.6}$ &
$-17.3^{-13.0}_{-21.3}$\\
\hline
Accident occurring on interstates I-70 or I-164 (dummy) &
$-.683^{-.570}_{-.797}$ &
$-.685^{-.575}_{ -.794}$ &
$-.715^{-.602}_{-.829}$ &
$-.715^{-.593}_{-.836}$ &
$-.734^{-.617}_{-.850}$\\
\hline
Pavement quality index (PQI) average$^{\,\rm d}$ &
$-.0122^{-.0189}_{-.00550}$ &
$-.0122^{-.00562}_{-.0188}$ &
$-.0140^{-.00627}_{-.0217}$ &
$-.0143^{-.00643}_{-.0221}$ &
$-.0163^{-.00850}_{-.0240}$\\
\hline
Logarithm of road segment length (in miles) &
$.791^{.832}_{.751}$ &
$.791^{.829 }_{.754}$ &
$.929^{.978}_{.880}$ &
$.939^{.993}_{.886}$ &
$.887^{.929}_{.845}$\\
\hline
Number of ramps on the viewing side per lane per mile &
$.226^{.300}_{.153}$ &
$.227^{.306}_{.149}$ &
$.298^{.387}_{.209}$ &
$.304^{.394}_{.214}$ &
$.317^{.404}_{.230}$\\
\hline
Number of lanes on a roadway & -- &
-- &
-- &
-- &
$1.19^{2.04}_{.386}$\\
\hline
Median configuration is depressed (dummy) &
$.184^{.288}_{.0795}$ &
$.183^{.282}_{.0839}$ &
$.201^{.319}_{.0820}$ &
$.202^{.325}_{.0781}$ &
--\\
\hline
Median barrier presence (dummy) &
$-1.43^{-1.22}_{-1.64}$ &
$-1.43^{-1.14}_{-1.72}$ &
-- &
-- &
$-1.69^{-1.00}_{-2.46}$\\
\hline
Width of the interior shoulder is less that 5 feet (dummy) &
$.323^{.443}_{.202}$ &
$.323^{.434}_{.211}$ &
$.435^{.572}_{.297}$ &
$.437^{.569}_{.307}$ &
$.374^{.505}_{.243}$\\
\hline
Outside shoulder width (in feet) &
$-.0480^{-.0196}_{-.0764}$ &
$-.0478^{-.0207}_{-.0749}$ &
$-.0532^{-.0176}_{-.0887}$ &
$-.0532^{-.020}_{-.0867}$ &
$-.0537^{-.0214}_{-.0862}$\\
\hline
Outside barrier absence (dummy) & -- &
-- &
$-.245^{-.117}_{-.373}$ &
$-.245^{-.101}_{-.389}$ &
$-.264^{-.124}_{-.403}$\\
\hline
Average annual daily traffic (AADT) &
$\displaystyle{-4.07^{-3.17\vphantom{O^o}}_{-4.97}}\atop\displaystyle{{}\times10^{-5}}$ &
$\displaystyle{-4.14^{-3.31\vphantom{O^o}}_{-5.04}}\atop\displaystyle{{}\times10^{-5}}$ &
$\displaystyle{-1.93^{-3.21\vphantom{O^o}}_{-6.50}}\atop\displaystyle{{}\times10^{-5}}$ &
$\displaystyle{-1.91^{-3.16\vphantom{O^o}}_{-5.83}}\atop\displaystyle{{}\times10^{-5}}$ &
$\displaystyle{-3.78^{-2.02\vphantom{O^o}}_{-5.26}}\atop\displaystyle{{}\times10^{-5}}$\\
\hline
Logarithm of average annual daily traffic &
$1.89^{2.17}_{1.61}$ &
$1.91^{2.16}_{1.67}$ &
$1.52^{1.88}_{1.15}$ &
$1.52^{1.86}_{1.15}$ &
$1.95^{2.34}_{1.49}$\\
\hline
Number of bridges per mile & -- &
-- &
-- &
-- &
$-.0214^{-.00164}_{-.0428}$\\
\hline
Maximum of reciprocal values of horizontal curve radii (in $1/{\rm mile}$) &
$-.140^{-.0710}_{-.209}$ &
$-.141^{-.0734}_{-.208}$ &
$-.134^{-.0559}_{-.213}$ &
$-.138^{-.0593}_{-.217}$ &
$-.106^{-.0289}_{-.183}$\\
\hline
Percentage of single unit trucks (daily average) &
$1.23^{1.84}_{.624}$ &
$1.23^{1.82}_{.646}$ &
$1.32^{1.96}_{.693}$ &
$1.32^{1.96}_{.691}$ &
$1.29^{1.90}_{.688}$\\
\hline
Number of changes per vertical profile along a roadway segment &
$.0555^{.0930}_{.0180}$ &
$.0562^{.0903}_{.0226}$ &
-- &
-- &
--\\
\hline
Over-dispersion parameter $\alpha$ in NB models &
$.144^{.183}_{.105}$ &
$.150^{.192}_{.114}$ &
$.130^{.168}_{.0925}$ &
$.142^{.185}_{.105}$ &
$.114^{.147}_{.0847}$\\
\hline
\end{tabular}

%% file: frequency_year_ZINB-MSNB_table_b.tex
\begin{tabular}{|l|c|c|c|c|c|}
\hline
${}_{{}_{{}\atop{\displaystyle\mbox{\bf Variable}}}}$
 & \multicolumn{2}{|c|}{\bf\boldmath ZINB-$\tau^{\,\rm a}$} &
   \multicolumn{2}{|c|}{\bf\boldmath ZINB-$\gamma^{\,\rm b}$} & {\bf MSNB}
\\
\cline{2-3}\cline{4-5}
 & {\bf by MLE} & {\bf by MCMC} & {\bf by MLE} & {\bf by MCMC}
 & {\bf by MCMC$^{\,\rm c}$}
\\
\hline
\multicolumn{6}{|c|}{$\tau$- and $\gamma$-parameters
in Equations~(\ref{EQ_Q_TAU}) and~(\ref{EQ_Q_GAMMA})}\\
\hline
The model parameter $\tau$ in Equation~(\ref{EQ_Q_TAU}) &
$-1.72^{-1.45}_{-2.00}$ &
$-1.73^{-1.50}_{-1.98}$ &
-- &
-- &
--\\
\hline
Intercept (constant term) &
-- &
-- &
$23.1^{41.3}_{4.99}$ &
$26.5^{47.0}_{10.9}$ &
--\\
\hline
Logarithm of road segment length (in miles) &
-- &
-- &
$-1.34^{-.942}_{-1.73}$ &
$-1.4^{-1.03 }_{-1.83}$ &
--\\
\hline
Median barrier presence (dummy) &
-- &
-- &
$3.97^{4.86}_{3.08}$ &
$4.16^{5.20}_{3.27}$ &
--\\
\hline
Average annual daily traffic (AADT) &
-- &
-- &
$\displaystyle{9.23^{15.1\vphantom{O^o}}_{3.35}}\atop\displaystyle{{}\times10^{-5}}$ &
$\displaystyle{10.5^{17.4\vphantom{O^o}}_{5.72}}\atop\displaystyle{{}\times10^{-5}}$ &
--\\
\hline
Logarithm of average annual daily traffic &
-- &
-- &
$-2.88^{-.901}_{-4.86}$ &
$-3.28^{-1.59}_{-5.57}$ &
--\\
\hline
\hline
Mean accident rate ($\lambda_{t,n}$ for NB), averaged over all values of ${\bf X}_{t,n}$ &
-- &
$3.38$ &
-- &
$3.42$ &
$3.88$\\
\hline
Standard deviation of accident rate
($\sqrt{\lambda_{t,n}(1+\alpha\lambda_{t,n})}\,$ for NB), & & & & &\\
averaged over all values of explanatory variables ${\bf X}_{t,n}$ &
-- &
$2.14$ &
-- &
$2.15$ &
$2.13$\\
\hline
Total number of free model parameters ($\beta$-s, $\gamma$-s, $\alpha$ and $\tau$) &
$16$ &
$16$ &
$19$ &
$19$ &
$16$\\
\hline
Posterior average of the log-likelihood (LL) &
-- &
$-2510.68^{-2506.13}_{-2517.12}$ &
$--$ &
$-2436.34^{-2431.12}_{-2443.54}$ &
$-2124.82^{-2096.30}_{-2153.91}$\\
\hline
Max$(LL)$:\quad true maximum value of log-likelihood (LL) for MLE; & & & & &\\
maximum observed value of LL for Bayesian-MCMC &
$\displaystyle{-2502.67}\atop{\rm(true)}$ &
$\displaystyle{-2503.21}\atop{\rm(observed)}$ &
$\displaystyle{-2426.54}\atop{\rm(true)}$ &
$\displaystyle{-2427.41}\atop{\rm(observed)}$ &
$\displaystyle{-2049.45}\atop{\rm(observed)}$\\
\hline
Logarithm of marginal likelihood of data ($\ln[f({\bf Y}|{\cal M})]$) &
-- &
$-2519.90^{-2516.95}_{-2521.59}$ &
-- &
$-2447.33^{-2443.93}_{-2448.86}$ &
$-2184.21^{-2186.70}_{-2169.56}$\\
\hline
Goodness-of-fit p-value &
-- &
$0.005$ &
-- &
$0.177$ &
$0.191$\\
\hline
Maximum of the potential scale reduction factors (PSRF)$^{\,\rm e}$ &
-- &
$1.01006$ &
-- &
$1.02200$ &
$1.02117$\\
\hline
Multivariate potential scale reduction factor (MPSRF)$^{\,\rm e}$ &
-- &
$1.01023$ &
-- &
$1.02302$ &
$1.02189$\\
%\hline
%Average candidate acceptance rate in Metropolis-Hasting sampling &
%-- &
%$\!.296\leq {\rm rate}\leq.297\!$ &
%-- &
%$\!.283\leq {\rm rate}\leq.297\!$ &
%$\!.295\leq {\rm rate}\leq.297\!$\\
\hline
\multicolumn{6}{l}{$^{\rm a}$ Standard (conventional) \mbox{ZINB-$\tau$}
model estimated by maximum likelihood estimation (MLE) and Markov Chain
Monte Carlo (MCMC) simulations.}\\
\multicolumn{6}{l}{$^{\rm b}$ Standard \mbox{ZINB-$\gamma$}
model estimated by maximum likelihood estimation (MLE) and Markov Chain
Monte Carlo (MCMC) simulations.}\\
\multicolumn{6}{l}{$^{\rm c}$ Two-state Markov switching negative
binomial (MSNB) model where all reported parameters are for the unsafe state $s=1$.}\\
\multicolumn{6}{l}{$^{\rm d}$ The pavement quality index (PQI) is a composite
measure of overall pavement quality evaluated on a 0 to 100 scale.}\\
\multicolumn{6}{l}{$^{\rm e}$ PSRF/MPSRF are calculated separately/jointly
for all continuous model parameters. PSRF and MPSRF are close to 1 for
converged MCMC chains.}
\end{tabular}

%% file: frequency_statistics_table.tex
\tabcolsep=0.45em
\begin{tabular}{|l|c|c|c|c|c|}
\hline
${\bf Variable}$ &
${\bf Mean}$ &
${\bf Std}^{\,\rm a}$ &
${\bf Min}^{\,\rm a}$ &
${\bf Median}$ &
${\bf Max}^{\,\rm a}$
\\
\hline
Accident occurring on interstates I-70 or I-164 (dummy) &
$.155$ &
$.363$ &
$0$ &
$0$ &
$1.00$
\\
\hline
Pavement quality index (PQI) average$^{\,\rm b}$ &
$88.6$ &
$5.96$ &
$69.0$ &
$90.3$ &
$98.5$
\\
\hline
Road segment length (in miles) &
$.886$ &
$1.48$ &
$.00900$ &
$.356$ &
$11.5$
\\
\hline
Logarithm of road segment length (in miles) &
$-.901$ &
$1.22$ &
$-4.71$ &
$-1.03$ &
$2.44$
\\
\hline
Total number of ramps on the road viewing and opposite sides &
$.725$ &
$1.79$ &
$0$ &
$0$ &
$16$
\\
\hline
Number of ramps on the viewing side per lane per mile &
$.138$ &
$.408$ &
$0$ &
$0$ &
$3.27$
\\
\hline
Median configuration is depressed (dummy) &
$.630$ &
$.484$ &
$0$ &
$1.00$ &
$1.00$
\\
\hline
Median barrier presence (dummy) &
$.161$ &
$.368$ &
$0$ &
$0$ &
$1$
\\
\hline
Interior shoulder presence (dummy) &
$.928$ &
$.258$ &
$0$ &
$1$ &
$1$
\\
\hline
Width of the interior shoulder is less that 5 feet (dummy) &
$.696$ &
$.461$ &
$0$ &
$1.00$ &
$1.00$
\\
\hline
Interior rumble strips presence (dummy) &
$.722$ &
$.448$ &
$0$ &
$1.00$ &
$1.00$
\\
\hline
Width of the outside shoulder is less that 12 feet (dummy) &
$.752$ &
$.432$ &
$0$ &
$1.00$ &
$1.00$
\\
\hline
Outside barrier absence (dummy) &
$.830$ &
$.376$ &
$0$ &
$1.00$ &
$1.00$
\\
\hline
Average annual daily traffic (AADT) &
$\displaystyle{3.03^{\vphantom{O^o}}\atop\displaystyle{{}\times10^4}}$ &
$\displaystyle{2.89^{\vphantom{O^o}}\atop\displaystyle{{}\times10^4}}$ &
$\displaystyle{.944^{\vphantom{O^o}}\atop\displaystyle{{}\times10^4}}$ &
$\displaystyle{1.65^{\vphantom{O^o}}\atop\displaystyle{{}\times10^4}}$ &
$\displaystyle{14.3^{\vphantom{O^o}}\atop\displaystyle{{}\times10^4}}$
\\
\hline
Logarithm of average annual daily traffic  &
$10.0$ &
$.623$ &
$9.15$ &
$9.71$ &
$11.9$
\\
\hline
Posted speed limit (in mph) &
$63.1$ &
$3.89$ &
$50.0$ &
$65.0$ &
$65.0$
\\
\hline
Number of bridges per mile &
$1.76$ &
$8.14$ &
$0$ &
$0$ &
$124$
\\
\hline
Maximum of reciprocal values of horizontal curve radii (in $1/{\rm mile}$) &
$.650$ &
$.632$ &
$0$ &
$.589$ &
$2.26$
\\
\hline
Maximum of reciprocal values of vertical curve radii (in $1/{\rm mile}$) &
$2.38$ &
$3.59$ &
$0$ &
$0$ &
$14.9$
\\
\hline
Number of vertical curves per mile &
$1.50$ &
$4.03$ &
$0$ &
$0$ &
$50.0$
\\
\hline
Percentage of single unit trucks (daily average) &
$.0859$ &
$.0678$ &
$.00975$ &
$.0683$ &
$.322$
\\
\hline
Winter season (dummy) &
$.242$ &
$.428$ &
$0$ &
$0$ &
$1.00$
\\
\hline
Spring season (dummy) &
$.254$ &
$.435$ &
$0$ &
$0$ &
$1.00$
\\
\hline
Summer season (dummy) &
$.254$ &
$.435$ &
$0$ &
$0$ &
$1.00$
\\
%=== Additional variables to full NB in the paper, from P, ZIP, ZINB models ===
\hline
Maximal external angle of the horizontal curve &
$9.78$ &
$12.0$ &
$0$ &
$5.32$ &
$66.7$
\\
\hline
Outside shoulder width (in feet) &
$11.3$ &
$1.74$ &
$6.20$ &
$11.2$ &
$21.8$
\\
\hline
Number of changes per vertical profile along a roadway segment &
$.522$ &
$.908$ &
$0$ &
$0$ &
$6.00$
\\
\hline
Number of lanes on a roadway &
$2.09$ &
$.286$ &
$2.00$ &
$2.00$ &
$3.00$
\\
\hline
Number of ramps on the viewing side &
$.310$ &
$.865$ &
$0$ &
$0$ &
$8.00$
\\
\hline
Maximum absolute value of change in grade of a vertical curve &
$.697$ &
$1.24$ &
$0$ &
$0$ &
$7.41$
\\
\hline
Number of vertical curves per roadway section &
$.445$ &
$.611$ &
$0$ &
$0$ &
$3.00$
\\
\hline
\multicolumn{6}{l}{$^{\rm a}$ Standard deviation, minimum and maximum of a variable.}\\
\multicolumn{6}{l}{$^{\rm b}$ The pavement quality index (PQI) is a
measure of overall pavement quality evaluated on a 0 to 100 scale.}
\end{tabular}

%% file: frequency_week_Poisson_table_a.tex
\tabcolsep=0.3em
\begin{tabular}{|l|c|c|c|c|c|c|}
\hline
${}_{{{}\atop{\displaystyle\mbox{\bf Variable}}}}$ &
${}_{{{}\atop{\displaystyle\mbox{{\bf P-by-MLE}$^{\,\rm a}$}}}}$  &
${}_{{{}\atop{\displaystyle\mbox{{\bf P-by-MCMC}$^{\,\rm b}$}}}}$ &
\multicolumn{2}{|c|}{{\bf Restricted MSP}$^{\,\rm c}$} &
\multicolumn{2}{|c|}{{\bf Full MSP}$^{\,\rm d}$}
\\
\cline{4-7} & & & {\bf state }{\boldmath$s=0$} & {\bf state }{\boldmath$s=1$} &
{\bf state }{\boldmath$s=0$} & {\bf state }{\boldmath$s=1$}
\\
\hline
Intercept (constant term) &
$-21.1^{-19.0}_{-23.3}$ &
$-20.4^{-18.4}_{-22.5}$ &
$-20.4^{-18.4}_{-22.5}$ &
$-19.4^{-17.4}_{-21.6}$ &
$-20.1^{-18.1}_{-22.1}$ &
$-20.1^{-18.1}_{-22.1}$\\
\hline
Accident occurring on interstates I-70 or I-164 (dummy) &
$-.627^{-.639}_{-.715}$ &
$-.629^{-.541}_{-.717}$ &
$-.628^{-.541}_{-.716}$ &
$-.628^{-.541}_{-.716}$ &
$-.587^{-.507}_{-.667}$ &
$-.587^{-.507}_{-.667}$\\
\hline
Pavement quality index (PQI) average$^{\,\rm e}$ &
$-.0132^{-.00681}_{-.0195}$ &
$-.0194^{-.0142}_{-.0245}$ &
$-.0193^{-.0143}_{-.0244}$ &
$-.0193^{-.0143}_{-.0244}$ &
$-.0206^{-.0160}_{-.0252}$ &
--\\
\hline
Road segment length (in miles) &
$.0678^{.0940}_{.0417}$ &
$.0722^{.0980}_{.0466}$ &
$.0721^{.0979}_{.0462}$ &
$.0721^{.0979}_{.0462}$ &
$.0754^{.0996}_{.0511}$ &
$.0754^{.0996}_{.0511}$\\
\hline
Logarithm of road segment length (in miles) &
$.872^{.934}_{.810}$ &
$.862^{.923}_{.800}$ &
$.862^{.923}_{.801}$ &
$.862^{.923}_{.801}$ &
$.865^{.923}_{.807}$ &
$.865^{.923}_{.807}$\\
\hline
Total number of ramps on the road viewing and opposite sides &
$-.0203^{-.00766}_{-.0329}$ &
$-.0246^{-.0123}_{-.0369}$ &
$-.0246^{-.0123}_{-.0369}$ &
$-.0246^{-.0123}_{-.0369}$ &
$-.0150^{-.00109}_{-.0288}$ &
$-.0345^{-.0186}_{-.0509}$\\
\hline
Number of ramps on the viewing side per lane per mile &
$.395^{.471}_{.320}$ &
$.402^{.477}_{.326}$ &
$.402^{.477}_{.327}$ &
$.402^{.477}_{.327}$ &
$.415^{.489}_{.340}$ &
$.415^{.489}_{.340}$\\
\hline
Median configuration is depressed (dummy) &
$.187^{.288}_{.0864}$ &
$.192^{.294}_{.0923}$ &
$.193^{.293}_{.0927}$ &
$.193^{.293}_{.0927}$ &
-- &
$.349^{.522}_{.180}$\\
\hline
Median barrier presence (dummy) &
$-3.05^{-2.42}_{-3.67}$ &
$-2.99^{-2.40}_{-3.66}$ &
$-3.00^{-2.41}_{-3.67}$ &
$-3.00^{-2.41}_{-3.67}$ &
$-3.11^{-2.52}_{-3.78}$ &
$-3.11^{-2.52}_{-3.78}$\\
\hline
Interior shoulder presence (dummy) &
$-1.11^{-.445}_{-1.77}$ &
$-.980^{.326}_{-2.27}$ &
$-.982^{.320}_{-2.32}$ &
$-.982^{.320}_{-2.32}$ &
$-1.12^{.476}_{-1.82}$ &
$-1.12^{.476}_{-1.82}$\\
\hline
Width of the interior shoulder is less that 5 feet (dummy) &
$.371^{.471}_{.271}$ &
$.387^{.487}_{.288}$ &
$.387^{.487}_{.289}$ &
$.387^{.487}_{.289}$ &
$.374^{.473}_{.277}$ &
$.374^{.473}_{.277}$\\
\hline
Interior rumble strips presence (dummy) &
$-.187^{-.0734}_{-.300}$ &
$-.172^{.970}_{-1.30}$ &
$-.172^{.967}_{-1.32}$ &
$-.172^{.967}_{-1.32}$ &
-- &
--\\
\hline
Width of the outside shoulder is less that 12 feet (dummy) &
$.282^{.376}_{.189}$ &
$.272^{.366}_{.179}$ &
$.273^{.367}_{.180}$ &
$.273^{.367}_{.180}$ &
$.276^{.369}_{.185}$ &
$.276^{.369}_{.185}$\\
\hline
Outside barrier absence (dummy) &
$-.246^{-.139}_{-.354}$ &
$-.254^{-.146}_{-.360}$ &
$-.254^{-.147}_{-.360}$ &
$-.254^{-.147}_{-.360}$ &
$-.280^{-.174}_{-.384}$ &
$-.280^{-.174}_{-.384}$\\
\hline
Average annual daily traffic (AADT) &
$\displaystyle{-3.99^{-3.16\vphantom{O^o}}_{-4.83}}\atop\displaystyle{{}\times10^{-5}}$ &
$\displaystyle{-3.97^{-3.15\vphantom{O^o}}_{-4.84}}\atop\displaystyle{{}\times10^{-5}}$ &
$\displaystyle{-3.95^{-3.13\vphantom{O^o}}_{-4.82}}\atop\displaystyle{{}\times10^{-5}}$ &
$\displaystyle{-3.95^{-3.13\vphantom{O^o}}_{-4.82}}\atop\displaystyle{{}\times10^{-5}}$ &
$\displaystyle{-3.64^{-2.87\vphantom{O^o}}_{-4.45}}\atop\displaystyle{{}\times10^{-5}}$ &
$\displaystyle{-3.64^{-2.87\vphantom{O^o}}_{-4.45}}\atop\displaystyle{{}\times10^{-5}}$\\
\hline
Logarithm of average annual daily traffic &
$2.06^{2.29}_{1.83}$ &
$2.03^{2.27}_{1.80}$ &
$2.02^{2.26}_{1.80}$ &
$2.02^{2.26}_{1.80}$ &
$1.94^{2.16}_{1.73}$ &
$1.94^{2.16}_{1.73}$\\
\hline
Posted speed limit (in mph) &
$.0151^{.0234}_{.00672}$ &
$.0149^{.0232}_{.00662}$ &
$.0149^{.0232}_{.00658}$ &
$.0149^{.0232}_{.00658}$ &
$.0252^{.0315}_{.0189}$ &
--\\
\hline
Number of bridges per mile &
$-.0212^{-.00413}_{-.0382}$ &
$-.0242^{-.00787}_{-.0415}$ &
$-.0243^{-.00792}_{-.0415}$ &
$-.0243^{-.00792}_{-.0415}$ &
$-.0254^{-.00907}_{-.0427}$ &
$-.0254^{-.00907}_{-.0427}$\\
\hline
Maximal external angle of the horizontal curve &
$.003363^{.00669}_{.000576}$ &
$.00395^{.00696}_{.000919}$ &
$.00395^{.00696}_{.000917}$ &
$.00395^{.00696}_{.000917}$ &
$.00602^{.00922}_{.00277}$ &
--\\
\hline
Maximum of reciprocal values of horizontal curve radii (in $1/{\rm mile}$) &
$-.247^{-.169}_{-.325}$ &
$-.249^{.172}_{-.327}$ &
$-.249^{.172}_{-.327}$ &
$-.249^{.172}_{-.327}$ &
$-.274^{-.208}_{-.341}$ &
$-.274^{-.208}_{-.341}$\\
\hline
Maximum of reciprocal values of vertical curve radii (in $1/{\rm mile}$) &
$.0196^{.0281}_{.0112}$ &
$.0176^{.0259}_{.00930}$ &
$.0176^{.0259}_{.00930}$ &
$.0176^{.0259}_{.00930}$ &
$.0182^{.0265}_{.00998}$ &
$.0182^{.0265}_{.00998}$\\
\hline
Number of vertical curves per mile &
$-.0588^{-.0248}_{-.0929}$ &
$-.0622^{-.0292}_{-.0968}$ &
$-.0623^{-.0292}_{-.0969}$ &
$-.0623^{-.0292}_{-.0969}$ &
$-.0644^{-.0315}_{-.0989}$ &
$-.0644^{-.0315}_{-.0989}$\\
\hline
Percentage of single unit trucks (daily average) &
$1.29^{1.76}_{.814}$ &
$1.14^{1.60}_{.684}$ &
$1.14^{1.60}_{.681}$ &
$1.14^{1.60}_{.681}$ &
-- &
$1.83^{2.47}_{1.19}$\\
\hline
\end{tabular}

%% file: frequency_week_Poisson_table_b.tex
\tabcolsep=0.45em
\begin{tabular}{|l|c|c|c|c|c|c|}
\hline
${}_{{{}\atop{\displaystyle\mbox{\bf Variable}}}}$ &
${}_{{{}\atop{\displaystyle\mbox{{\bf P-by-MLE}$^{\,\rm a}$}}}}$  &
${}_{{{}\atop{\displaystyle\mbox{{\bf P-by-MCMC}$^{\,\rm b}$}}}}$ &
\multicolumn{2}{|c|}{{\bf Restricted MSP}$^{\,\rm c}$} &
\multicolumn{2}{|c|}{{\bf Full MSP}$^{\,\rm d}$}
\\
\cline{4-7} & & & {\bf state }{\boldmath$s=0$} & {\bf state }{\boldmath$s=1$} &
{\bf state }{\boldmath$s=0$} & {\bf state }{\boldmath$s=1$}
\\
\hline
Winter season (dummy) &
$.185^{.254}_{.115}$ &
$.185^{.254}_{.116}$ &
$-.0627^{.181}_{-.173}$ &
$-.0627^{.181}_{-.173}$ &
-- &
$-.364^{.487}_{-.232}$\\
\hline
Spring season (dummy) &
$-.156^{.0817}_{-.231}$ &
$-.156^{.0821}_{-.231}$ &
$-.131^{.0689}_{-.230}$ &
$-.131^{.0689}_{-.230}$ &
-- &
--\\
\hline
Summer season (dummy) &
$-.168^{.0932}_{-.243}$ &
$-.168^{.0936}_{-.243}$ &
$-.0571^{.134}_{-.149}$ &
$-.0571^{.134}_{-.149}$ &
-- &
$-.345^{.147}_{-.568}$\\
\hline
\hline
Mean accident rate ($\lambda_{t,n}$), averaged over all values of ${\bf X}_{t,n}$ &
-- &
$.0661$ &
$.0570$ &
$.1540$ &
$.0533$ &
$.1100$\\
\hline
Standard deviation of accident rate ($\lambda_{t,n}$), averaged over all & & & & & &\\
values of explanatory variables ${\bf X}_{t,n}$ &
-- &
$.1900$ &
$.1770$ &
$.2900$ &
$.1730$ &
$.2390$\\
\hline
Markov transition probability of jump $0\to1$ ($p_{0\rightarrow1}$) &
-- & -- &
\multicolumn{2}{|c|}{$.0705^{.113}_{.0389}$} &
\multicolumn{2}{|c|}{$.163^{.239}_{.0989}$}\\
\hline
Markov transition probability of jump $1\to0$ ($p_{1\rightarrow0}$) &
-- & -- &
\multicolumn{2}{|c|}{$.662^{.840}_{.439}$} &
\multicolumn{2}{|c|}{$.632^{.779}_{.476}$}\\
\hline
Unconditional probabilities of states 0 and 1 (${\bar p}_0$ and ${\bar p}_1$) &
-- & -- &
\multicolumn{2}{|c|}{$.902^{.947}_{.829}$\quad 
and\quad $.0981^{.171}_{.0528}$} &
\multicolumn{2}{|c|}{$.794^{.871}_{.708}$\quad 
and\quad $.206^{.292}_{.129}$}\\
\hline
Total number of free model parameters ($\beta$-s and $\alpha$-s) &
$26$ &
$26$ &
\multicolumn{2}{|c|}{$27$} &
\multicolumn{2}{|c|}{$25$}\\
\hline
Posterior average of the log-likelihood (LL) &
-- &
$-16381.08^{-16367.39}_{-16381.08}$ &
\multicolumn{2}{|c|}{$-16035.97^{-16023.36}_{-16047.89}$} &
\multicolumn{2}{|c|}{$-15964.02^{-15947.44}_{-15983.66}$}\\
%\hline
%max$(LL)$ & $\displaystyle{-16081.2}\atop{\rm(true)}$ &
%$\displaystyle{-16086.3}\atop{\rm(observed)}$ &
%\multicolumn{2}{|c|}{$-15786.6\,$(observed)} &
%\multicolumn{2}{|c|}{$-15744.8\,$(observed)}\\
\hline
Max$(LL)$:\quad true maximum value of log-likelihood (LL) for MLE; & & &
\multicolumn{2}{|c|}{} & \multicolumn{2}{|c|}{} \\
maximum observed value of LL for Bayesian-MCMC &
$-16355.68\,{\rm(true)}\!$ &
$-16362.30\,{\rm(observ.)}\!$ &
\multicolumn{2}{|c|}{$-15990.70\,$(observed)} &
\multicolumn{2}{|c|}{$-15928.03\,$(observed)}\\
\hline
Logarithm of marginal likelihood of data ($\ln[f({\bf Y}|{\cal M})]$) &
-- &
$-16384.97^{-16381.71}_{-16386.24}$ &
\multicolumn{2}{|c|}{$-16056.91^{-16050.68}_{-16059.76}$} &
\multicolumn{2}{|c|}{$-16001.15^{-15992.86}_{-16003.65}$}\\
\hline
Goodness-of-fit p-value &
-- & $0.296$ &
\multicolumn{2}{|c|}{$0.404$} &
\multicolumn{2}{|c|}{$0.393$}\\
\hline
Maximum of the potential scale reduction factors (PSRF)$^{\,\rm f}$ &
-- &
$1.02205$ &
\multicolumn{2}{|c|}{$1.00711$} &
\multicolumn{2}{|c|}{$1.00759$}\\
\hline
Multivariate potential scale reduction factor (MPSRF)$^{\,\rm f}$ &
-- &
$1.02361$ &
\multicolumn{2}{|c|}{$1.00776$} &
\multicolumn{2}{|c|}{$1.00792$}\\
%\hline
%Average candidate acceptance rate in Metropolis-Hasting sampling &
%-- &
%$\!.295\leq {\rm rate}\leq.297\!$ &
%\multicolumn{2}{|c|}{$.296\leq {\rm rate}\leq.297$} &
%\multicolumn{2}{|c|}{$.295\leq {\rm rate}\leq.297$}\\
%\hline
%acc./rd.sect. & \multicolumn{6}{|c|}{$5769/335$}\\
\hline
\multicolumn{7}{l}{$^{\rm a}$ Standard (conventional) Poisson
estimated by maximum likelihood estimation (MLE).}\\
\multicolumn{7}{l}{$^{\rm b}$ Standard Poisson
estimated by Markov Chain Monte Carlo (MCMC) simulations.}\\
\multicolumn{7}{l}{$^{\rm c}$ Restricted two-state Markov switching Poisson
(MSP) model with only the intercept and over-dispersion parameters
allowed to vary between states.}\\
\multicolumn{7}{l}{$^{\rm d}$ Full two-state Markov switching Poisson
(MSP) model with all parameters allowed to vary between states.}\\
\multicolumn{7}{l}{$^{\rm e}$ The pavement quality index (PQI) is a composite
measure of overall pavement quality evaluated on a 0 to 100 scale.}\\
\multicolumn{7}{l}{$^{\rm f}$ PSRF/MPSRF are calculated separately/jointly
for all continuous model parameters. PSRF and MPSRF are close to 1 for
converged MCMC chains.}
\end{tabular}

%% file: frequency_week_NB_table_a.tex
\tabcolsep=0.3em
\begin{tabular}{|l|c|c|c|c|c|c|}
\hline
${}_{{{}\atop{\displaystyle\mbox{\bf Variable}}}}$ &
${}_{{{}\atop{\displaystyle\mbox{{\bf NB-by-MLE}$^{\,\rm a}$}}}}$  &
${}_{{{}\atop{\displaystyle\mbox{{\bf NB-by-MCMC}$^{\,\rm b}$}}}}$ &
\multicolumn{2}{|c|}{{\bf Restricted MSNB}$^{\,\rm c}$} &
\multicolumn{2}{|c|}{{\bf Full MSNB}$^{\,\rm d}$}
\\
\cline{4-7} & & & {\bf state }{\boldmath$s=0$} & {\bf state }{\boldmath$s=1$} &
{\bf state }{\boldmath$s=0$} & {\bf state }{\boldmath$s=1$}
\\
\hline
Intercept (constant term) &
$-21.3^{-18.7}_{-23.9}$ &
$-20.6^{-18.5}_{-22.7}$ &
$-20.9^{-18.7}_{-23.0}$ &
$-19.9^{-17.8}_{-22.1}$ &
$-20.7^{-18.7}_{-22.8}$ &
$-20.7^{-18.7}_{-22.8}$\\
\hline
Accident occurring on interstates I-70 or I-164 (dummy) &
$-.655^{-.562}_{-.748}$ &
$-.657^{-.565}_{-.750}$ &
$-.656^{-.564}_{-.748}$ &
$-.656^{-.564}_{-.748}$ &
$-.660^{-.568}_{-.752}$ &
$-.660^{-.568}_{-.752}$\\
\hline
Pavement quality index (PQI) average$^{\,\rm e}$ &
$-.0132^{-.00581}_{-.0205}$ &
$-.0189^{-.0134}_{-.0244}$ &
$-.0195^{-.0141}_{-.0248}$ &
$-.0195^{-.0141}_{-.0248}$ &
$-.0220^{-.0166}_{-.0273}$ &
$-.0125^{-.00700}_{-.0180}$\\
\hline
Road segment length (in miles) &
$.0512^{.0809}_{.0215}$ &
$.0546^{.0826}_{.0266}$ &
$.0538^{.0812}_{.0264}$ &
$.0538^{.0812}_{.0264}$ &
$.0395^{.0625}_{.0165}$ &
$.0395^{.0625}_{.0165}$\\
\hline
Logarithm of road segment length (in miles) &
$.909^{.974}_{.845}$ &
$.903^{.964}_{.842}$ &
$.900^{.961}_{.840}$ &
$.900^{.961}_{.840}$ &
$.913^{.973}_{.853}$ &
$.913^{.973}_{.853}$\\
\hline
Total number of ramps on the road viewing and opposite sides &
$-.0172^{-.00174}_{-.0327}$ &
$-.021^{-.00624}_{-.0358}$ &
$-.0187^{-.00423}_{-.0331}$ &
$-.0187^{-.00423}_{-.0331}$ &
-- &
$-.0264^{-.00656}_{-.0464}$\\
\hline
Number of ramps on the viewing side per lane per mile &
$.394^{.479}_{.309}$ &
$.400^{.479}_{.319}$ &
$.397^{.475}_{.317}$ &
$.397^{.475}_{.317}$ &
$.359^{.429}_{.289}$ &
$.359^{.429}_{.289}$\\
\hline
Median configuration is depressed (dummy) &
$.210^{.314}_{.106}$ &
$.214^{.318}_{.111}$ &
$.211^{.315}_{.108}$ &
$.211^{.315}_{.108}$ &
$.209^{.313}_{.107}$ &
$.209^{.313}_{.107}$\\
\hline
Median barrier presence (dummy) &
$-3.02^{-2.38}_{-3.67}$ &
$-2.99^{-2.40}_{-3.67}$ &
$-3.01^{-2.42}_{-3.69}$ &
$-3.01^{-2.42}_{-3.69}$ &
$-3.01^{-2.42}_{-3.69}$ &
$-3.01^{-2.42}_{-3.69}$\\
\hline
Interior shoulder presence (dummy) &
$-1.15^{-.486}_{-1.81}$ &
$-1.06^{.135}_{-2.26}$ &
$-1.02^{.148}_{-2.23}$ &
$-1.02^{.148}_{-2.23}$ &
$-1.16^{-.523}_{-1.87}$ &
$-1.16^{-.523}_{-1.87}$\\
\hline
Width of the interior shoulder is less that 5 feet (dummy) &
$.373^{.477}_{.270}$ &
$.384^{.491}_{.279}$ &
$.386^{.492}_{.281}$ &
$.386^{.492}_{.281}$ &
$.380^{.486}_{.275}$ &
$.380^{.486}_{.275}$\\
\hline
Interior rumble strips presence (dummy) &
$-.166^{-.0382}_{-.293}$ &
$-.142^{.857}_{-1.16}$ &
$-.163^{.836}_{-1.14}$ &
$-.163^{.836}_{-1.14}$ &
-- &
--\\
\hline
Width of the outside shoulder is less that 12 feet (dummy) &
$.281^{.380}_{.182}$ &
$.272^{.370}_{.174}$ &
$.268^{.366}_{.170}$ &
$.268^{.366}_{.170}$ &
$.267^{.365}_{.170}$ &
$.267^{.365}_{.170}$\\
\hline
Outside barrier absence (dummy) &
$-.249^{-.139}_{-.358}$ &
$-.255^{-.142}_{-.366}$ &
$-.255^{-.142}_{-.366}$ &
$-.255^{-.142}_{-.366}$ &
$-.251^{-.140}_{-.362}$ &
$-.251^{-.140}_{-.362}$\\
\hline
Average annual daily traffic (AADT) &
$\displaystyle{-4.09^{-3.04\vphantom{O^o}}_{-5.15}}\atop\displaystyle{{}\times10^{-5}}$ &
$\displaystyle{-4.09^{-3.24\vphantom{O^o}}_{-4.95}}\atop\displaystyle{{}\times10^{-5}}$ &
$\displaystyle{-4.07^{-3.22\vphantom{O^o}}_{-4.94}}\atop\displaystyle{{}\times10^{-5}}$ &
$\displaystyle{-4.07^{-3.22\vphantom{O^o}}_{-4.94}}\atop\displaystyle{{}\times10^{-5}}$ &
$\displaystyle{-3.90^{-3.11\vphantom{O^o}}_{-4.72}}\atop\displaystyle{{}\times10^{-5}}$ &
$\displaystyle{-4.53^{-3.61\vphantom{O^o}}_{-5.48}}\atop\displaystyle{{}\times10^{-5}}$\\
\hline
Logarithm of average annual daily traffic  &
$2.08^{2.36}_{1.80}$ &
$2.06^{2.30}_{1.83}$ &
$2.07^{2.30}_{1.83}$ &
$2.07^{2.30}_{1.83}$ &
$2.07^{2.30}_{1.84}$ &
$2.07^{2.30}_{1.84}$\\
\hline
Posted speed limit (in mph) &
$.0154^{.0244}_{.00643}$ &
$.0150^{.0241}_{.00589}$ &
$.0161^{.0251}_{.00697}$ &
$.0161^{.0251}_{.00697}$ &
$.0161^{.0252}_{.00712}$ &
$.0161^{.0252}_{.00712}$\\
\hline
Number of bridges per mile &
$-.0213^{-.00187}_{-.0407}$ &
$-.0241^{-.00721}_{-.0419}$ &
$-.0233^{-.00648}_{-.0410}$ &
$-.0233^{-.00648}_{-.0410}$ &
-- &
$-.0607^{-.0232}_{-.102}$\\
\hline
Maximum of reciprocal values of horizontal curve radii (in $1/{\rm mile}$) &
$-.182^{-.122}_{-.242}$ &
$-.179^{-.118}_{-.241}$ &
$-.178^{-.117}_{-.239}$ &
$-.178^{-.117}_{-.239}$ &
$-.175^{-.114}_{-.237}$ &
$-.175^{-.114}_{-.237}$\\
\hline
Maximum of reciprocal values of vertical curve radii (in $1/{\rm mile}$) &
$.0191^{.0285}_{.00972}$ &
$.0177^{.027}_{.00843}$ &
$.0183^{.0275}_{.00917}$ &
$.0183^{.0275}_{.00917}$ &
$.0184^{.0274}_{.00925}$ &
$.0184^{.0274}_{.00925}$\\
\hline
Number of vertical curves per mile &
$-.0535^{-.0180}_{-.0889}$ &
$-.057^{-.0233}_{-.0924}$ &
$-.0586^{-.0249}_{-.0940}$ &
$-.0586^{-.0249}_{-.0940}$ &
$-.0565^{-.0231}_{-.0917}$ &
$-.0565^{-.0231}_{-.0917}$\\
\hline
Percentage of single unit trucks (daily average) &
$1.38^{1.88}_{.886}$ &
$1.25^{1.75}_{.758}$ &
$1.19^{1.68}_{.701}$ &
$1.19^{1.68}_{.701}$ &
$.726^{1.28}_{.171}$ &
$2.57^{3.39}_{1.77}$\\
\hline
\end{tabular}

%% file: frequency_week_NB_table_b.tex
\tabcolsep=0.45em
\begin{tabular}{|l|c|c|c|c|c|c|}
\hline
${}_{{{}\atop{\displaystyle\mbox{\bf Variable}}}}$ &
${}_{{{}\atop{\displaystyle\mbox{{\bf NB-by-MLE}$^{\,\rm a}$}}}}$  &
${}_{{{}\atop{\displaystyle\mbox{{\bf NB-by-MCMC}$^{\,\rm b}$}}}}$ &
\multicolumn{2}{|c|}{{\bf Restricted MSNB}$^{\,\rm c}$} &
\multicolumn{2}{|c|}{{\bf Full MSNB}$^{\,\rm d}$}
\\
\cline{4-7} & & & {\bf state }{\boldmath$s=0$} & {\bf state }{\boldmath$s=1$} &
{\bf state }{\boldmath$s=0$} & {\bf state }{\boldmath$s=1$}
\\
\hline
Winter season (dummy) &
$.148^{.226}_{.0698}$ &
$.148^{.226}_{.0689}$ &
$-.116^{.0563}_{-.261}$ &
$-.116^{.0563}_{-.261}$ &
$-.159^{-.0494}_{-.269}$ &
--\\
\hline
Spring season (dummy) &
$-.173^{-.0878}_{-.258}$ &
$-.173^{-.0899}_{-.257}$ &
$-.0932^{.0547}_{-.209}$ &
$-.0932^{.0547}_{-.209}$ &
-- &
--\\
\hline
Summer season (dummy) &
$-.179^{-.0921}_{-.266}$ &
$-.180^{-.0963}_{-.263}$ &
$-.0332^{.111}_{-.146}$ &
$-.0332^{.111}_{-.146}$ &
-- &
$-.549^{-.293}_{-.883}$\\
\hline
Over-dispersion parameter $\alpha$ in NB models &
$.957^{1.07}_{.845}$ &
$.968^{1.09}_{.849}$ &
$.537^{.677}_{.392}$ &
$1.24^{1.51}_{.986}$ &
$.443^{.595}_{.300}$ &
$1.16^{1.39}_{.945}$\\
\hline
\hline
Mean accident rate ($\lambda_{t,n}$ for NB), averaged over all values of ${\bf X}_{t,n}$ &
-- &
$.0663$ &
$.0558$ &
$.1440$ &
$.0533$ &
$.1130$\\
\hline
Standard deviation of accident rate
($\sqrt{\lambda_{t,n}(1+\alpha\lambda_{t,n})}\,$ for NB), & & & & & &\\
averaged over all values of explanatory variables ${\bf X}_{t,n}$ & -- &
$.2050$ &
$.1810$ &
$.3350$ &
$.1760$ &
$.2820$\\
\hline
Markov transition probability of jump $0\to1$ ($p_{0\rightarrow1}$) &
-- & -- &
\multicolumn{2}{|c|}{$.0933^{.147}_{.0531}$} &
\multicolumn{2}{|c|}{$.158^{.225}_{.100}$}\\
\hline
Markov transition probability of jump $1\to0$ ($p_{1\rightarrow0}$) &
-- & -- &
\multicolumn{2}{|c|}{$.651^{.820}_{.463}$} &
\multicolumn{2}{|c|}{$.627^{.773}_{.474}$}\\
\hline
Unconditional probabilities of states 0 and 1 (${\bar p}_0$ and ${\bar p}_1$) &
-- & -- &
\multicolumn{2}{|c|}{$.873^{.929}_{.797}$\quad and\quad $.127^{.203}_{.0713}$} &
\multicolumn{2}{|c|}{$.798^{.868}_{.718}$\quad and\quad $.202^{.282}_{.132}$}\\
\hline
Total number of free model parameters ($\beta$-s and $\alpha$-s) &
$26$ &
$26$ &
\multicolumn{2}{|c|}{$28$} &
\multicolumn{2}{|c|}{$28$}\\
\hline
Posterior average of the log-likelihood (LL) & -- &
$-16097.2^{-16091.3}_{-16105.0}$ &
\multicolumn{2}{|c|}{$-15821.8^{-15807.9}_{-15835.2}$} &
\multicolumn{2}{|c|}{$-15778.0^{-15672.9}_{-15794.9}$}\\
\hline
Max$(LL)$:\quad true maximum value of log-likelihood (LL) for MLE; & & &
\multicolumn{2}{|c|}{} & \multicolumn{2}{|c|}{} \\
maximum observed value of LL for Bayesian-MCMC &
$-16081.2\,{\rm(true)}\!$ &
$-16086.3\,{\rm(observ.)}\!$ &
\multicolumn{2}{|c|}{$-15786.6\,$(observed)} &
\multicolumn{2}{|c|}{$-15744.8\,$(observed)}\\
\hline
Logarithm of marginal likelihood of data ($\ln[f({\bf Y}|{\cal M})]$) &
-- &
$-16108.6^{-16105.7}_{-16110.7}$ &
\multicolumn{2}{|c|}{$-15850.2^{-15840.1}_{-15849.5}$} &
\multicolumn{2}{|c|}{$-15809.4^{-15801.7}_{-15811.9}$}\\
\hline
Goodness-of-fit p-value &
-- & $0.701$ &
\multicolumn{2}{|c|}{$0.729$} &
\multicolumn{2}{|c|}{$0.647$}\\
\hline
Maximum of the potential scale reduction factors (PSRF)$^{\,\rm f}$ & -- &
$1.00874$ &
\multicolumn{2}{|c|}{$1.00754$} &
\multicolumn{2}{|c|}{$1.00939$}\\
\hline
Multivariate potential scale reduction factor (MPSRF)$^{\,\rm f}$ &
-- &
$1.00928$ &
\multicolumn{2}{|c|}{$1.00925$} &
\multicolumn{2}{|c|}{$1.01002$}\\
%\hline
%Average candidate acceptance rate in Metropolis-Hasting sampling &
%-- &
%$\!.295\leq {\rm rate}\leq.297\!$ &
%\multicolumn{2}{|c|}{$.296\leq {\rm rate}\leq.297$} &
%\multicolumn{2}{|c|}{$.295\leq {\rm rate}\leq.297$}\\
\hline
\multicolumn{7}{l}{$^{\rm a}$ Standard (conventional) negative binomial
estimated by maximum likelihood estimation (MLE).}\\
\multicolumn{7}{l}{$^{\rm b}$ Standard negative binomial
estimated by Markov Chain Monte Carlo (MCMC) simulations.}\\
\multicolumn{7}{l}{$^{\rm c}$ Restricted two-state Markov switching negative
binomial (MSNB) model with only the intercept and over-dispersion parameters
allowed to vary between states.}\\
\multicolumn{7}{l}{$^{\rm d}$ Full two-state Markov switching negative
binomial (MSNB) model with all parameters allowed to vary between states.}\\
\multicolumn{7}{l}{$^{\rm e}$ The pavement quality index (PQI) is a composite
measure of overall pavement quality evaluated on a 0 to 100 scale.}\\
\multicolumn{7}{l}{$^{\rm f}$ PSRF/MPSRF are calculated separately/jointly
for all continuous model parameters. PSRF and MPSRF are close to 1 for
converged MCMC chains.}
\end{tabular}

%% file: frequency_week_corr_table.tex
\begin{tabular}{|l|c|c|c|}
\hline
%& {\bf All year} & {\bf Winter} (Nov.--Mar.) & {\bf Summer} (May--Sept.)\\
& {\bf All year} & {\bf Winter} & {\bf Summer}\\
& & (Nov.--Mar.) & (May--Sept.)\\
\hline
Precipitation (inch) & $0.031$ & -- & $0.144$\\
\hline
Temperature ($^oF$) & $-0.518$ & $-0.591$ & $0.201$\\
\hline
Snowfall (inch) & $0.602$ & $0.577$ & --\\
\cline{2-4}
$\quad\quad\quad>0.2$ (dummy) & $0.651$ & $0.638$ & --\\
\hline
Fog / Frost (dummy) & $0.223$ & (frost) $0.539$ & (fog) $0.051$\\
\hline
Visibility distance (mile) & $-0.221$ & $-0.232$ & $-0.126$\\
\hline
\end{tabular}

%% file: ch_results_sev.tex
%
% This is the severity model estimation results chapter
%

\chapter{Severity model estimation results}
\label{RESULTS_SEV}
In this chapter we present model estimation results for accident severities. We estimate a standard multinomial logit (ML) model and a Markov switching multinomial logit (MSML) model. We compare the performance of these models in fitting the accident severity data.

The severity outcome of an accident is determined by the injury level sustained by the most injured individual (if any) involved into the accident. In this study we consider three accident severity outcomes: ``fatality'', ``injury'' and ``PDO (property damage only)'', which we number as $i=1,2,3$ respectively ($I=3$). We use data from 811720 accidents that were observed in Indiana in 2003-2006, and we use weekly time periods, $t=1,2,3,\ldots,T=208$ in total.\footnote{A week is from Sunday to Saturday, there are 208 full weeks in the 2003-2006 time interval.} The state $s_t$ can change every week. To increase the predictive power of our models, we consider accidents separately for each combination of accident type (1-vehicle and 2-vehicle) and roadway class (interstate highways, US routes, state routes, county roads, streets). We do not consider accidents with more than two vehicles involved.\footnote{Among 811720 accidents 241011 (29.7\%) are 1-vehicle, 525035 (64.7\%) are 2-vehicle, and only 45674 (5.6\%) are accidents with more than two vehicles involved.} Thus, in total, there are ten roadway-class-accident-type combinations that we consider. For each roadway-class-accident-type combination the following two types of accident frequency models are estimated:
\begin{itemize}
    \item
    First, we estimate a standard single-state multinomial logit (ML) model, which is specified by equations~(\ref{EQ_MLE_L_1}) and~(\ref{EQ_MLE_ML}). We estimate this model, first, by the maximum likelihood estimation (MLE), and, second, by the Bayesian inference approach and MCMC simulations [for details on MLE modeling of accident severities see \citep{M_06}; see also footnote~\ref{MLE_LIMDEP} on page~\pageref{MLE_LIMDEP}]. We refer to this model as ``ML-by-MLE'' if estimated by MLE, and as ``ML-by-MCMC'' if estimated by MCMC. As one expects, for our choice of a non-informative prior distribution, the estimated ML-by-MCMC model turned out to be very similar to the corresponding ML-by-MLE model (estimated for the same roadway-class-accident-type combination).
    \item
    Second, we estimate a two-state Markov switching multinomial logit (MSML) model, specified by equation~(\ref{EQ_MS_ML}), by the Bayesian-MCMC methods. To choose the explanatory variables for the MSML model, we start with using the variables that enter the standard ML model (see footnote~\ref{VARIABLES_CHOICE} on page~\pageref{VARIABLES_CHOICE}). Then, we consecutively construct and use $60\%$, $85\%$ and $95\%$ Bayesian credible intervals for evaluation of the statistical significance of each $\beta$-parameter. As a result, in the final model some components of $\betabf_{(0)}$ and $\betabf_{(1)}$ are restricted to zero or restricted to be the same in the two states (see footnote~\ref{BETA_RESTRICT} on page~\pageref{BETA_RESTRICT}). We refer to this model as ``MSML''.
\end{itemize}
Note that the two states, and thus the MSML models, do not have to exist for every roadway-class-accident-type combination. For example, they will not exist if all estimated model parameters turn out to be statistically the same in the two states, $\betabf_{(0)}=\betabf_{(1)}$ (which suggests the two states are identical and the MSML models reduce to the corresponding standard ML models). Also, the two states will not exist if all estimated state variables $s_t$ turn out to be close to zero, resulting in $p_{0\rightarrow1}\ll p_{1\rightarrow0}$, compare to equation~(\ref{EQ_MS_P_RESTRICT_1}), then the less frequent state $s_t=1$ is not realized and the process stays in state $s_t=0$.

Turning to the estimation results, our findings show that two states of roadway safety and the appropriate MSML models exist for severity outcomes of 1-vehicle accidents occurring on all roadway classes (interstate highways, US routes, state routes, county roads, streets), and for severity outcomes of 2-vehicle accidents occurring on streets. The model estimation results for these roadway-class-accident-type combinations, where Markov switching across two states exists, are given in Tables~\ref{T_interst_one-veh}--\ref{T_street_two-veh}. We do not find existence of two states of roadway safety in the cases of 2-vehicle accidents on interstate highways, US routes, state routes and county roads (in these cases all estimated state variables $s_t$ were found to be close to zero, and, therefore, MSML models reduced to standard non-switching ML models). The standard ML models estimated for these roadway-class-accident-type combinations are given in Tables~\ref{T_interst_two-veh}--\ref{T_county_two-veh} in the Appendix. In Tables~\ref{T_interst_one-veh}--\ref{T_street_two-veh} and Tables~\ref{T_interst_two-veh}--\ref{T_county_two-veh} posterior (or MLE) estimates of all continuous model parameters ($\beta$-s, $p_{0\rightarrow1}$ and $p_{1\rightarrow0}$) are given together with their $95\%$ confidence intervals (if MLE) or $95\%$ credible intervals (if Bayesian-MCMC), refer to the superscript and subscript numbers adjacent to parameter posterior/MLE estimates, and also see footnote~\ref{MLE_NORMALITY} on page~\pageref{MLE_NORMALITY}. Table~\ref{T_sev_expl-stat} gives description and summary statistics of all accident characteristic variables ${\bf X}_{t,n}$ except the intercept.

Because we are mostly interested in MSML models, below let us focus on and discuss only model estimation results for roadway-class-accident-type combinations that exhibit existence of two states of roadway safety. These roadway-class-accident-type combinations (six combinations in total) include cases of 1-vehicle accidents occurring on interstate highways, US routes, state routes, county roads, streets, and 2-vehicle accidents occurring on streets, see Tables~\ref{T_interst_one-veh}--\ref{T_street_two-veh}.

\begin{figure}[t]
\vspace{14.1truecm}
\includegraphics{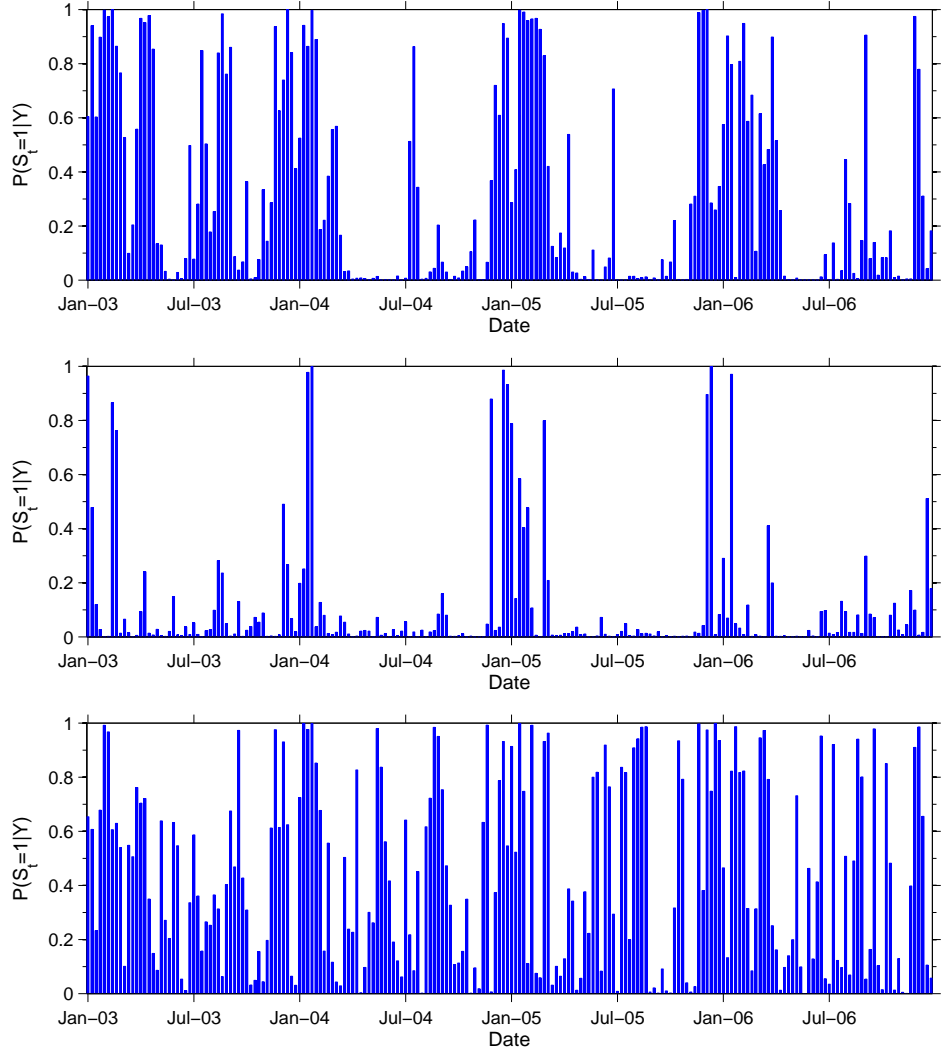}
\caption{Weekly posterior probabilities $P(s_t=1|{\bf Y})$
for the MSML models estimated for severity of 1-vehicle accidents
on interstate highways (top plot), US routes (middle plot) and
state routes (bottom plot).
\label{FIGURE_sev_year_1}
}
\end{figure}

\begin{figure}[t]
\vspace{14.1truecm}
\includegraphics{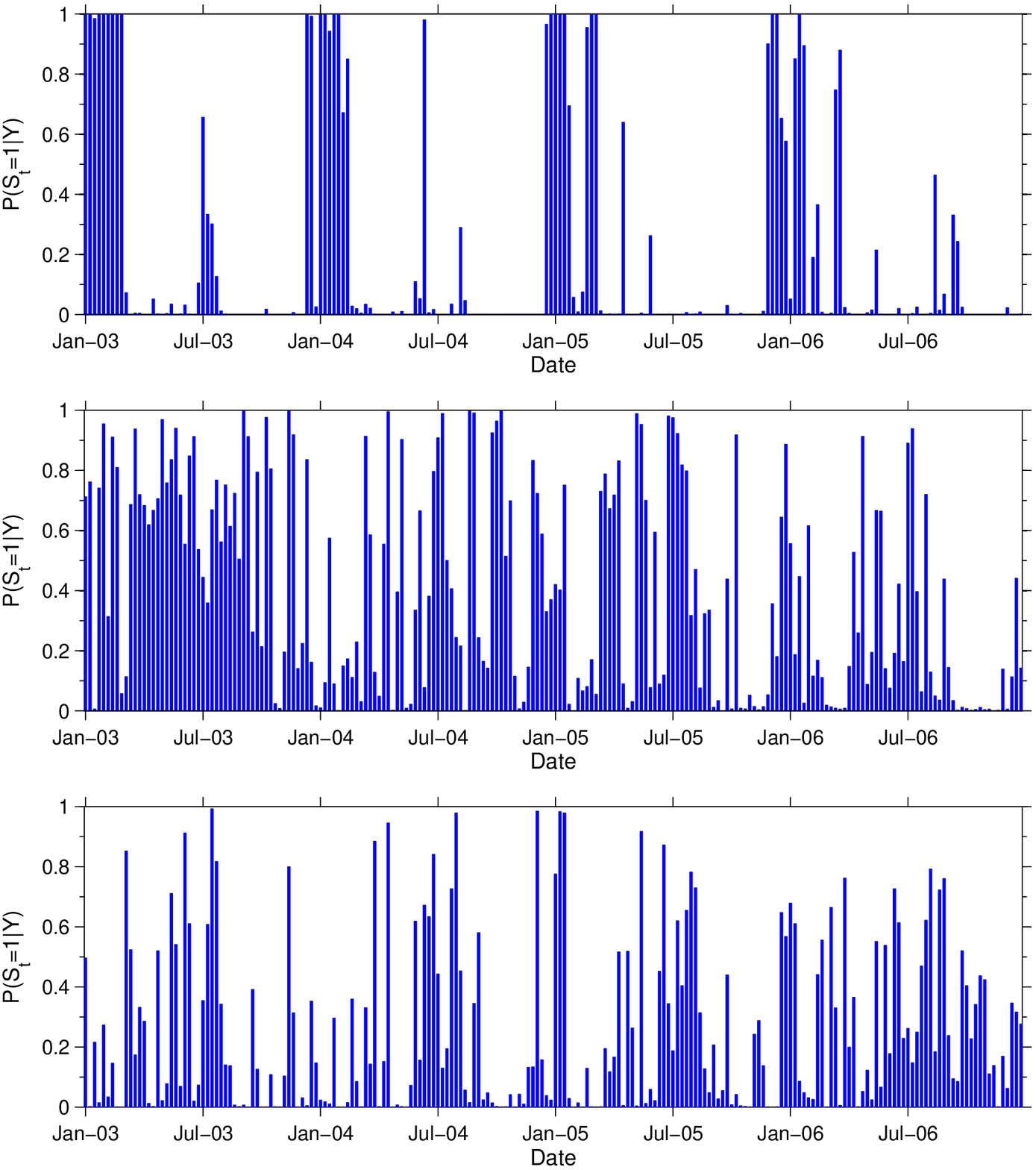}
\caption{Weekly posterior probabilities $P(s_t=1|{\bf Y})$
for the MSML models estimated for severity of 1-vehicle accidents occurring on county roads (top plot), streets (middle plot) and for 2-vehicle accidents occurring on streets (bottom plot).
\label{FIGURE_sev_year_2}
}
\end{figure}

The top, middle and bottom plots in Figure~\ref{FIGURE_sev_year_1} show weekly posterior probabilities $P(s_t=1|{\bf Y})$ of the less frequent state $s_t=1$ for the MSML models estimated for severity of 1-vehicle accidents occurring on interstate highways, US routes and state routes respectively.\footnote{Note that these posterior probabilities are equal to the posterior expectations of $s_t$, $P(s_t=1|{\bf Y})=1\times P(s_t=1|{\bf Y})+0\times P(s_t=0|{\bf Y})=E(s_t|{\bf Y})$.} The top, middle and bottom plots in Figure~\ref{FIGURE_sev_year_2} show weekly posterior probabilities $P(s_t=1|{\bf Y})$ of the less frequent state $s_t=1$ for the MSML models estimated for severity of 1-vehicle accidents occurring on county roads, streets and for 2-vehicle accidents occurring on streets respectively.

\begin{table}[p]
\centering
\caption{Estimation results for multinomial logit models of severity outcomes of one-vehicle accidents on Indiana interstate highways}
\label{T_interst_one-veh}
\begin{scriptsize}
\input{severity_interst_one-veh_table.tex}

\end{scriptsize}
\end{table}

\begin{table}[p]
\centering
\caption{Estimation results for multinomial logit models of severity outcomes of one-vehicle accidents on Indiana US routes}
\label{T_US_one-veh}
\begin{scriptsize}
\input{severity_US_one-veh_table.tex}

\end{scriptsize}
\end{table}

\begin{table}[p]
\centering
\caption{Estimation results for multinomial logit models of severity outcomes of one-vehicle accidents on Indiana state routes}
\label{T_state_one-veh}
\begin{scriptsize}
\input{severity_state_one-veh_table.tex}

\end{scriptsize}
\end{table}

\begin{table}[p]
\centering
\caption{Estimation results for multinomial logit models of severity outcomes of one-vehicle accidents on Indiana county roads}
\label{T_county_one-veh}
\begin{scriptsize}
\input{severity_county_one-veh_table.tex}

\end{scriptsize}
\end{table}

\begin{table}[p]
\centering
\caption{Estimation results for multinomial logit models of severity outcomes of one-vehicle accidents on Indiana streets}
\label{T_street_one-veh}
\begin{scriptsize}
\input{severity_street_one-veh_table.tex}

\end{scriptsize}
\end{table}

\begin{landscape}
\begin{table}[p]
\centering
\caption{Estimation results for multinomial logit models of severity outcomes of two-vehicle accidents on Indiana streets}
\label{T_street_two-veh}
\begin{scriptsize}
\input{severity_street_two-veh_table_a.tex}

\end{scriptsize}
\end{table}
%\addtocounter{table}{-1}
\begin{table}[p]
\centering
%\caption{(Continued)}
\centerline{Table \arabic{chapter}.\arabic{table}: (Continued)}\vspace{1.7ex}
\begin{scriptsize}
\input{severity_street_two-veh_table_b.tex}

\end{scriptsize}
\end{table}
\end{landscape}

\begin{landscape}
\begin{table}[p]
\centering
\caption{Explanations and summary statistics for variables and parameters listed in Tables~\ref{T_interst_one-veh}--\ref{T_street_two-veh} and in Tables~\ref{T_interst_two-veh}--\ref{T_county_two-veh}}
\label{T_sev_expl-stat}
\begin{scriptsize}
\input{severity_expl-stat_table_a.tex}

\end{scriptsize}
\end{table}
%\addtocounter{table}{-1}
\begin{table}[p]
\centering
%\caption{(Continued)}
\centerline{Table \arabic{chapter}.\arabic{table}: (Continued)}\vspace{1.7ex}
\begin{scriptsize}
\input{severity_expl-stat_table_b.tex}

\end{scriptsize}
\end{table}
%\addtocounter{table}{-1}
\begin{table}[p]
\centering
%\caption{(Continued)}
\centerline{Table \arabic{chapter}.\arabic{table}: (Continued)}\vspace{1.7ex}
\begin{scriptsize}
\input{severity_expl-stat_table_c.tex}

\end{scriptsize}
\end{table}
\end{landscape}

From Tables~\ref{T_interst_one-veh}--\ref{T_street_two-veh} we find that in all cases when the two states and Markov switching multinomial logit (MSML) models exist, these models are strongly favored by the empirical data over the corresponding standard multinomial logit (ML) models. Indeed, for example, from lines ``marginal $LL$'' in Tables~\ref{T_interst_one-veh}--\ref{T_street_two-veh} we see that the MSML models provide considerable, ranging from $40.49$ to $206.41$, improvements of the logarithm of the marginal likelihood of the data as compared to the corresponding ML models.\footnote{In addition, we find that DIC (deviance information criterion) favors the MSML models over the corresponding ML models by DIC value improvement ranging from $168.33$ to $450.52$. However, we prefer to rely on the Bayes factor approach instead of the DIC (see footnote~\ref{DIC} on page~\pageref{DIC}).} Thus, from equation~(\ref{EQ_BAYES_FACTOR}) we find that, given the accident severity data, the posterior probabilities of the MSML models are larger than the probabilities of the corresponding ML models by factors ranging from $e^{40.49}$ to $e^{206.41}$. Note that we use equation~(\ref{EQ_L_MARGINAL}) for calculation of the values and the $95\%$ confidence intervals of the logarithms of the marginal likelihoods. The confidence intervals are found by bootstrap simulations (see footnote~\ref{BOOTSTRAP} on page~\pageref{BOOTSTRAP}).

Note that a classical statistics approach for model comparison, based on the maximum likelihood estimation (MLE), also favors the MSML models over the standard ML models. For example, refer to line ``max$(LL)$'' in Table~\ref{T_interst_one-veh} given for the case of 1-vehicle accidents on interstate highways. The MLE gave the maximum log-likelihood value $-8465.79$ for the standard ML model. The maximum log-likelihood value observed during our MCMC simulations for the MSML model is equal to $-8358.97$. An imaginary MLE, at its convergence, would give a MSML log-likelihood value that would be even larger than this observed value. Therefore, if estimated by the MLE, the MSML model would provide large, at least $106.82$ improvement in the maximum log-likelihood value over the corresponding ML model. This improvement would come with only modest increase in the number of free continuous model parameters ($\beta$-s) that enter the likelihood function (refer to Table~\ref{T_interst_one-veh} under ``\# free par.''). Similar arguments hold for comparison of MSML and ML models estimated for other roadway-class-accident-type combinations where two states of roadway safety exist (see Tables~\ref{T_US_one-veh}--\ref{T_street_two-veh}).

To evaluate the goodness-of-fit for a model, we use the posterior (or MLE) estimates of all continuous model parameters ($\beta$-s, $\alpha$, $p_{0\rightarrow1}$, $p_{1\rightarrow0}$) and generate $10^4$ artificial data sets under the hypothesis that the model is true (see footnote~\ref{S_GENERATION} on page~\pageref{S_GENERATION}). We find the distribution of $\chi^2$, given by equation~(\ref{CHI_SQUARED_LOGIT}), and calculate the goodness-of-fit p-value for the observed value of $\chi^2$. The resulting p-values for our multinomial logit models are given in Tables~\ref{T_interst_one-veh}--\ref{T_street_two-veh}. These p-values are around \mbox{$20$--$80\%$}. Therefore, all models fit the data well.

Now, refer to Table~\ref{T_sev_corr}. The first six rows of this table list time-correlation coefficients between posterior probabilities $P(s_t=1|{\bf Y})$ for the six MSML models that exist and are estimated for six roadway-class-accident-type combinations (1-vehicle accidents on interstate highways, US routes, state routes, county roads, streets, and 2-vehicle accidents on streets).\footnote{See footnote~\ref{WEIGHTED_CORR} on page~\pageref{WEIGHTED_CORR} for details on computation of correlation coefficients.} We see that the states for 1-vehicle accidents on all high-speed roads (interstate highways, US routes, state routes and county roads) are correlated with each other. The values of the corresponding correlation coefficients are positive and range from $0.263$ to $0.688$ (see Table~\ref{T_sev_corr}). This result suggests an existence of common (unobservable) factors that can cause switching between states of roadway safety for 1-vehicle accidents on all high-speed roads.

\begin{table}[p]
\centering
\caption{Correlations of the posterior probabilities $P(s_t=1|{\bf Y})$ with each other and with weather-condition variables (for the MSML models of accident severities)}
\label{T_sev_corr}
\begin{scriptsize}
\input{severity_corr_table.tex}

\end{scriptsize}
\end{table}

The remaining rows of Table~\ref{T_sev_corr} show correlation coefficients between posterior probabilities $P(s_t=1|{\bf Y})$ and weather-condition variables. These correlations were found by using daily and hourly historical weather data in Indiana, available at the Indiana State Climate Office at Purdue University (www.agry.purdue.edu/climate). For these correlations, the precipitation and snowfall amounts are daily amounts in inches averaged over the week and across Indiana weather observation stations (see footnote~\ref{SNOWFALL} on page~\pageref{SNOWFALL}). The temperature variable is the mean daily air temperature $(^oF)$ averaged over the week and across the weather stations. The wind gust variable is the maximal instantaneous wind speed (mph) measured during the 10-minute period just prior to the observational time. Wind gusts are measured every hour and averaged over the week and across the weather stations. The effect of fog/frost is captured by a dummy variable that is equal to one if and only if the difference between air and dewpoint temperatures does not exceed $5^oF$ (in this case frost can form if the dewpoint is below the freezing point $32^oF$, and fog can form otherwise). The fog/frost dummies are calculated for every hour and are averaged over the week and across the weather stations. Finally, visibility distance variable is the harmonic mean of hourly visibility distances, which are measured in miles every hour and are averaged over the week and across the weather stations (see footnote~\ref{HARMONIC_MEAN} on page~\pageref{HARMONIC_MEAN}).

From the results given in Table~\ref{T_sev_corr} we find that for 1-vehicle accidents on all high-speed roads (interstate highways, US routes, state routes and county roads), the less frequent state $s_t=1$ is positively correlated with extreme temperatures (low during winter and high during summer), rain precipitations and snowfalls, strong wind gusts, fogs and frosts, low visibility distances. It is reasonable to expect that roadway safety is different during bad weather as compared to better weather, resulting in the two-state nature of roadway safety.

The results of Table~\ref{T_sev_corr} suggest that Markov switching for road safety on streets is very different from switching on all other roadway classes. In particular, the states of roadway safety on streets exhibit low correlation with states on other roads. In addition, only streets exhibit Markov switching in the case of 2-vehicle accidents. Finally, states of roadway safety on streets show little correlation with weather conditions. A possible explanation of these differences is that streets are mostly located in urban areas and they have traffic moving at speeds lower that those on other roads.

Next, we consider the estimation results for the stationary unconditional probabilities $\bar p_0$ and $\bar p_1$ of states $s_t=0$ and $s_t=1$ for MSML models [see equations~(\ref{EQ_MS_P_BAR})]. These transition probabilities are listed in lines ``${\bar p}_0$ and ${\bar p}_1$'' of Tables~\ref{T_interst_one-veh}--\ref{T_street_two-veh}. We find that the ratio $\bar p_1/\bar p_0$ is approximately equal to $0.46$, $0.13$, $0.74$, $0.25$, $0.65$ and $0.36$ in the cases of 1-vehicle accidents on interstate highways, US routes, state routes, county roads, streets, and 2-vehicle accidents on streets respectively. Thus, for some roadway-class-accident-type combinations (for example, 1-vehicle accidents on US routes) the less frequent state $s_t=1$ is quite rare, while for other combinations (for example, 1-vehicle accidents on state routes) state $s_t=1$ is only slightly less frequent than state $s_t=0$.

Finally, we set model coefficients $\betabf_{(0)}$ and $\betabf_{(1)}$ to their posterior means, calculate the probabilities of fatality and injury outcomes in states $0$ and $1$ by using equation~(\ref{EQ_MLE_ML}), and average these probabilities over all values of the explanatory variables ${\bf X}_{t,n}$ observed in the data sample. We compare these probabilities across the two states of roadway safety, $s_t=0$ and $s_t=1$, for MSML models [refer to lines ``$\langle P_{t,n}^{(i)}\rangle_X$'' in Tables~\ref{T_interst_one-veh}--\ref{T_street_two-veh}]. We find that in many cases these averaged probabilities of fatality and injury outcomes do not differ very significantly across the two states of roadway safety (the only significant differences are for fatality probabilities in the cases of 1-vehicle accidents on US routes, county roads and streets). This means that in many cases states $s_t=0$ and $s_t=1$ are approximately equally dangerous as far as accident severity is concerned. We discuss this result in the next chapter (which includes a discussion of all our results).

%% file: severity_interst_one-veh_table.tex
\tabcolsep=0.15em

\begin{tabular}{|l|c|c|c|c|c|c|c|c|}
\hline
& \multicolumn{2}{|c|}{} & \multicolumn{2}{|c|}{} &
\multicolumn{4}{|c|}{\bf MSML}
\\
\cline{6-9}
${}^{{}^{\bf\displaystyle ~Variable}}$ &
\multicolumn{2}{|c|}{${}^{{}^{\bf\displaystyle ML\mbox{\bf-}by\mbox{\bf-}MLE}}$} & \multicolumn{2}{|c|}{${}^{{}^{\bf\displaystyle ML\mbox{\bf-}by\mbox{\bf-}MCMC}}$} &
\multicolumn{2}{|c|}{{\bf state }{\boldmath$s=0$}} &
\multicolumn{2}{|c|}{{\bf state }{\boldmath$s=1$}}
\\
\cline{2-9}
& {\bf fatality} & {\bf injury} & {\bf fatality} & {\bf injury}
& {\bf fatality} & {\bf injury} & {\bf fatality} & {\bf injury}
\\
\hline
~intercept & $-11.9^{-10.1}_{-13.7}$ &
$-3.69^{-3.53}_{-3.84}$ &
$-12.4^{-10.6}_{-14.5}$ &
$-3.72^{-3.56}_{-3.88}$ &
$-12.2^{-10.5}_{-14.4}$ &
$-3.98^{-3.79}_{-4.17}$ &
$-12.2^{-10.5}_{-14.4}$ &
$-3.22^{-2.98}_{-3.45}$\\
\hline
~sum & $.235^{.329}_{.142}$ &
$.235^{.329}_{.142}$ &
$.237^{.329}_{.143}$ &
$.237^{.329}_{.143}$ &
$.176^{.293}_{.0551}$ &
$.176^{.293}_{.0551}$ &
$.176^{.293}_{.0551}$ &
$.615^{.959}_{.282}$\\
\hline
~thday & $-.798^{-.115}_{-1.48}$ &
-- &
$-.853^{-.206}_{-1.59}$ &
-- &
$-.872^{-.225}_{-1.61}$ &
-- &
$-.872^{-.225}_{-1.61}$ &
--\\
\hline
~cons & $-.418^{-.213}_{-.623}$ &
$-.418^{-.213}_{-.623}$ &
$-.425^{-.224}_{-.632}$ &
$-.425^{-.224}_{-.632}$ &
$-.566^{-.319}_{-.822}$ &
$-.566^{-.319}_{-.822}$ &
$-.566^{-.319}_{-.822}$ &
--\\
\hline
~light & $-.392^{-.0368}_{-.748}$ &
$.137^{.224}_{.0501}$ &
$-.387^{-.0301}_{-.740}$ &
$.143^{.230}_{.0568}$ &
$-.378^{-.0236}_{-.729}$ &
$.139^{.226}_{.0522}$ &
$-.378^{-.0236}_{-.729}$ &
$.139^{.226}_{.0522}$\\
\hline
\end{tabular}

%% file: severity_US_one-veh_table.tex
\tabcolsep=0.3em

\begin{tabular}{|l|c|c|c|c|c|c|c|c|}
\hline
& \multicolumn{2}{|c|}{} & \multicolumn{2}{|c|}{} &
\multicolumn{4}{|c|}{\bf MSML}
\\
\cline{6-9}
${}^{{}^{\bf\displaystyle ~Variable}}$ &
\multicolumn{2}{|c|}{${}^{{}^{\bf\displaystyle ML\mbox{\bf-}by\mbox{\bf-}MLE}}$} & \multicolumn{2}{|c|}{${}^{{}^{\bf\displaystyle ML\mbox{\bf-}by\mbox{\bf-}MCMC}}$} &
\multicolumn{2}{|c|}{{\bf state }{\boldmath$s=0$}} &
\multicolumn{2}{|c|}{{\bf state }{\boldmath$s=1$}}
\\
\cline{2-9}
& {\bf fatality} & {\bf injury} & {\bf fatality} & {\bf injury}
& {\bf fatality} & {\bf injury} & {\bf fatality} & {\bf injury}
\\
\hline
~intercept & $-6.51^{-5.00}_{-8.03}$ &
$-2.13^{-1.79}_{-2.47}$ &
$-6.62^{-5.16}_{-8.14}$ &
$-2.12^{-1.78}_{-2.47}$ &
$-5.72^{-4.69}_{-6.92}$ &
$-2.05^{-1.71}_{-2.40}$ &
$-5.72^{-4.69}_{-6.92}$ &
$-2.79^{-2.37}_{-3.23}$\\
\hline
~sum & $.514^{.894}_{.134}$ &
$.200^{.305}_{.0947}$ &
$.509^{.883}_{.124}$ &
$.200^{.305}_{.0951}$ &
$.190^{.300}_{.0789}$ &
$.190^{.300}_{.0789}$ &
$.190^{.300}_{.0789}$ &
--\\
\hline
~light & $-.498^{-.142}_{-.855}$ &
$.194^{.287}_{.101}$ &
$-.492^{-.136}_{-.848}$ &
$.203^{.296}_{.110}$ &
$-.493^{-.136}_{-.857}$ &
$.197^{.290}_{.105}$ &
-- &
$.197^{.290}_{.105}$\\
\hline
~snow & $-1.17^{-.170}_{-2.18}$ &
-- &
$-1.30^{-.357}_{-2.47}$ &
-- &
$-1.10^{-.151}_{-2.27}$ &
$.165^{.317}_{.0115}$ &
$-1.10^{-.151}_{-2.27}$ &
$.165^{.317}_{.0115}$\\
\hline
~nojun & $.701^{1.25}_{.149}$ &
$.217^{.335}_{.0994}$ &
$.727^{1.31}_{.199}$ &
$.213^{.331}_{.0968}$ &
$.787^{1.36}_{.259}$ &
$.214^{.332}_{.0965}$ &
$.787^{1.36}_{.259}$ &
$.214^{.332}_{.0965}$\\
\hline
\end{tabular}

%% file: severity_state_one-veh_table.tex
\tabcolsep=0.22em

\begin{tabular}{|l|c|c|c|c|c|c|c|c|}
\hline
& \multicolumn{2}{|c|}{} & \multicolumn{2}{|c|}{} &
\multicolumn{4}{|c|}{\bf MSML}
\\
\cline{6-9}
${}^{{}^{\bf\displaystyle ~Variable}}$ &
\multicolumn{2}{|c|}{${}^{{}^{\bf\displaystyle ML\mbox{\bf-}by\mbox{\bf-}MLE}}$} & \multicolumn{2}{|c|}{${}^{{}^{\bf\displaystyle ML\mbox{\bf-}by\mbox{\bf-}MCMC}}$} &
\multicolumn{2}{|c|}{{\bf state }{\boldmath$s=0$}} &
\multicolumn{2}{|c|}{{\bf state }{\boldmath$s=1$}}
\\
\cline{2-9}
& {\bf fatality} & {\bf injury} & {\bf fatality} & {\bf injury}
& {\bf fatality} & {\bf injury} & {\bf fatality} & {\bf injury}
\\
\hline
~intercept & $-3.98^{-3.66}_{-4.30}$ &
$-1.67^{-1.53}_{-1.80}$ &
$-4.03^{-3.71}_{-4.36}$ &
$-1.71^{-1.58}_{-1.85}$ &
$-3.44^{-3.10}_{-3.79}$ &
$-1.68^{-1.54}_{-1.81}$ &
$-4.96^{-4.15}_{-5.96}$ &
$-1.68^{-1.54}_{-1.81}$\\
\hline
~sum & $.232^{.307}_{.156}$ &
$.232^{.307}_{.156}$ &
$.232^{.307}_{.157}$ &
$.232^{.307}_{.157}$ &
$.238^{.314}_{.163}$ &
$.238^{.314}_{.163}$ &
$.238^{.314}_{.163}$ &
$.238^{.314}_{.163}$\\
\hline
~$X_{12}$ & $-.390^{-.302}_{-.478}$ &
$-.390^{-.302}_{-.478}$ &
$-.395^{-.306}_{-.483}$ &
$-.395^{-.306}_{-.483}$ &
-- &
$-.385^{-.296}_{-.474}$ &
$-2.05^{-.954}_{-3.62}$ &
$-3.85^{-.296}_{-.474}$\\
\hline
~light & $-.646^{-.408}_{-.884}$ &
$.193^{.261}_{.125}$ &
$-.641^{-.404}_{-.879}$ &
$.199^{.267}_{.132}$ &
$-.689^{-.448}_{-.931}$ &
-- &
$-.689^{-.448}_{-.931}$ &
$.277^{.378}_{.177}$\\
\hline
~precip & $-.854^{.466}_{-1.24}$ &
-- &
$-.868^{-.494}_{-1.27}$ &
-- &
$-.829^{-.448}_{-1.24}$ &
-- &
$-.829^{-.448}_{-1.24}$ &
--\\
\hline
\end{tabular}

%% file: severity_county_one-veh_table.tex
\tabcolsep=0.15em

\begin{tabular}{|l|c|c|c|c|c|c|c|c|}
\hline
& \multicolumn{2}{|c|}{} & \multicolumn{2}{|c|}{} &
\multicolumn{4}{|c|}{\bf MSML}
\\
\cline{6-9}
${}^{{}^{\bf\displaystyle ~Variable}}$ &
\multicolumn{2}{|c|}{${}^{{}^{\bf\displaystyle ML\mbox{\bf-}by\mbox{\bf-}MLE}}$} & \multicolumn{2}{|c|}{${}^{{}^{\bf\displaystyle ML\mbox{\bf-}by\mbox{\bf-}MCMC}}$} &
\multicolumn{2}{|c|}{{\bf state }{\boldmath$s=0$}} &
\multicolumn{2}{|c|}{{\bf state }{\boldmath$s=1$}}
\\
\cline{2-9}
& {\bf fatality} & {\bf injury} & {\bf fatality} & {\bf injury}
& {\bf fatality} & {\bf injury} & {\bf fatality} & {\bf injury}
\\
\hline
~intercept & $-6.39^{-5.78}_{-7.00}$ &
$-1.62^{-1.53}_{-1.71}$ &
$-6.49^{-5.89}_{-7.12}$ &
$-1.65^{-1.56}_{-1.75}$ &
$-6.16^{-5.59}_{-6.73}$ &
$-1.81^{-1.70}_{-1.93}$ &
$-7.51^{-6.75}_{-8.29}$ &
$-2.13^{-1.99}_{-2.26}$\\
\hline
~sum & $.151^{.201}_{.100}$ &
$.151^{.201}_{.100}$ &
$.149^{.200}_{.0988}$ &
$.149^{.200}_{.0988}$ &
$.142^{.194}_{.0891}$ &
$.142^{.194}_{.0891}$ &
-- &
$.142^{.194}_{.0891}$\\
\hline
~wday & $-.281^{-.108}_{-.453}$ &
$-.0987^{-.0541}_{-.143}$ &
$-.275^{-.102}_{-.446}$ &
$-.0952^{-.0505}_{-.140}$ &
$-.146^{-.0934}_{-.198}$ &
$-.146^{-.0934}_{-.198}$ &
$-.146^{-.0934}_{-.198}$ &
--\\
\hline
~dayt & $-.456^{-.263}_{-.649}$ &
-- &
$-.443^{-.252}_{-.637}$ &
-- &
$-.492^{-.281}_{-.709}$ &
-- &
-- &
--\\
\hline
~$X_{12}$ & $-.642^{-.160}_{-1.13}$ &
$-.169^{-.0733}_{-.264}$ &
$-.667^{-.207}_{-1.18}$ &
$-.169^{-.0746}_{-.264}$ &
$-.689^{-.227}_{-1.20}$ &
$-.207^{-.0941}_{-.320}$ &
$-.689^{-.227}_{-1.20}$ &
--\\
\hline
\end{tabular}

%% file: severity_street_one-veh_table.tex
\tabcolsep=0.15em

\begin{tabular}{|l|c|c|c|c|c|c|c|c|}
\hline
& \multicolumn{2}{|c|}{} & \multicolumn{2}{|c|}{} &
\multicolumn{4}{|c|}{\bf MSML}
\\
\cline{6-9}
${}^{{}^{\bf\displaystyle ~Variable}}$ &
\multicolumn{2}{|c|}{${}^{{}^{\bf\displaystyle ML\mbox{\bf-}by\mbox{\bf-}MLE}}$} & \multicolumn{2}{|c|}{${}^{{}^{\bf\displaystyle ML\mbox{\bf-}by\mbox{\bf-}MCMC}}$} &
\multicolumn{2}{|c|}{{\bf state }{\boldmath$s=0$}} &
\multicolumn{2}{|c|}{{\bf state }{\boldmath$s=1$}}
\\
\cline{2-9}
& {\bf fatality} & {\bf injury} & {\bf fatality} & {\bf injury}
& {\bf fatality} & {\bf injury} & {\bf fatality} & {\bf injury}
\\
\hline
~intercept & $-8.60^{-7.61}_{-9.57}$ &
$-3.87^{-3.67}_{-4.07}$ &
$-8.68^{-7.75}_{-9.76}$ &
$-3.393^{-3.74}_{-4.14}$ &
$-8.87^{-7.93}_{-9.99}$ &
$-3.94^{-3.73}_{-4.14}$ &
$-7.94^{-6.96}_{-9.08}$ &
$-3.94^{-3.73}_{-4.14}$\\
\hline
~wint & $-.192^{-.129}_{-.256}$ &
$-.192^{-.129}_{-.256}$ &
$-.187^{-.124}_{-.251}$ &
$-.187^{-.124}_{-.251}$ &
$-.159^{-.0641}_{-.262}$ &
$-.159^{-.0641}_{-.262}$ &
$-.159^{-.0641}_{-.262}$ &
$-.217^{-.0574}_{-.375}$\\
\hline
~jobend & $.141^{.208}_{.0730}$ &
$.141^{.208}_{.0730}$ &
$.142^{.209}_{.0750}$ &
$.142^{.209}_{.0750}$ &
-- &
$.144^{.212}_{.0765}$ &
-- &
$.144^{.212}_{.0765}$\\
\hline
~cons & $-.270^{-.0532}_{-.487}$ &
$-.270^{-.0532}_{-.487}$ &
$-.279^{-.0644}_{-.496}$ &
$-.279^{-.0644}_{-.496}$ &
$-2.22^{-.393}_{-5.07}$ &
-- &
$-2.22^{-.393}_{-5.07}$ &
$-.598^{-.202}_{-1.02}$\\
\hline
~day & $-.779^{-.524}_{-1.03}$ &
$.0654^{.119}_{.0123}$ &
$-.776^{-.526}_{-1.03}$ &
$.0784^{.131}_{.0257}$ &
$-.768^{-.516}_{-1.02}$ &
-- &
$-.768^{-.516}_{-1.02}$ &
$.139^{.251}_{.0329}$\\
\hline
\end{tabular}

%% file: severity_street_two-veh_table_a.tex
\begin{tabular}{|l|c|c|c|c|c|c|c|c|}
\hline
& \multicolumn{2}{|c|}{} & \multicolumn{2}{|c|}{} &
\multicolumn{4}{|c|}{\bf MSML}
\\
\cline{6-9}
${}^{{}^{\bf\displaystyle Variable}}$ &
\multicolumn{2}{|c|}{${}^{{}^{\bf\displaystyle ML\mbox{\bf-}by\mbox{\bf-}MLE}}$} & \multicolumn{2}{|c|}{${}^{{}^{\bf\displaystyle ML\mbox{\bf-}by\mbox{\bf-}MCMC}}$} &
\multicolumn{2}{|c|}{{\bf state }{\boldmath$s=0$}} &
\multicolumn{2}{|c|}{{\bf state }{\boldmath$s=1$}}
\\
\cline{2-9}
& {\bf fatality} & {\bf injury} & {\bf fatality} & {\bf injury}
& {\bf fatality} & {\bf injury} & {\bf fatality} & {\bf injury}
\\
\hline
intercept & $-10.6^{-9.58}_{-11.6}$ &
$-2.86^{-2.71}_{-3.02}$ &
$-10.7^{-9.68}_{-11.7}$ &
$-2.95^{-2.79}_{-3.10}$ &
$-13.1^{-11.0}_{-16.2}$ &
$-3.00^{-2.87}_{-3.12}$ &
$-13.1^{-11.0}_{-16.2}$ &
$-3.00^{-2.87}_{-3.12}$\\
\hline
wint & $-.135^{-.101}_{-.169}$ &
$-.135^{-.101}_{-.169}$ &
$-.134^{-.0999}_{-.168}$ &
$-.134^{-.0999}_{-.168}$ &
-- &
$-.130^{-.0939}_{-.165}$ &
-- &
$-.130^{-.0939}_{-.165}$\\
\hline
wday & $-.896^{-.546}_{-1.25}$ &
$-.104^{-.0699}_{-.138}$ &
$-.892^{-.539}_{-1.24}$ &
$-.102^{-.0679}_{-.136}$ &
$-.835^{-.481}_{-1.18}$ &
$-.0980^{-.0639}_{-.132}$ &
$-.835^{-.481}_{-1.18}$ &
$-.0980^{-.0639}_{-.132}$\\
\hline
morn & $-.0550^{-.0117}_{-.0983}$ &
$-.0550^{-.0117}_{-.0983}$ &
$-.485^{-.00559}_{-.0916}$ &
$-.485^{-.00559}_{-.0916}$ &
-- &
$-.0659^{-.0130}_{-.121}$ &
-- &
--\\
\hline
$X_{12}$ & $-.0801^{-.0188}_{-.142}$ &
$-.0801^{-.0188}_{-.142}$ &
$-.0598^{-.00109}_{-.120}$ &
$-.0598^{-.00109}_{-.120}$ &
-- &
-- &
-- &
--\\
\hline
\end{tabular}

%% file: severity_street_two-veh_table_b.tex
\begin{tabular}{|l|c|c|c|c|c|c|c|c|}
\hline
& \multicolumn{2}{|c|}{} & \multicolumn{2}{|c|}{} &
\multicolumn{4}{|c|}{\bf MSML}
\\
\cline{6-9}
${}^{{}^{\bf\displaystyle Variable}}$ &
\multicolumn{2}{|c|}{${}^{{}^{\bf\displaystyle ML\mbox{\bf-}by\mbox{\bf-}MLE}}$} & \multicolumn{2}{|c|}{${}^{{}^{\bf\displaystyle ML\mbox{\bf-}by\mbox{\bf-}MCMC}}$} &
\multicolumn{2}{|c|}{{\bf state }{\boldmath$s=0$}} &
\multicolumn{2}{|c|}{{\bf state }{\boldmath$s=1$}}
\\
\cline{2-9}
& {\bf fatality} & {\bf injury} & {\bf fatality} & {\bf injury}
& {\bf fatality} & {\bf injury} & {\bf fatality} & {\bf injury}
\\
\hline
driver & -- &
$.172^{.257}_{.0856}$ &
-- &
$.172^{.257}_{.0859}$ &
$2.07^{5.08}_{.216}$ &
$.164^{.237}_{.0900}$ &
$2.07^{5.08}_{.216}$ &
--\\
\hline
$X_{27}$ & -- &
$-.0165^{-.00346}_{-.0296}$ &
-- &
$-.0163^{-.00333}_{-.0293}$ &
-- &
$-.0203^{-.00678}_{-.0341}$ &
-- &
$-.0203^{-.00678}_{-.0341}$\\
\hline
nosig & -- &
$-.186^{-.150}_{-.223}$ &
-- &
$-.194^{-.158}_{-.230}$ &
-- &
$-.194^{-.158}_{-.230}$ &
-- &
$-.194^{-.158}_{-.230}$\\
\hline
singSUV & -- &
$-.0860^{-.0584}_{-.114}$ &
-- &
$-.0854^{-.0579}_{-.113}$ &
-- &
$-.0864^{-.0588}_{-.114}$ &
-- &
$-.0864^{-.0588}_{-.114}$\\
\hline
oldvage & -- &
$.0205^{.0236}_{.0174}$ &
-- &
$.0205^{.0236}_{.0174}$ &
$.0205^{.0235}_{.0175}$ &
$.0205^{.0235}_{.0175}$ &
$.0205^{.0235}_{.0175}$ &
$.0205^{.0235}_{.0175}$\\
\hline
\end{tabular}

%% file: severity_expl-stat_table_a.tex
\begin{tabular}{|l|l|c|c|c|c|c|}
\hline
${\bf Variable}$ &
${\bf Description}$ &
${\bf Mean}$ &
${\bf Std}^{\,\rm a}$ &
${\bf Min}^{\,\rm a}$ &
${\bf Median}$ &
${\bf Max}^{\,\rm a}$
\\
\hline
age0 &
Age of the driver at fault is less than $18$ years old (dummy) &
$.0846$ &
$.278$ &
$0$ &
$0$ &
$1.00$
\\
\hline
age0o &
Age of the oldest driver involved into the accident is less than $18$ years old (dummy) &
$.0103$ &
$.101$ &
$0$ &
$0$ &
$1.00$
\\
\hline
cons &
Construction at the accident location (dummy) &
$.0272$ &
$.163$ &
$0$ &
$0$ &
$1.00$
\\
\hline
curve &
Roadway is at curve (dummy) &
$.0459$ &
$.209$ &
$0$ &
$0$ &
$1.00$
\\
\hline
dark &
Dark time with no street lights (dummy) &
$.0439$ &
$.205$ &
$0$ &
$0$ &
$1.00$
\\
\hline
darklamp &
Dark and street lights on (dummy) &
$.130$ &
$.337$ &
$0$ &
$0$ &
$1.00$
\\
\hline
day &
Daylight (dummy) &
$.784$ &
$.412$ &
$0$ &
$1.00$ &
$1.00$
\\
\hline
\end{tabular}

%% file: severity_expl-stat_table_b.tex
\begin{tabular}{|l|l|c|c|c|c|c|}
\hline
${\bf Variable}$ &
${\bf Description}$ &
${\bf Mean}$ &
${\bf Std}^{\,\rm a}$ &
${\bf Min}^{\,\rm a}$ &
${\bf Median}$ &
${\bf Max}^{\,\rm a}$
\\
\hline
nosig &
No any traffic control device for the vehicle at fault (dummy) &
$.233$ &
$.423$ &
$0$ &
$0$ &
$1.00$
\\
\hline
olddrv &
The driver at fault is older than the other driver, if a 2-vehicle accident (dummy) &
$47.3$ &
$16.5$ &
$15.0$ &
$$ &
$99.0$
\\
\hline
oldvage &
Age of the oldest vehicle involved (in years) &
$10.2$ &
$5.07$ &
$-1.00$ &
$$ &
$41.0$
\\
\hline
othUS &
License state of the vehicle at fault is a U.S. state except Indiana and its neighboring 
& & & & &\\
&
states (IL, KY, OH, MI) (dummy) &
$.0272$ &
$.148$ &
$0$ &
$0$ &
$1.00$
\\
\hline
precip &
Precipitation: rain/freezing rain/snow/sleet/hail (dummy) &
$.172$ &
$.377$ &
$0$ &
$0$ &
$1.00$
\\
\hline
priv &
Road traveled by the vehicle at fault is a private drive (dummy) &
$.0289$ &
$.168$ &
$0$ &
$0$ &
$1.00$
\\
\hline
\end{tabular}

%% file: severity_expl-stat_table_c.tex
\begin{tabular}{|l|l|c|c|c|c|c|}
\hline
${\bf Variable}$ &
${\bf Description}$ &
${\bf Mean}$ &
${\bf Std}^{\,\rm a}$ &
${\bf Min}^{\,\rm a}$ &
${\bf Median}$ &
${\bf Max}^{\,\rm a}$
\\
\hline
$X_{12}$ &
Roadway type (dummy: 1 if urban, 0 if rural) &
$.829$ &
$.377$ &
$0$ &
$1.00$ &
$1.00$
\\
\hline
$X_{27}$ &
Number of occupants in the vehicle at fault &
$1.45$ &
$1.18$ &
$0$ &
$$ &
$70.0$
\\
\hline
$X_{29}$ &
Speed limit (used if known and the same for all vehicles involved) &
$36.7$ &
$9.86$ &
$5.00$ &
$$ &
$75.0$
\\
\hline
$X_{33}$ &
At least one of the vehicles involved was on fire (dummy) &
$.00505$ &
$.0709$ &
$0$ &
$0$ &
$1.00$
\\
\hline
$X_{34}$ &
Age of the driver at fault (in years) &
$37.0$ &
$9.86$ &
$3.00$ &
$$ &
$99.0$
\\
\hline
$X_{35}$ &
Gender of the driver at fault (dummy: 1 if female, 0 if male) &
$.449$ &
$.497$ &
$0$ &
$0$ &
$1.00$
\\
\hline
\hline
$\langle P_{t,n}^{(i)}\rangle_X$ &
Probability of $i^{\rm th}$ severity outcome averaged over all values
of explanatory variables ${\bf X}_{t,n}$ &
-- &
-- &
-- &
-- &
--
\\
\end{tabular}

%% file: severity_corr_table.tex
\begin{tabular}{|l|c|c|c|c|c|c|}
\hline
& 1-vehicle,  & 1-vehicle, & 1-vehicle,   & 1-vehicle,   & 1-vehicle, & 2-vehicle,\\
& interstates & US routes  & state routes & county roads & streets    & streets  \\
\hline
1-vehicle, interstates & $1$ & $0.418$ & $0.293$ & $0.606$ & $-0.013$ & $-0.173$\\
\hline
1-vehicle, US routes & $0.418$ & $1$ & $0.263$ & $0.688$ & $-0.070$ & $-0.155$\\
\hline
1-vehicle, state routes & $0.293$ & $0.263$ & $1$ & $0.409$ & $-0.047$ & $-0.035$\\
\hline
1-vehicle, county roads & $0.606$ & $0.688$ & $0.409$ & $1$ & $-0.022$ & $-0.051$\\
\hline
1-vehicle, streets & $-0.013$ & $-0.070$ & $-0.047$ & $-0.022$ & $1$ & $0.115$\\
\hline
2-vehicle, streets & $-0.173$ & $-0.155$ & $-0.035$ & $-0.051$ & $0.115$ & $1$\\
\hline
\hline
\multicolumn{7}{|c|}{All year}\\
\hline
Precipitation (inch) & $-0.139$ & $-0.060$ & $0.096$ & $-0.037$ & $0.067$ & $0.146$\\
\hline
Temperature ($^oF$) & $-0.606$ & $-0.439$ & $-0.234$ & $-0.665$ & $0.231$ & $0.220$\\
\hline
Snowfall (inch) & $0.479$ & $0.635$ & $0.319$ & $0.723$ & $0.003$ & $-0.100$\\
\cline{2-7}
$\quad\quad\quad>0.0$ (dummy) & $0.695$ & $0.412$ & $0.382$ & $0.695$ & $-0.142$ & $-0.131$\\
\cline{2-7}
$\quad\quad\quad>0.1$ (dummy) & $0.532$ & $0.585$ & $0.328$ & $0.847$ & $-0.046$ & $-0.161$\\
\hline
Wind gust (mph) & $0.108$ & $0.100$ & $0.087$ & $0.206$ & $0.164$ & $0.051$\\
\hline
Fog / Frost (dummy) & $0.093$ & $0.164$ & $0.193$ & $0.167$ & $0.047$ & $0.119$\\
\hline
Visibility distance (mile) & $-0.228$ & $-0.221$ & $-0.172$ & $-0.298$ & $-0.019$ & $-0.081$\\
\hline
\hline
\multicolumn{7}{|c|}{Winter (November - March)}\\
\hline
Precipitation (inch) & $-0.134$ & $-0.037$ & $0.027$ & $-0.053$ & $0.065$ & $0.356$\\
\hline
Temperature ($^oF$) & $-0.595$ & $-0.479$ & $-0.397$ & $-0.735$ & $-0.008$ & $0.236$\\
\hline
Snowfall (inch) & $0.439$ & $0.592$ & $0.375$ & $0.645$ & $0.157$ & $-0.110$\\
\cline{2-7}
$\quad\quad\quad>0.0$ (dummy) & $0.596$ & $0.282$ & $0.475$ & $0.607$ & $0.115$ & $-0.142$\\
\cline{2-7}
$\quad\quad\quad>0.1$ (dummy) & $0.445$ & $0.518$ & $0.370$ & $0.789$ & $0.112$ & $-0.210$\\
\hline
Wind gust (mph) & $0.302$ & $0.134$ & $0.122$ & $0.353$ & $0.237$ & $0.071$\\
\hline
Frost (dummy) & $0.537$ & $0.544$ & $0.440$ & $0.716$ & $0.052$ & $-0.225$\\
\hline
Visibility distance (mile) & $-0.251$ & $-.304$ & $-0.249$ & $-0.380$ & $-0.155$ & $-0.109$\\
\hline
\hline
\multicolumn{7}{|c|}{Summer (May - September)}\\
\hline
Precipitation (inch) & $0.000$ & $0.006$ & $0.259$ & $0.096$ & $0.047$ & $-0.063$\\
\hline
Temperature ($^oF$) & $0.179$ & $0.149$ & $0.113$ & $0.037$ & $0.062$ & $0.155$\\
\hline
Snowfall (inch) & -- & -- & -- & -- & -- & --\\
\cline{2-7}
$\quad\quad\quad>0.0$ (dummy) & -- & -- & -- & -- & -- & --\\
\cline{2-7}
$\quad\quad\quad>0.1$ (dummy) & -- & -- & -- & -- & -- & --\\
\hline
Wind gust (mph) & $-0.126$ & $-.009$ & $0.164$ & $0.029$ & $0.121$ & $0.034$\\
\hline
Fog (dummy) & $0.203$ & $0.193$ & $0.275$ & $0.101$ & $-0.076$ & $-0.011$\\
\hline
Visibility distance (mile) & $-0.139$ & $-0.124$ & $-0.062$ & $-0.009$ & $0.077$ & $-0.094$\\
\hline
\end{tabular}

%% file: ch_conclusions.tex
%
%  This is the summary chapter.
%

\chapter{Summary and conclusions}\label{SUMMARY}

In this final chapter we give our major conclusions for the two-state Markov switching models estimated for annual accident frequencies, weekly accident frequencies, and for accident severities.
\begin{itemize}
\item
    Our conclusions for the Markov switching models of annual accident frequencies, specified in Section~\ref{Markov_freq_year} and estimated in Section~\ref{RESULTS_FREQ_YEAR}, are as follows. First, these models provide a far superior statistical fit for accident frequencies as compared to the standard zero-inflated models. Second, the Markov switching models explicitly consider transitions between the zero-accident state and the unsafe state over time, and permit a direct empirical estimation of what states roadway segments are in at different time periods. In particular, we found evidence that some roadway segments changed their states over time (see the bottom-right plot in Figure~\ref{FIGURE_freq_year_1}). Third, note that the Markov switching models avoid a theoretically implausible assumption that some roadway segments are always safe because, in these models, any segment has a non-zero probability of being in the unsafe state. Indeed, the long-term unconditional mean of the accident rate for the $n^{\rm th}$ roadway segment is equal to $\bar p^{(n)}_{1}\langle\lambda_{t,n}\rangle_t$, where $\bar p^{(n)}_{1}=p^{(n)}_{0\rightarrow1}/(p^{(n)}_{0\rightarrow1}+p^{(n)}_{1\rightarrow0})$ is the stationary probability of being in the unsafe state $s_{t,n}=1$ and $\langle\lambda_{t,n}\rangle_t$ is the time average of the accident rate in the unsafe state [refer to equations~(\ref{EQ_MLE_LAMBDA_1}) and~(\ref{EQ_MS_P_BAR})]. This long-term mean is always above zero (see the bottom plot in Figure~\ref{FIGURE_freq_year_2}), even for segments that seem to be in the zero-accident state over the whole observed five-year time interval of our empirical data. Finally, we conclude that two-state Markov switching count data models are likely to be a better alternative to zero-inflated models, in order to account for excess of zeros observed in accident frequency data.
\item
    Our conclusions for the Markov switching models of weekly accident frequencies, specified in Section~\ref{Markov_freq_week} and estimated in Section~\ref{RESULTS_FREQ_WEEK}, are as follows. Our empirical finding that two states exist and that these states are correlated with weather conditions has important implications. For example, multiple states of roadway safety can potentially exist due to slow and/or inadequate adjustment by drivers (and possibly by roadway maintenance services) to adverse conditions and other unpredictable, unidentified, and/or unobservable variables that influence roadway safety. All these variables are likely to interact and change over time, resulting in transitions from one state to another. As discussed earlier, the empirical findings show that the less frequent state is significantly less safe than the other, more frequent state. The estimation results of the full MSNB/MSP models show that explanatory variables ${\bf X}_{t,n}$ exert different influences on roadway safety in different states as indicated by the fact that some of the parameter estimates for the two states of the full MSNB/MSP models are significantly different. Thus, the states not only differ by average accident frequencies, but also differ in the magnitude and/or direction of the effects that various variables exert on accident frequencies. This again underscores the importance of the two-state approach.\footnote{One might also consider a threshold model in which the state value is a function of explanatory variables [similar to threshold autoregressive models used in econometrics \citep{T_02}]. This interesting possibility is beyond the scope of this study.}
\item
    Our conclusions for the Markov switching models of accident severities, specified in Section~\ref{Markov_sever} and estimated in Chapter~\ref{RESULTS_SEV}, are as follows. We found that two states of roadway safety and Markov switching multinomial logit (MSML) models exist for severity of 1-vehicle accidents occurring on high-speed roads (interstate highways, US routes, state routes, county roads), but not for 2-vehicle accidents on these roads. One of possible explanations of this result is that 1- and 2-vehicle accidents may differ in their nature. For example, on one hand, severity of 1-vehicle accidents may frequently be determined by driver-related factors (speeding, falling a sleep, driving under the influence, etc). Drivers' behavior might exhibit a two-state pattern. In particular, drivers might be overconfident and/or have difficulties in adjustments to bad weather conditions. On the other hand, severity of a 2-vehicle accident might crucially depend on the actual physics involved in the collision between the two cars (for example, head-on and side impacts are more dangerous than rear-end collisions). As far as slow-speed streets are concerned, in this case both 1- and 2-vehicle accidents exhibit two-state nature for their severity. Further studies are needed to understand these results. In this study, the important result is that in all cases when two states of roadway safety exist, the two-state MSML models provide a superior statistical fit for accident severity outcomes as compared to the standard ML models.

    We found that in many cases states $s_t=0$ and $s_t=1$ are approximately equally dangerous as far as accident severity is concerned. This result holds despite the fact that state $s_t=1$ is correlated with adverse weather conditions. A likely and simple explanation of this finding is that during bad weather both number of serious accidents (fatalities and injuries) and number of minor accidents (PDOs) increase, so that their relative fraction stays approximately constant. In addition, most drivers are rational and they are likely to take some precautions while driving during bad weather. From the results of modeling annual accident frequencies, we know that the total number of accidents significantly increases during adverse weather conditions. Thus, driver's precautions are probably not sufficient to avoid increases in accident rates during bad weather.
\end{itemize}

We can speculate that one of the major causes of the existence of different states of roadway safety can be slow and inadequate adjustment by some drivers to sudden worsening of weather and roadway conditions (such as snow or ice on a roadway). Of course, apart from weather conditions, there can be additional unpredictable and unidentified factors that influence road safety. All these factors are likely to interact and change in time, resulting in unobserved heterogeneity in accident data. Markov switching between states of roadway safety intends to account for these factors and for the resulting unobserved heterogeneity.\footnote{The Markov property of the switching serves as a reasonable approximation, which helps to simplify our analysis. For example, the Markov property holds reasonably well for changes of weather conditions in time.}
Examples of other statistical models that intend to account for unobserved heterogeneity, include finite mixture models, random parameters (mixed) models, and random effects models \citep[][]{SAMM_98,WKM_03,PL_08}. A theoretical advantage of Markov switching models over other models is that the former allows for an explicit identification of the states of roadway safety at different time periods. Another advantage of Markov switching models is that they explicitly consider how various explanatory variables exert different influences on road safety in different states. For example, in the case of the MSNB and MSP models of accident frequencies estimated in this study, the states differ not only by the values of the average accident frequency ($\lambda$), but also by the values of the model coefficients ($\beta$-s) in the two states.

As far as practical application of Markov switching models for prediction of averaged accident rates is concerned, this prediction depends on whether it is conditional or unconditional. For probabilities conditioned on the previous state, one uses the transition probabilities.\footnote{For example, if the previous state was zero, $s_{t-1,n}=0$, then the probabilities of the current state $s_{t,n}$ being zero and one are equal to the transition probabilities $p^{(n)}_{0\rightarrow0}$ and $p^{(n)}_{0\rightarrow1}$ respectively, refer to equation~(\ref{EQ_MS_P}).} For all unconditional expectations and long-term predictions, one uses unconditional probabilities ($\bar p^{(n)}_0$ and $\bar p^{(n)}_1$), given by equation (\ref{EQ_MS_P_BAR}). In particular, the long-term probability of being in a state is equal to the unconditional probability of this state. Please note that, even if the current state is known (zero or one), then in a long run, all expectations converge to the unconditional expectations exponentially fast (this is a property of Markov processes). Because researchers and practitioners are usually interested in a long-term improvement of safety, using the unconditional probabilities is more appropriate for predictions and decision making.

A determination of the roadway safety state value (zero or one) during a specific time period $t$ is complicated by the unobservability of the state variable. As a result, we rely on Bayesian inference in this case -- we use an accident data, estimate a Markov switching model for this data, and find the posterior probabilities for the state values at time $t$. These posterior probabilities should be used for inference about the state values.

In terms of future work on Markov switching models for accident frequencies and severities, additional empirical studies (for other accident data samples) and multi-state models (with more than two states of roadway safety) are two areas that would further demonstrate the potential of the approach.

%% file: ch_bibliography.tex
%
%  This is the bibliography for the thesis.
%

%\bibliography{all}

%% file: ch_appendix.tex
%
%  This is the Appendix.
%

\appendix

%BEGIN OUR FORMAT
\renewcommand{\thechapter}{A}
\refstepcounter{chapter}
\makeatletter
\renewcommand{\theequation}{\thechapter.\@arabic\c@equation}
\makeatother
\pagestyle{empty}
\addtocounter{page}{-1}
%\chapter*{A\lowercase{ppendix}}
\chapter*{\textnormal{\normalsize{Appendix}}}
\addcontentsline{toc}{chapter}{APPENDIX}
\addtocounter{page}{+1}
%END OUR FORMAT

%\chapter{}
%\label{APPEND}

\begin{table}[h]
\centering
\caption{Estimation results for multinomial logit models of severity outcomes of two-vehicle accidents on Indiana interstate highways}
\label{T_interst_two-veh}
\begin{scriptsize}

\input{severity_interst_two-veh_table.tex}

\end{scriptsize}
\end{table}

\begin{table}[t]
\centering
\caption{Estimation results for multinomial logit models of severity outcomes of two-vehicle accidents on Indiana US routes}
\label{T_US_two-veh}
\begin{scriptsize}
\input{severity_US_two-veh_table.tex}

\end{scriptsize}
\end{table}

\begin{table}[t]
\centering
\caption{Estimation results for multinomial logit models of severity outcomes of two-vehicle accidents on Indiana state routes}
\label{T_state_two-veh}
\begin{scriptsize}
\input{severity_state_two-veh_table.tex}

\end{scriptsize}
\end{table}

\begin{table}[t]
\centering
\caption{Estimation results for multinomial logit models of severity outcomes of two-vehicle accidents on Indiana county roads}
\label{T_county_two-veh}
\begin{scriptsize}
\input{severity_county_two-veh_table.tex}

\end{scriptsize}
\end{table}

%% file: severity_interst_two-veh_table.tex
\begin{tabular}{|l|c|c|c|c|}
\hline
& \multicolumn{2}{|c|}{$\bf\displaystyle ML\mbox{\bf-}by\mbox{\bf-}MLE$} & \multicolumn{2}{|c|}{$\bf\displaystyle ML\mbox{\bf-}by\mbox{\bf-}MCMC$}\\
\cline{2-5}
${}^{{}^{\bf\displaystyle Variable}}$ & {\bf fatality} & {\bf injury} &
{\bf fatality} & {\bf injury}\\
\hline
intercept & $-11.3^{-9.00}_{-13.5}$ &
$-3.50^{-3.17}_{-3.84}$ &
$-12.0^{-9.75}_{-14.6}$ &
$-3.57^{-3.23}_{-3.90}$\\
\hline
nigh & $1.36^{2.05}_{.665}$ &
$.583^{.796}_{.370}$ &
$1.35^{2.02}_{.599}$ &
$.594^{.805}_{.379}$\\
\hline
driv & $.736^{1.28}_{.196}$ &
$.139^{.244}_{.0344}$ &
$.725^{1.26}_{.187}$ &
$.136^{.240}_{.0309}$\\
\hline
dark & $.365^{.510}_{.220}$ &
$.365^{.510}_{.220}$ &
$.355^{.499}_{.209}$ &
$.355^{.499}_{.209}$\\
\hline
veh & $-.815^{-.499}_{-1.13}$ &
$-.815^{-.499}_{-1.13}$ &
$-.825^{-.518}_{-1.15}$ &
$-.825^{-.518}_{-1.15}$\\
\hline
hl20 & $1.81^{2.72}_{.894}$ &
$.701^{.810}_{.591}$ &
$2.43^{3.83}_{1.36}$ &
$.749^{.863}_{.637}$\\
\hline
moto & $2.60^{3.16}_{2.03}$ &
$2.60^{3.16}_{2.03}$ &
$2.59^{3.18}_{2.03}$ &
$2.59^{3.18}_{2.03}$\\
\hline
$X_{29}$ & $.0629^{.0997}_{.0261}$ &
$.0144^{.0199}_{.00890}$ &
$.0646^{.103}_{.0298}$ &
$.0146^{.0201}_{.00906}$\\
\hline
$X_{33}$ & $2.95^{3.95}_{1.94}$ &
$1.28^{1.82}_{.743}$ &
$2.88^{3.86}_{1.76}$ &
$1.28^{1.82}_{.734}$\\
\hline
$X_{35}$ & $.168^{.285}_{.0500}$ &
$.168^{.285}_{.0500}$ &
$.169^{.053}_{.286}$ &
$.169^{.053}_{.286}$\\
\hline
oldvage & $.0323^{.0416}_{.0230}$ &
$.0323^{.0416}_{.0230}$ &
$.0323^{.0416}_{.0230}$ &
$.0323^{.0416}_{.0230}$\\
\hline
maxpass & $.0563^{.0855}_{.0271}$ &
$.0563^{.0855}_{.0271}$ &
$.0568^{.0866}_{.0276}$ &
$.0568^{.0866}_{.0276}$\\
\hline
mm & -- &
$-.208^{.0911}_{-.325}$ &
-- &
$-.208^{.0914}_{-.325}$\\
\hline
$\langle P_{t,n}^{(i)}\rangle_X$ & -- &
-- &
$.00443$ &
$.149$\\
\hline
$p_{0\rightarrow1}$ & \multicolumn{2}{|c|}{--} &
\multicolumn{2}{|c|}{--}\\
\hline
$p_{1\rightarrow0}$ & \multicolumn{2}{|c|}{--} &
\multicolumn{2}{|c|}{--}\\
\hline
${\bar p}_0$ and ${\bar p}_1$ & \multicolumn{2}{|c|}{--} &
\multicolumn{2}{|c|}{--}\\
\hline
\hline
$\#$ free par. & \multicolumn{2}{|c|}{$19$} &
\multicolumn{2}{|c|}{$19$}\\
\hline
averaged $LL\!$ & \multicolumn{2}{|c|}{--} &
\multicolumn{2}{|c|}{$-6704.58^{-6699.51}_{-6711.54}$}\\
\hline
max$(LL)$ & \multicolumn{2}{|c|}{$-6704.47\,{\rm(true)}\!$} &
\multicolumn{2}{|c|}{$-6696.12\,{\rm(observed)}\!$}\\
\hline
marginal $LL\!$ & \multicolumn{2}{|c|}{--} &
\multicolumn{2}{|c|}{$6717.06^{-6711.07}_{-6717.28}$}\\
\hline
Good.-of-fit & \multicolumn{2}{|c|}{--} &
\multicolumn{2}{|c|}{$0.536$}\\
\hline
max(PSRF) & \multicolumn{2}{|c|}{--} &
\multicolumn{2}{|c|}{$1.00326$}\\
\hline
MPSRF & \multicolumn{2}{|c|}{--} &
\multicolumn{2}{|c|}{$1.00567$}\\
\hline
%accept. rate & \multicolumn{2}{|c|}{--} &
%\multicolumn{2}{|c|}{$.293\leq {\rm rate}\leq.299$}\\
%\hline
$\#$ observ. & \multicolumn{4}{|c|}
{accid.=fatal.+inj.+PDO:\quad$15656=72+2329+13255$}\\
\hline
\end{tabular}

%% file: severity_US_two-veh_table.tex
\begin{tabular}{|l|c|c|c|c|}
\hline
& \multicolumn{2}{|c|}{$\bf\displaystyle ML\mbox{\bf-}by\mbox{\bf-}MLE$} & \multicolumn{2}{|c|}{$\bf\displaystyle ML\mbox{\bf-}by\mbox{\bf-}MCMC$}\\
\cline{2-5}
${}^{{}^{\bf\displaystyle Variable}}$ & {\bf fatality} & {\bf injury} &
{\bf fatality} & {\bf injury}\\
\hline
intercept & $-10.3^{-8.78}_{-11.7}$ &
$-3.06^{-2.78}_{-3.34}$ &
$-10.4^{-8.91}_{-11.8}$ &
$-3.11^{-2.83}_{-3.40}$\\
\hline
wint & $-.0962^{-.0290}_{-.163}$ &
$-.0962^{-.0290}_{-.163}$ &
$-.0952^{-.0287}_{-.162}$ &
$-.0952^{-.0287}_{-.162}$\\
\hline
wday & $.0761^{-.00950}_{-.143}$ &
$.0761^{-.00950}_{-.143}$ &
$-.0725^{-.00654}_{-.139}$ &
$-.0725^{-.00654}_{-.139}$\\
\hline
dayt & $-.427^{-.110}_{-.744}$ &
$-.126^{-.0668}_{-.185}$ &
$-.422^{-.105}_{-.737}$ &
$-.121^{-.0619}_{-.179}$\\
\hline
$X_{12}$ & $-1.35^{-.955}_{-1.75}$ &
$-.313^{-.241}_{-.385}$ &
$-1.36^{-.972}_{-1.77}$ &
$-.320^{-.248}_{-.392}$\\
\hline
dark & $.546^{.931}_{.161}$ &
$.115^{.229}_{-.00220}$ &
$.543^{.926}_{.156}$ &
$.115^{.227}_{-.00229}$\\
\hline
snow & $-.259^{-.0903}_{-.428}$ &
$-.259^{-.0903}_{-.428}$ &
$-.262^{-.0952}_{-.431}$ &
$-.262^{-.0952}_{-.431}$\\
\hline
driv & $.0600^{.118}_{-.00240}$ &
$.0600^{.118}_{-.00240}$ &
$.0556^{.112}_{-.00157}$ &
$.0556^{.112}_{-.00157}$\\
\hline
nojun & $.302^{.582}_{.0216}$ &
$-.214^{-.158}_{-.269}$ &
$.0303^{.583}_{.0263}$ &
$-.213^{-.158}_{-.269}$\\
\hline
driver & $.426^{.571}_{.280}$ &
$.426^{.571}_{.280}$ &
$.428^{.573}_{.285}$ &
$.428^{.573}_{.285}$\\
\hline
hl10 & $.541^{.835}_{.247}$ &
$.652^{.718}_{.586}$ &
$.564^{.867}_{.268}$ &
$.687^{.756}_{.618}$\\
\hline
moto & $3.98^{4.62}_{3.35}$ &
$1.88^{2.24}_{1.51}$ &
$3.97^{4.60}_{3.31}$ &
$1.88^{2.25}_{1.51}$\\
\hline
vage & $.0483^{.0709}_{.0258}$ &
-- &
$.0482^{.0705}_{.0254}$ &
--\\
\hline
$X_{29}$ & $.0749^{.0999}_{.0498}$ &
$.0231^{.0268}_{.0194}$ &
$.0757^{.101}_{.0511}$ &
$.0233^{.0270}_{.0196}$\\
\hline
priv & $-1.13^{-.540}_{-1.73}$ &
$-1.13^{-.540}_{-1.73}$ &
$-1.18^{-.607}_{-1.81}$ &
$-1.18^{-.607}_{-1.81}$\\
\hline
$X_{33}$ & $2.98^{3.64}_{2.32}$ &
$1.40^{1.76}_{1.03}$ &
$2.97^{3.62}_{2.28}$ &
$1.39^{1.76}_{1.03}$\\
\hline
singTR & $1.14^{1.44}_{.843}$ &
-- &
$1.15^{1.44}_{.843}$ &
--\\
\hline
maxpass & $.0776^{.0979}_{.0572}$ &
$.0776^{.0979}_{.0572}$ &
$.0784^{.0991}_{.0583}$ &
$.0784^{.0991}_{.0583}$\\
\hline
olddrv & $.0198^{.0287}_{.0110}$ &
$.0230^{.0283}_{.0177}$ &
$.0199^{.0286}_{.0110}$ &
$.00481^{.00648}_{.00314}$\\
\hline
mm & $.316^{.598}_{.0343}$ &
$.00480^{.00650}_{.00320}$ &
$.321^{.602}_{.0417}$ &
$-.230^{-.172}_{-.289}$\\
\hline
oldvage & -- &
$-.234^{-.175}_{-.292}$ &
-- &
$.0230^{.0283}_{.0177}$\\
\hline
$\langle P_{t,n}^{(i)}\rangle_X$ & -- &
-- &
$.00759$ &
$.255$\\
\hline
$p_{0\rightarrow1}$ & \multicolumn{2}{|c|}{--} &
\multicolumn{2}{|c|}{--}\\
\hline
$p_{1\rightarrow0}$ & \multicolumn{2}{|c|}{--} &
\multicolumn{2}{|c|}{--}\\
\hline
${\bar p}_0$ and ${\bar p}_1$ & \multicolumn{2}{|c|}{--} &
\multicolumn{2}{|c|}{--}\\
\hline
\hline
$\#$ free par. & \multicolumn{2}{|c|}{$32$} &
\multicolumn{2}{|c|}{$32$}\\
\hline
averaged $LL\!$ & \multicolumn{2}{|c|}{--} &
\multicolumn{2}{|c|}{$-16535.45^{-16528.62}_{-16544.16}$}\\
\hline
max$(LL)$ & \multicolumn{2}{|c|}{$-16527.94\,{\rm(true)}\!$} &
\multicolumn{2}{|c|}{$--16522.89\,{\rm(observed)}\!$}\\
\hline
marginal $LL\!$ & \multicolumn{2}{|c|}{--} &
\multicolumn{2}{|c|}{$-16549.59^{16544.60}_{16551.83}$}\\
\hline
Good.-of-fit & \multicolumn{2}{|c|}{--} &
\multicolumn{2}{|c|}{$0.372$}\\
\hline
max(PSRF) & \multicolumn{2}{|c|}{--} &
\multicolumn{2}{|c|}{$1.00275$}\\
\hline
MPSRF & \multicolumn{2}{|c|}{--} &
\multicolumn{2}{|c|}{$1.00358$}\\
\hline
%accept. rate & \multicolumn{2}{|c|}{--} &
%\multicolumn{2}{|c|}{$.293\leq {\rm rate}\leq.300$}\\
%\hline
$\#$ observ. & \multicolumn{4}{|c|}
{accid.=fatal.+inj.+PDO:\quad$28259=222+7285+21022$}\\
\hline
\end{tabular}

%% file: severity_state_two-veh_table.tex
\begin{tabular}{|l|c|c|c|c|}
\hline
& \multicolumn{2}{|c|}{$\bf\displaystyle ML\mbox{\bf-}by\mbox{\bf-}MLE$} & \multicolumn{2}{|c|}{$\bf\displaystyle ML\mbox{\bf-}by\mbox{\bf-}MCMC$}\\
\cline{2-5}
${}^{{}^{\bf\displaystyle Variable}}$ & {\bf fatality} & {\bf injury} &
{\bf fatality} & {\bf injury}\\
\hline
intercept & $-13.1^{-11.6}_{-14.5}$ &
$-3.65^{-3.37}_{-3.94}$ &
$-13.2^{-11.8}_{-14.6}$ &
$-3.75^{-3.47}_{-4.03}$\\
\hline
wint & $-.0668^{-.00790}_{-.126}$ &
$-.0668^{-.00790}_{-.126}$ &
$-.0669^{-.00888}_{-.126}$ &
$-.0669^{-.00888}_{-.126}$\\
\hline
wday & $-.133^{-.0737}_{-.192}$ &
$-.133^{-.0737}_{-.192}$ &
$-.132^{-.0727}_{-.191}$ &
$-.132^{-.0727}_{-.191}$\\
\hline
$X_{12}$ & $-.787^{-.448}_{-1.13}$ &
$-.251^{-.189}_{-.313}$ &
$-.796^{-.462}_{-1.14}$ &
$-.262^{-.201}_{-.324}$\\
\hline
dark & $1.07^{1.35}_{.794}$ &
$.248^{.338}_{.158}$ &
$1.07^{1.34}_{.787}$ &
$.248^{.338}_{.158}$\\
\hline
wall & $-2.01^{-.0430}_{-3.98}$ &
-- &
$-2.56^{-.708}_{-5.48}$ &
--\\
\hline
nojun & $.385^{.627}_{.142}$ &
$-.170^{-.121}_{-.219}$ &
$.383^{.627}_{.142}$ &
$-.172^{-.123}_{-.221}$\\
\hline
curve & $1.01^{1.30}_{.715}$ &
$.234^{.323}_{.145}$ &
$1.00^{1.29}_{.705}$ &
$.239^{.327}_{.150}$\\
\hline
driver & $1.07^{1.68}_{.450}$ &
$.422^{.542}_{.301}$ &
$1.11^{1.78}_{.521}$ &
$.418^{.539}_{.299}$\\
\hline
hl20 & $1.21^{1.64}_{.777}$ &
$.725^{.810}_{.640}$ &
$1.22^{1.70}_{.780}$ &
$.885^{.981}_{.789}$\\
\hline
moto & $2.92^{3.51}_{2.33}$ &
$1.97^{2.25}_{1.68}$ &
$2.92^{3.50}_{2.31}$ &
$1.97^{2.27}_{1.69}$\\
\hline
$X_{29}$ & $.0942^{.115}_{.0734}$ &
$.0246^{.0277}_{.0215}$ &
$.0950^{.116}_{.0749}$ &
$.0249^{.0280}_{.0218}$\\
\hline
priv & $-.856^{-.378}_{-1.33}$ &
$-.856^{-.378}_{-1.33}$ &
$-.881^{-.421}_{-1.39}$ &
$-.881^{-.421}_{-1.39}$\\
\hline
$X_{33}$ & $3.10^{3.65}_{2.55}$ &
$1.26^{1.58}_{.947}$ &
$3.10^{3.64}_{2.54}$ &
$1.27^{1.59}_{.950}$\\
\hline
$X_{35}$ & $.380^{.739}_{.0206}$ &
-- &
$.384^{.743}_{.0324}$ &
--\\
\hline
singTR & $1.00^{1.28}_{0.726}$ &
$-.114^{-.0215}_{-.206}$ &
$1.00^{1.27}_{.722}$ &
$-.113^{-.0224}_{-.206}$\\
\hline
voldo & $.255^{.309}_{.201}$ &
$.255^{.309}_{.201}$ &
$.254^{.308}_{.200}$ &
$.254^{.308}_{.200}$\\
\hline
maxpass & $.0536^{.0683}_{.0389}$ &
$.0536^{.0683}_{.0389}$ &
$.0544^{.0693}_{.0398}$ &
$.0544^{.0693}_{.0398}$\\
\hline
olddrv & $.0212^{.0284}_{.0140}$ &
$.00450^{.00600}_{.00310}$ &
$.0212^{.0284}_{.0140}$ &
$.0.450^{.00595}_{.00306}$\\
\hline
mm & $.625^{.962}_{.288}$ &
$-.177^{-.125}_{-.229}$ &
$.633^{.975}_{.305}$ &
$-.177^{-.124}_{-.230}$\\
\hline
nocons & -- &
$.280^{.427}_{.133}$ &
-- &
$.280^{.428}_{.136}$\\
\hline
driver & -- &
$.454^{.743}_{.166}$ &
-- &
$.460^{.745}_{.170}$\\
\hline
$\langle P_{t,n}^{(i)}\rangle_X$ & -- &
-- &
$.00843$ &
$.257$\\
\hline
$p_{0\rightarrow1}$ & \multicolumn{2}{|c|}{--} &
\multicolumn{2}{|c|}{--}\\
\hline
$p_{1\rightarrow0}$ & \multicolumn{2}{|c|}{--} &
\multicolumn{2}{|c|}{--}\\
\hline
${\bar p}_0$ and ${\bar p}_1$ & \multicolumn{2}{|c|}{--} &
\multicolumn{2}{|c|}{--}\\
\hline
\hline
$\#$ free par. & \multicolumn{2}{|c|}{$35$} &
\multicolumn{2}{|c|}{$35$}\\
\hline
averaged $LL\!$ & \multicolumn{2}{|c|}{--} &
\multicolumn{2}{|c|}{$-21088.31^{-21081.09}_{-21097.38}$}\\
\hline
max$(LL)$ & \multicolumn{2}{|c|}{$-21096.20\,{\rm(true)}\!$} &
\multicolumn{2}{|c|}{$-21074.01\,{\rm(observed)}\!$}\\
\hline
marginal $LL\!$ & \multicolumn{2}{|c|}{--} &
\multicolumn{2}{|c|}{$-21103.71^{-21097.88}_{-21105.96}$}\\
\hline
Good.-of-fit & \multicolumn{2}{|c|}{--} &
\multicolumn{2}{|c|}{$0.635$}\\
\hline
max(PSRF) & \multicolumn{2}{|c|}{--} &
\multicolumn{2}{|c|}{$1.00141$}\\
\hline
MPSRF & \multicolumn{2}{|c|}{--} &
\multicolumn{2}{|c|}{$1.00176$}\\
\hline
%accept. rate & \multicolumn{2}{|c|}{--} &
%\multicolumn{2}{|c|}{$.294\leq {\rm rate}\leq.299$}\\
%\hline
$\#$ observ. & \multicolumn{4}{|c|}
{accid.=fatal.+inj.+PDO:\quad$36136=311+9276+26549$}\\
\hline
\end{tabular}

%% file: severity_county_two-veh_table.tex
\begin{tabular}{|l|c|c|c|c|}
\hline
& \multicolumn{2}{|c|}{$\bf\displaystyle ML\mbox{\bf-}by\mbox{\bf-}MLE$} & \multicolumn{2}{|c|}{$\bf\displaystyle ML\mbox{\bf-}by\mbox{\bf-}MCMC$}\\
\cline{2-5}
${}^{{}^{\bf\displaystyle Variable}}$ & {\bf fatality} & {\bf injury} &
{\bf fatality} & {\bf injury}\\
\hline
intercept & $-10.6^{-9.49}_{-11.8}$ &
$-3.50^{-3.29}_{-3.72}$ &
$-10.7^{-9.61}_{-11.9}$ &
$-3.58^{-3.37}_{-3.80}$\\
\hline
wint & $-.145^{-.0756}_{-.214}$ &
$-.145^{-.0756}_{-.214}$ &
$-.146^{-.0774}_{-.216}$ &
$-.146^{-.0774}_{-.216}$\\
\hline
sund & $.192^{.290}_{.0945}$ &
$.192^{.290}_{.0945}$ &
$.190^{.287}_{.0927}$ &
$.190^{.287}_{.0927}$\\
\hline
morn & $-.108^{-.0276}_{-.188}$ &
$-.108^{-.0276}_{-.188}$ &
$-.101^{-.0215}_{-.181}$ &
$-.101^{-.0215}_{-.181}$\\
\hline
$X_{12}$ & $-1.48^{-.647}_{-2.31}$ &
$-.160^{-.0794}_{-.242}$ &
$-1.56^{-.777}_{-2.50}$ &
$-1.65^{-.0841}_{-.246}$\\
\hline
darklamp & $-.197^{-.0239}_{-.371}$ &
$-.197^{-.0239}_{-.371}$ &
$-.204^{-.0342}_{-.377}$ &
$-.204^{-.0342}_{-.377}$\\
\hline
way4 & $.249^{.342}_{.216}$ &
$.249^{.342}_{.216}$ &
$.279^{.342}_{.215}$ &
$.279^{.342}_{.215}$\\
\hline
driver & $.247^{.370}_{.125}$ &
$.247^{.370}_{.125}$ &
$.258^{.382}_{.137}$ &
$.258^{.382}_{.137}$\\
\hline
hl20 & $1.58^{2.11}_{1.04}$ &
$.914^{.993}_{.836}$ &
$1.60^{2.18}_{1.07}$ &
$.957^{1.04}_{.875}$\\
\hline
moto & $4.04^{4.67}_{3.40}$ &
$2.19^{2.58}_{1.80}$ &
$4.04^{4.67}_{3.38}$ &
$2.21^{2.61}_{1.82}$\\
\hline
$X_{29}$ & $.0813^{.101}_{.0615}$ &
$.0287^{.0320}_{.0253}$ &
$.0820^{.102}_{.0627}$ &
$.0290^{.0324}_{.0257}$\\
\hline
$X_{33}$ & $2.82^{3.58}_{2.06}$ &
$1.18^{1.56}_{.794}$ &
$2.77^{3.51}_{1.96}$ &
$1.17^{1.56}_{.787}$\\
\hline
singSUV & $.471^{.778}_{.163}$ &
-- &
$.471^{.780}_{.166}$ &
--\\
\hline
oldvage & $.0390^{.0630}_{.0151}$ &
$.0215^{.0269}_{.0162}$ &
$.0387^{.0621}_{.0145}$ &
$.0217^{.0270}_{.0163}$\\
\hline
age0 & -- &
$.142^{.230}_{.0534}$ &
-- &
$.143^{.231}_{.0552}$\\
\hline
singTR & -- &
$-.174^{-.0454}_{-.303}$ &
-- &
$-.173^{-.0461}_{-.302}$\\
\hline
maxpass & -- &
$.0176^{.0286}_{.00670}$ &
-- &
$.0179^{.0288}_{.00685}$\\
\hline
age0o & -- &
$-.575^{-.335}_{-.815}$ &
-- &
$-.585^{-.347}_{-.829}$\\
\hline
mm & -- &
$-.258^{-.194}_{-.322}$ &
-- &
$-.258^{-.194}_{-.322}$\\
\hline
$\langle P_{t,n}^{(i)}\rangle_X$ & -- &
-- &
$.00662$ &
$.247$\\
\hline
$p_{0\rightarrow1}$ & \multicolumn{2}{|c|}{--} &
\multicolumn{2}{|c|}{--}\\
\hline
$p_{1\rightarrow0}$ & \multicolumn{2}{|c|}{--} &
\multicolumn{2}{|c|}{--}\\
\hline
${\bar p}_0$ and ${\bar p}_1$ & \multicolumn{2}{|c|}{--} &
\multicolumn{2}{|c|}{--}\\
\hline
\hline
$\#$ free par. & \multicolumn{2}{|c|}{$26$} &
\multicolumn{2}{|c|}{$26$}\\
\hline
averaged $LL\!$ & \multicolumn{2}{|c|}{--} &
\multicolumn{2}{|c|}{$-14423.80^{-14417.75}_{-14431.72}$}\\
\hline
max$(LL)$ & \multicolumn{2}{|c|}{$-14411.12\,{\rm(true)}\!$} &
\multicolumn{2}{|c|}{$-14412.78\,{\rm(observed)}\!$}\\
\hline
marginal $LL\!$ & \multicolumn{2}{|c|}{--} &
\multicolumn{2}{|c|}{$-14434.79^{-14431.73}_{-14437.04}$}\\
\hline
Good.-of-fit & \multicolumn{2}{|c|}{--} &
\multicolumn{2}{|c|}{$0.370$}\\
\hline
max(PSRF) & \multicolumn{2}{|c|}{--} &
\multicolumn{2}{|c|}{$1.00141$}\\
\hline
MPSRF & \multicolumn{2}{|c|}{--} &
\multicolumn{2}{|c|}{$1.00225$}\\
\hline
%accept. rate & \multicolumn{2}{|c|}{--} &
%\multicolumn{2}{|c|}{$.294\leq {\rm rate}\leq.298$}\\
%\hline
$\#$ observ. & \multicolumn{4}{|c|}
{accid.=fatal.+inj.+PDO:\quad$25597=173+6315+19109$}\\
\hline
\end{tabular}

%% file: ch_vita.tex
%
% This is vita of Nataliya Malyshkina
%

\begin{vita}

Nataliya V. Malyshkina was born in Ekaterinburg (Yekaterinburg), Russia on September 6, 1978. Based on excellent results of entrance examinations, she was admitted as a student to Ural State University of Railroad Transportation at the age of 15 (normal admission age in Russia is 17). In 1999 she graduated with a Master Diploma from the Department of Railroad Transportation Planning and Operations at this university. She joined this department as a full-time teacher and lecturer immediately after the graduation. In August 2005 Nataliya joined the School of Civil Engineering at Purdue University as a graduate student and research assistant. In December 2006 she received her Master of Science in Civil Engineering from Purdue University. She has an affinity for statistics, econometrics, microeconomics, mathematical and numerical modeling, programming. Although Nataliya's recent work has mostly been focused on roadway safety, her research interests are broad and include transportation systems analysis, modeling and planning, transportation economics and management, traffic operations and control. Nataliya's hobbies include classical literature and music, chess, swimming, bicycling, hiking.

\end{vita}